\documentclass[
aip,
amsmath,
amssymb, 
preprint]{revtex4-1}

\usepackage{graphicx}
\usepackage{dcolumn}
\usepackage{bm}

\usepackage[utf8]{inputenc}
\usepackage[T1]{fontenc}
\usepackage{mathptmx}
\usepackage[cal=cm]{mathalfa}
\usepackage{bbold}
\usepackage{dsfont}
\usepackage{cases}
\usepackage{xcolor}
\usepackage{ulem}
\usepackage{framed}
\usepackage{longtable}

\usepackage{enumitem}
\usepackage{mathtools}

\newcommand{\revS}[1]{{\color{black}{#1}}}

\newcommand*{\ampd}[2]{
    t^{#1}(#2)
}

\newcommand*{\lhs}[1]{
    \eta^{#1}
}
\newcommand*{\ampdr}[2]{
    t^{#1}_\Re(#2)
}

\newcommand*{\ampdi}[2]{
    t^{#1}_\Im(#2)
}

\newcommand*{\muld}[2]{
    {\bar{t}}^{#1}(#2)
}

\begin{document}

\preprint{CPP-RICC2}

\title{Damped (linear) response theory within the resolution-of-identity coupled cluster singles and approximate doubles (RI-CC2) method.}

\author{Daniil A. Fedotov}
\affiliation{DTU Chemistry, Technical University of Denmark, Kemitorvet Bldg 207, DK-2800 Kongens Lyngby, Denmark}%
\author{Sonia Coriani}
\email{soco@kemi.dtu.dk}
\affiliation{DTU Chemistry, Technical University of Denmark, Kemitorvet Bldg 207, DK-2800 Kongens Lyngby, Denmark}
\author{Christof H{\"a}ttig}
\email{christof.haettig@rub.de}
\affiliation{Arbeitsgruppe Quantenchemie, Ruhr-Universit{\"a}t, Bochum D-44780, Germany}

\date{\today}

\begin{abstract}
An implementation of 
a complex solver for the solution of the response equations required to compute the complex response functions of damped response theory is presented for the resolution-of-identity (RI) coupled cluster singles and approximate doubles CC2 method.
The implementation uses a partitioned formulation that avoids the storage of double excitation amplitudes to make it applicable to large molecules.
The solver is the keystone element for the development of 
the damped coupled cluster response
formalism for linear and nonlinear effects in resonant frequency regions at the RI-CC2 level of theory.
Illustrative results are reported
for the 
one-photon absorption cross section 
of C$_{60}$,
the electronic circular dichroism of
$n$-helicenes ($n$ = 5,6,7),
and the $C_6$ dispersion coefficients of
a set of selected organic molecules and fullerenes.
\end{abstract}

\maketitle

\section{Introduction}
Damped response theory~\cite{DampedResponse}
and the conceptually equivalent complex polarization propagator (CPP) approach~\cite{Norman:CPP:2001,Norman:CPP:2005,Jensen:CPP:2005,Norman:CPP:Xray:2006,Norman:CPP:Xray:2006:1,CPP:XTPA:Norman2016,Lanczos1,Lanczos2,CPP:RIXS:CC,Norman:PCCP:2011} are increasingly popular frameworks to compute 
resonance convergent response functions, and thereby simulate a variety of spectroscopic effects. They have proven particular convenient in cases where traditional stick-spectra-based approaches are impractical due to a large density of states -- a prototypical example being the absorption spectrum of large molecules in extended basis sets over a broad frequency range.
They are also advantageous in that they give access to molecular (stick) properties at (additional) resonant conditions, like in two-photon absorption (TPA)~\cite{CPP:TPA,CPP:XTPA:Norman2016} and resonant inelastic x-ray scattering (RIXS).~\cite{CPP:RIXS:ADC,CPP:RIXS:CC} {In addition, the CPP approach can be used to compute,  e.g.,  polarizabilities at imaginary frequencies that are needed for the calculation of} $C_6$ dispersion coefficients.\cite{Lanczos1,CPP:C6:Jiemchooroj,CPP:ADC}

Damped response/CPP frameworks have been successfully implemented at various levels of theory, from Hartree-Fock and Time-Dependent Density Functional theory,~\cite{Norman:CPP:2001,Norman:CPP:2005,Jensen:CPP:2005,CPP:Joanna:solver2} to Multiconfigurational Self-Consistent Field,~\cite{Norman:CPP:2001,Norman:CPP:2005} Algebraic Diagrammatic Construction (ADC)~\cite{CPP:ADC,MaxADC} and Coupled-Cluster (CC) Theory.~\cite{Lanczos1,Lanczos2,CPP:Kauczor:CC,CPP:RIXS:CC}
Extensions to solvated environments {(embedding and solvation models)}~\cite{CPP:MCD:PE,CPP:PE:MCD} 
and the relativistic domain~\cite{CPP:4components,damped_4components} have also been presented.

Applications to date include linear properties like one-photon absorption (OPA) and electronic circular dichroism (ECD) in different frequency regions (from UV to X-ray),~\cite{Norman:CPP:Xray:2006,Norman:CPP:Xray:2006:1,Lanczos2,CPP:Kauczor:CC,CPP:ECD,CVS-CPP:1}
and $C_6$ dispersion coefficients computed from  polarizabilities at imaginary frequencies,~\cite{Lanczos1,CPP:C6:Jiemchooroj,CPP:ADC}
to non-linear effects like magnetic-field induced (MCD) and nuclear-spin induced (NSCD) circular dichroism,~\cite{CPP:MCD,CPP:MCD:Nucleic,CPP:NSCD}
magneto-chiral dichroism (MChD) and birefringence (MChB) dispersion,~\cite{CPP:MCHD}
two-photon absorption in both UV-vis and X-ray regimes,~\cite{CPP:TPA,CPP:XTPA:Norman2016} and
Resonant Inelastic X-ray Scattering.~\cite{CPP:RIXS:ADC,CPP:RIXS:CC,CVS-CPP:1,NandaPCCP} 

A keystone element of all damped response/CPP frameworks is the solution of the (linear) response equations for a complex, or damped, frequency.~\cite{Norman:PCCP:2011} 
An implementation of a complex linear response solver within a coupled cluster framework was presented by \citeauthor{CPP:Kauczor:CC}\cite{CPP:Kauczor:CC}
for the response of the cluster amplitudes, and later extended
to the response of the Lagrange multipliers by 
\citeauthor{CPP:RIXS:CC},\cite{CPP:RIXS:CC}  in both cases using an algorithm that assumes the storage of the amplitudes or multipliers for all excitation classes (\textit{vide infra}).
Specific strategies to eliminate convergence issues in the X-ray frequency range have been discussed by \citeauthor{CVS-CPP:1}\cite{CVS-CPP:1} and by \citeauthor{NandaPCCP}\cite{NandaPCCP}
Here, we extend the complex solver of Ref.~\citenum{CPP:Kauczor:CC} to the case of the
Resolution-of-Identity (RI) Coupled Cluster Singles and Approximate method CC2 as implemented in the Turbomole package,~\cite{turbomole2014,Turbomole2020} 
{which employs a partitioned formulation which avoids the storage of amplitudes and multipliers for double excitation. 
This is important for large scale applications of CC2 which would otherwise be hampered by I/O and storage demands.} 
As illustrative results we report UV-vis one-photon absorption (OPA) spectra of 
C$_{60}$, the electronic circular dichroism (ECD) 
spectra of three helicenes, 
and the ground-state $C_6$ dispersion coefficients of a set of organic molecules previously studied in the literature with other ab initio methods.


\section{Theory}

\subsection{The CC complex linear response function: definitions and properties of interest}
\label{Properties}

In CC damped linear response theory,~\cite{Lanczos1,Lanczos2,CPP:Kauczor:CC,CPP:RIXS:CC} we compute the complex polarizability as:
\begin{equation}
\begin{aligned}
\langle\langle x; y\rangle\rangle_{\omega+i\gamma}
= \frac{1}{2}{\hat{C}}^{\pm\omega}\big\{
\lhs{x} \ampd{y}{\omega+i\gamma} + 
\lhs{y} \ampd{x}{-\omega-i\gamma}+
{\mathbf{F}}\ampd{y}{\omega+i\gamma}\ampd{x}{-\omega-i\gamma}
\big\}
~.
\end{aligned}
\label{cmplxLR}
\end{equation}
{where $\hat{C}$
is a symmetrization operator defined as
$\hat{C}^{\pm\omega}f(\omega) = {f(\omega) + f^*(-\omega)}$. Note that the symmetrization operator only turns the sign of the real frequency $\omega$.
}
We refer to, {\it{e.g.}}, Ref.~\citenum{Christiansen:1998} for the general definitions of the $\mathbf{F}$ matrix and $\lhs{y}$ vectors in CC response theory.
The solution of the response equations yielding the amplitudes $\ampd{y}{\omega+i\gamma}$ within the RI-CC2 framework is discussed in the next section.
{Here we only note that $\mathbf{F}$ and  
{(for real operators $x$ and $y$)} also $\lhs{y}$
are purely real while the amplitude responses fulfil the symmetry:
\begin{equation}
    \ampd{x}{\omega + i\gamma}^\ast = \ampd{x}{\omega -i\gamma} ~.
\end{equation}}%
If both operators are real and only diagonal components are considered,
the real and imaginary parts of the complex dipole-dipole polarizability in Eq.~\eqref{cmplxLR} are
\begin{multline}
 \Re\langle\langle x; x \rangle\rangle_{\omega+i\gamma} =
\lhs{x}_{\Re} \ \ampdr{x}{\omega+i\gamma} + 
\lhs{x}_{\Re} \ \ampdr{x}{-\omega-i\gamma} \\
+ 
{\mathbf{F}}\ampdr{x}{\omega+i\gamma}
\ampdr{x}{-\omega-i\gamma}
- {\mathbf{F}}\ampdi{x}{-\omega-i\gamma}\ampdi{x}{\omega+i\gamma} 
\end{multline}
\begin{multline}
\Im\langle\langle x; x \rangle\rangle_{\omega+i\gamma}
=
\lhs{x}_{\Re} \ \ampdi{x}{\omega+i\gamma} + 
\lhs{x}_{\Re} \ \ampdi{x}{-\omega-i\gamma}
\\
+ 
{\mathbf{F}}\ampdi{x}{\omega+i\gamma}\ampdr{x}{-\omega-i\gamma}+
{\mathbf{F}}\ampdi{x}{-\omega-i\gamma}\ampdr{x}{\omega+i\gamma}
\end{multline}
where we have explicitly split the complex response amplitudes into real and imaginary parts
\begin{equation}
    \ampd{x}{\omega+i\gamma} = \ampdr{x}{\omega+i\gamma}
    + i~\ \ampdi{x}{\omega+i\gamma}~.
\end{equation}
The imaginary {part of the} polarizability can be used to compute, for instance, one-photon absorption (OPA) cross sections:
\begin{equation}
\label{XASgen}
    \sigma_{\rm{OPA}}(\omega)
    \propto \omega \ \Im \langle\langle \mu_\alpha; \mu_\alpha\rangle\rangle_{\omega+i\gamma}
\end{equation}
where $\mu_\alpha$ 
is the $\alpha$-component of the electric dipole operator, and the incident  frequency $\omega$ is chosen within the
specific region of interest, {\it e.g.} UV-vis or X-ray.
The polarizability dispersion profiles,
illustrating the variation of the dipole polarizability over a given frequency range, 
can conversely be obtained from the real part of the complex dipole polarizability.

If one of the two operators in the linear response function, say
${\mathcal{X}}$,
is purely imaginary,
we have
\begin{equation}
    \ampd{\mathcal{X}}{\omega+i\gamma}^\ast = -\ampd{\mathcal{X}}{\omega-i\gamma} 
\end{equation}
and it is the real part of the complex response function that yields the absorption 
component
\begin{equation}
\begin{aligned}
    \Re \langle\langle x; {\mathcal{X}} \rangle\rangle_{\omega+i \gamma}
    =& \tfrac{1}{2} \Big\{ 
    \lhs{x}_{\Re}
    \ 
    \ampdr{{\mathcal{X}}}{\omega+i\gamma}
    -\lhs{x}_{\Re} \ \ampdr{{\mathcal{X}}}{-\omega-i\gamma}\\&
    -\lhs{{\mathcal{X}}}_{\Im} \ \ampdi{x}{-\omega-i\gamma}
    +\lhs{{\mathcal{X}}}_{\Im} \ \ampdi{x}{\omega+i\gamma}\\&
    +{\mathbf{F}}\ampdr{x}{-\omega-i\gamma}\ampdr{{\mathcal{X}}}{\omega+i\gamma}
    +{\mathbf{F}}\ampdr{x}{\omega+i\gamma}\ampdr{{\mathcal{X}}}{-\omega-i\gamma}\\&
    -{\mathbf{F}}\ampdi{x}{-\omega-i\gamma}\ampdi{{\mathcal{X}}}{\omega+i\gamma}
    -{\mathbf{F}}\ampdi{x}{\omega+i\gamma}\ampdi{{\mathcal{X}}}{-\omega-i\gamma}
    \Big\}
 .
\end{aligned}
\end{equation}
 A prototypical case described by such a response function is the electronic circular dichroism (ECD) cross section --
 most often expressed as difference $\Delta\epsilon$ in
 extinction coefficients for left and right circularly polarized light -- 
in the length gauge (lg)
\begin{equation}
    \Delta\epsilon^{\rm lg} (\omega)
    \propto \omega \ \Re \langle\langle m_{\alpha};\mu_{\alpha}\rangle\rangle_{\omega+i\gamma}
\end{equation}
whereas the optical rotation dispersion (ORD) profile is given by the imaginary part:
\begin{equation}
    \sigma^{\rm lg}_{\textrm{ORD}}(\omega) \propto \omega \ \Im \langle\langle m_{\alpha};\mu_{\alpha}\rangle\rangle_{\omega+i \gamma}~~.
\end{equation}
{We note in passing that, in cases like ORD and ECD, the symmetric form of the (complex) polarizability requires solving 
the complex response equations for both imaginary and real operators.
Alternatively, one can resort to the asymmetric form}
\begin{align}
    \langle\!\langle {\mathcal{X}}; x \rangle\!\rangle_{\omega+i \gamma} = \frac{1}{2}\hat{C}^{\pm\omega} \left\{
    \bar{t}^x(\omega +i \gamma) \xi^{\mathcal{X}} + \eta^{\mathcal{X}} t^x(\omega +i \gamma)\right\}~,
\end{align}
which thus requires the solution of the CPP equations for the left response multipliers $\bar{t}^x(\omega+i \gamma)$ 
\begin{align}
    \bar{t}^x(\omega+i \gamma) \Big[ \mathbf{A} + (\omega +i \gamma) \mathbf{1} \Big] = -  \eta^x - \mathbf{F} t^x(\omega+i \gamma) 
\end{align}
along with those for the response amplitudes $t^x(\omega+i \gamma)$ for the real operator $x$.
This allows one to bypass the solution of the response vectors for the imaginary operator. The first-order 
(complex) Lagrange multipliers are also needed for higher-order response and transition properties, like the previously mentioned two-photon absorption, RIXS and MCD.~\cite{CPP:RIXS:CC,CPP:MCD:CC}

{The length gauge expressions of the {optical rotation (OR)} tensor and of the rotatory strengths within resonant response theory are gauge-origin dependent. The velocity-gauge forms, that involve two imaginary operators, the linear momentum  $p_\alpha$ and  magnetic moment operator $m_\alpha$, are, on the other hand, origin independent.~\cite{warnke_furche:ECD} 
Within CC theory, the `modified' velocity gauge expression of the OR tensor is typically used\cite{PEDERSEN:2004:modvel,RICC2_OR_Friese}
\begin{equation}
    G^{\rm{mv}}_{\alpha\alpha}(\omega) = 
    \omega^{-1}
    \left\{
     \langle\langle 
    p_\alpha;
    m_{\alpha}
    \rangle\rangle_{\omega}
    -
    \langle\langle 
    p_\alpha;
    m_{\alpha}
    \rangle\rangle_{0}
    \right\}
\end{equation}
which ensures that the thus-computed OR tensor is zero in the limit of zero frequency, as it should be according to exact theory.
We generalise the above expression to obtain the CPP optical rotatory dispersion and the electronic circular dichroism 
in the modified velocity gauge
\begin{equation}
\label{velo1ord}
    \sigma^{\rm{mv}}_{\textrm{ORD}}(\omega) \propto 
    \Re \left\{
     \langle\langle 
    p_\alpha;
    m_{\alpha}
\rangle\rangle_{\omega+i\gamma}
    -
    \langle\langle 
    p_\alpha;
    m_{\alpha}
    \rangle\rangle_{0}
    \right\}
\end{equation}
and 
\begin{equation}
\label{velo1ecd}
    \Delta\epsilon^{\rm{mv}}(\omega)
    \propto 
    \Im \left\{
     \langle\langle 
    p_\alpha;
    m_{\alpha}
\rangle\rangle_{\omega+i\gamma}
    -
    \langle\langle 
    p_\alpha;
    m_{\alpha}
    \rangle\rangle_{0}
    \right\}
    = 
     \Im\langle\langle 
    p_\alpha;
    m_{\alpha}
\rangle\rangle_{\omega+i\gamma}
\end{equation}
}
{Note that the correction to the ECD expression is redundant, since the imaginary part of the (real) response function is zero at the static limit.
An alternative choice of CPP expression is to use a life-time parameter that is scaled with the real frequency: 
\begin{equation}
\label{velo2ord}
\langle\langle p_\alpha,m_\alpha\rangle\rangle_{\omega(1+i\gamma)} -
\langle\langle p_\alpha,m_\alpha\rangle\rangle_{0}
\end{equation}
This is a slightly different approach than the one typically used with CPP, again with no correction for ECD. It would have the formal advantage of conserving the symmetry  \mbox{$\sigma_{\textrm{ORD}}(-\omega) = - \sigma_{\textrm{ORD}}(\omega)$}.
This alternative expression 
would only be advantageous over the first one in practical applications  where $\omega$ and $\gamma$ are of similar magnitude,  or when $\gamma > \omega$, for instance because one scans with $\omega$ through 0. If one is interested in computing, for instance, $\sigma_{\textrm{ECD}}(\omega)$ for the UV/Vis region with $\gamma$ of the order of 0.1 eV, the first expression is to be preferred. This is the case here, so all ECD results presented in the following are obtained according to Eq.~\eqref{velo1ord}.
}

Finally, within damped linear response theory, one can also straightforwardly compute the isotropic dipole-dipole polarizability at purely imaginary frequencies, $\overline{\alpha}(i\omega)$,
by setting the real frequency equal to zero and
$\gamma=\omega$ in Eq.~\eqref{cmplxLR}.
From the isotropic averaged polarizability at imaginary frequency, one can then obtain coefficients to describe the long-range part of London dispersion interactions, {\it{e.g.}} 
the
$C_6$ dispersion coefficients~\cite{CPP:C6:Jiemchooroj,Lanczos1,CPP:ADC}
\begin{equation}
\label{C6eq}
    C_6 = \frac{3\hbar}{\pi}
    \int_{0}^{\infty}
    \overline{\alpha}^{A}(i \omega)
    \overline{\alpha}^{B}(i \omega)
    d \omega
\end{equation}
where  $A$
and $B$ label the interacting systems.
The
$C_6$ dispersion coefficients
can be used, {\it e.g.}, to compute the long-range dispersion interaction energy between $A$ and $B$,
also known as Casimir-Polder potential,
according to the simplified expression valid in the van der Waals region,~\cite{CPP:ADC} as
$\Delta E (R_{AB}) = -\frac{\hbar}{\pi} \frac{C_6}{R^6_{AB}}$,
and to determine long-range dispersion interactions corrections to density functional theory.~\cite{Grimme:2006,Tkatchenko2009,Grimme:2010}

\subsection{The complex linear response equations for (RI-)CC2 }
The properties defined in the previous section entail the solution of complex response equations to obtain the real and imaginary components of the response amplitudes $\ampd{x}{\omega+i\gamma}$ and multipliers $\muld{x}{\omega+i\gamma}$:
\begin{align}
\label{rightlin2}
\left\{{\bf{A}} - {(\omega+i\gamma)}  {\bf{1}}\right\}
\ampd{x}{\omega+i\gamma} &= - {\xi}^x
    \\
\muld{x}{\omega + i\gamma}
\left\{{\bf{A}} + (\omega+i\gamma)  {\bf{1}}
\right\}
& = - {{\eta}}^x - {\bf{F}} \ampd{x}{\omega+i\gamma}
\label{leftlin2}
\end{align}
{where 
{\bf{A}} is the CC 
Jacobian.~\cite{Christiansen:1998}
See Refs.~\citenum{CC2},
~\citenum{RICC2ENE}, and ~\citenum{RICC2:Friese:2012} 
for specific definitions of the CC2 right-hand-side vectors $\xi^x$
and $\eta^x$,
and of the matrices {\bf{A}}
and {\bf{F}}.
%
%
}
We will in the following concentrate 
solely on the solution of Eq.~\eqref{rightlin2} {within RI-CC2} {without storing any double excitation amplitudes, multipliers, or trial vectors}.
For this, we start from the complex linear response equations in matrix form of Ref.~\citenum{CPP:Kauczor:CC} and
explicitly partition them in singles ($S$) and {doubles} ($D$) blocks. For ease of notation, we omit in the following the frequency argument on the response amplitudes,  and 
 write 
\begin{equation}
\begin{bmatrix}
\mathbf{A}_{SS} - 
\omega \mathbf{1}_{SS}   
& \mathbf{A}_{SD}  
& \gamma \mathbf{1}_{SS}
& \mathbf{0}                \\
\mathbf{A}_{DS}          & \mathbf{A}_{DD} - \omega \mathbf{1}_{DD} & \mathbf{0} & \gamma \mathbf{1}_{DD}          \\
-\gamma \mathbf{1}_{SS}         & \mathbf{0}    & \mathbf{A}_{SS} - \omega\mathbf{1}_{SS} & \mathbf{A}_{SD}          \\
\mathbf{0}  & -\gamma \mathbf{1}_{DD}         & \mathbf{A}_{DS}  & \mathbf{A}_{DD} - 
\omega \mathbf{1}_{DD} \\
\end{bmatrix}
\begin{bmatrix}
t^{x}_{\Re,S}   \\
t^{x}_{\Re,D}   \\
t^{x}_{\Im,S}   \\
t^{x}_{\Im,D}   \\
\end{bmatrix}
= - 
\begin{bmatrix}
\xi^{x}_{\Re,S}   \\
\xi^{x}_{\Re,D}   \\
\xi^{x}_{\Im,S}   \\
\xi^{x}_{\Im,D}   \\
\end{bmatrix}
\end{equation}
equivalent to the system of equations
\begin{subnumcases}{}
 (\mathbf{A}_{SS} - \omega\mathbf{1}_{SS}) \ t^{x}_{\Re,S} + 
 \gamma \ t^{x}_{\Im,S} =  -\xi^{x}_{\Re,S} - 
 \mathbf{A}_{SD} t^{x}_{\Re,D}  \label{eq:sep_var1}\\
 (\mathbf{A}_{DD} - \omega\mathbf{1}_{DD}) t^{x}_{\Re,D} + \gamma \ t^{x}_{\Im,D} =  -\xi^{x}_{\Re,D} - \mathbf{A}_{DS} t^{x}_{\Re,S}  \label{eq:sep_var2}\\
({\mathbf{A}}_{SS} - \omega \mathbf{1}_{SS}) \ t^{x}_{\Im,S} - \gamma \ t^{x}_{\Re,S} =  - \xi^{x}_{\Im,S} - 
\mathbf{A}_{SD} t^{x}_{\Im,D} \label{eq:sep_var3}\\
({\mathbf{A}}_{DD} - \omega{\mathbf{1}}_{DD}) \ 
t^{x}_{\Im,D} - 
\gamma \ t^{x}_{\Re,D} =  
- \xi^{x}_{\Im,D} - \mathbf{A}_{DS} t^{x}_{\Im,S}  
\label{eq:sep_var4}
\end{subnumcases} 
Assuming we work with canonical molecular orbitals, the doubles-doubles
block
$\mathbf{A}_{DD}$ of the CC2 Jacobian is diagonal, and so is in this case the doubles-doubles 
resolvent matrix,~\cite{CC2} {$\mathbf{R}_{DD} = -[\mathbf{A}_{DD}-(\omega+i\gamma)\mathbf{1}_{DD}]^{-1}$}. 
We therefore define
 \begin{align}
 \boldsymbol{\Delta} &= (\mathbf{A}_{DD} - \omega\mathbf{1}_{DD} )  ,
 \end{align}
with the diagonal elements
 \begin{align}
  \Delta_{ab}^{ij} &=
  (\epsilon_{a} - \epsilon_{i}
  + \epsilon_{b} - \epsilon_{j}-\omega)
 .
\end{align}
%
We isolate $t^{x}_{\Im,D}$ from Eq.~\eqref{eq:sep_var4}  and $t^x_{\Re,D}$ from 
Eq.~\eqref{eq:sep_var2},
and introduce each resulting expression
into the other,
to arrive at
\begin{equation}
\label{tRD_onlyS}
t^x_{\Re,D} =
-\frac{\Delta}{\gamma^2 + \Delta^2} \Big(
\xi^x_{\Re,D} + \mathbf{A}_{DS} t^x_{\Re,S} \Big)
+\frac{\gamma}{\gamma^2 + \Delta^2} \Big( \xi^x_{\Im,D}
 + \mathbf{A}_{DS} t^x_{\Im,S}  \Big)
\end{equation}
%
\begin{equation}
\label{tID_onlyS}
t^x_{\Im,D} =
-\frac{\Delta}{\gamma^2 + \Delta^2} \Big(
\xi^x_{\Im,D} + \mathbf{A}_{DS} t^x_{\Im,S} \Big)
-\frac{\gamma}{\gamma^2 + \Delta^2} \Big( \xi^x_{\Re,D}
 + \mathbf{A}_{DS} t^x_{\Re,S}  \Big)
%
\end{equation}
Inserting Eq.~\eqref{tRD_onlyS} into Eq.~\eqref{eq:sep_var1}
%
and Eq.~\eqref{tID_onlyS} into Eq.~\eqref{eq:sep_var3} we finally obtain
the effective CC2 CPP linear response equations in compact matrix form
\begin{equation}
\begin{split}
\begin{bmatrix}
\mathbf{A}_{SS}^{\rm{eff}}(\omega,\gamma) - \omega \mathbf{1}_{SS} &   -{\boldsymbol{\Gamma}}^{\text{eff}}_{SS}(\omega,\gamma)+\gamma\mathbf{1}_{SS}  \\
{\boldsymbol{\Gamma}}^{\text{eff}}_{SS}(\omega,\gamma)-\gamma\mathbf{1}_{SS} &  \mathbf{A}_{SS}^{\rm{eff}}(\omega,\gamma) -\omega \mathbf{1}_{SS}\end{bmatrix}
\begin{bmatrix}
t^x_{\Re,S}   \\
t^x_{\Im,S}
\end{bmatrix}
= -
\begin{bmatrix}
\xi^{x,\rm{eff}}_{\Re,S}
(\omega,\gamma)    \\
\xi^{x,\rm{eff}}_{\Im,S}
(\omega,\gamma)
\end{bmatrix}
\end{split}
\end{equation}
where
\begin{align}
\label{Aeff_new}
\mathbf{A}^{\rm{eff}}_{SS}(\omega,\gamma) = & \mathbf{A}_{SS}- \mathbf{A}_{SD} \frac{\Delta }{\Delta^2 +\gamma^2} \mathbf{A}_{DS}
\\
{\boldsymbol{\Gamma}}^{\text{eff}}_{SS}(\omega,\gamma) = &  -\mathbf{A}_{SD}\frac{\gamma}{\Delta^2 + \gamma^2} \mathbf{A}_{DS}
\end{align}
and
\begin{equation}
\label{xiR_new}
\begin{split}
\xi^{x,\rm{eff}}_{\Re,S}(\omega,\gamma) = \xi_{\Re,S}^{x} -\mathbf{A}_{SD}\frac{\Delta}{\Delta^2+\gamma^2}\xi_{\Re,D}^{x}
+ \mathbf{A}_{SD} \frac{\gamma}{\Delta^2+\gamma^2}\xi^x_{\Im,D}  
\end{split}
\end{equation}
%
\begin{equation}
\label{xiI_new}
\begin{split}
\xi^{x,\rm{eff}}_{\Im,S}
(\omega,\gamma) =   \xi^x_{\Im,S} - \mathbf{A}_{SD}\frac{\Delta}{\Delta^2+\gamma^2} 
\xi^x_{\Im,D}
-\mathbf{A}_{SD} \frac{\gamma}{\Delta^2+\gamma^2} \xi^x_{\Re,D}
\end{split}
\end{equation}
Thus, the CPP(-RI)-CC2 building blocks are the same as in the standard linear response case,~\cite{RICC2ENE,RICC2TMES} just with slightly different 
generalized values for the diagonal elements of the resolvent as scaling factors.
These scaling factors are exactly the same as used in the preconditioning step 
in Ref.~\citenum{CPP:Kauczor:CC}.

\section{Implementation}

\subsection{The iterative CPP solver}
The general strategy for the implementation of our solver consists in working exclusively with real trial vectors, 
generating two new vectors at each iteration from the real and imaginary parts of the preconditioned residual vectors,
and solving the complex linear response equation, Eq.~\eqref{Aeff_new}, in the reduced space.
%
In detail, the fundamental steps of the iterative solver are the following:
\begin{enumerate}
\item {\it{Generation of the start trial vectors}}
by preconditioning the effective RHS vectors
\begin{equation}
(\tilde{{b}}_{1})_{ai} = 
\frac{\epsilon_{a}-\epsilon_i - \omega}{(\epsilon_{a}-\epsilon_{i} - \omega)^2+\gamma^2}
\xi_{\Re,ai}^{x,\rm{eff}}
+ 
\frac{\gamma}{(\epsilon_{a}-\epsilon_i - \omega)^2 + \gamma^2}
\xi_{\Im,ai}^{x,\rm{eff}}
\end{equation}
and
\begin{equation}
(\tilde{{b}}_{2})_{ai} = 
\frac{\epsilon_{a}-\epsilon_{i} - \omega}{(\epsilon_{a}-\epsilon_i - \omega)^2+\gamma^2}
\xi_{\Im,ai}^{x,\rm{eff}}
- 
\frac{\gamma}{(\epsilon_{a}-\epsilon_i - \omega)^2 + \gamma^2}
\xi_{\Re,ai}^{x,\rm{eff}}
\end{equation}
%
followed by orthonormalization;

\item {\it{Computation of the linearly transformed vectors:}} 
\begin{equation}
{\sigma}_{1}^{\Re}=\mathbf{A}^{\rm{eff}}{b}_{1},\quad
{\sigma}_{2}^{\Re}=\mathbf{A}^{\rm{eff}}\boldsymbol{b}_{2},\quad
{\sigma}_{1}^{\Im}=\boldsymbol{\Gamma}^{\rm{eff}}
{b}_{1},\quad
{\sigma}_{2}^{\Im}=\boldsymbol{\Gamma}^{\rm{eff}}{b}_{2}
\end{equation}
%

 \item {\it Computation of the reduced-space building blocks:}
\begin{align}
&\mathbf{A}^{\rm{red}}_{ij}  = {b}_i^T {\sigma}_{j}^{\Re}~; \quad\quad
{\boldsymbol{\Gamma}}^{\rm{red}}_{ij}  =  {b}_i^T {\sigma}_j^{\Im}
\\
&  \xi^{x,\rm{red}}_{\Re,i}  ={{b}}_i^T {\xi}^{x,\rm{eff}}_{\Re}~; \quad 
  \xi_{\Im,i}^{x,\rm{red}} = {{b}}_i^T \xi^{x,\rm{eff}}_{\Im}
\end{align}
where $i$ and $j$ run on the number of trial vectors. Note that $n_{\rm{red}} = 2 n$, where $n$ is the iteration number;
\item {\it Construction and solution of the CPP equation in reduced space.}
\begin{equation}
\begin{split}
\begin{bmatrix}
\mathbf{A}^{\rm{red}} - \omega \mathbf{1} & -\boldsymbol{\Gamma}^{\rm{red}} + \gamma\mathbf{1} \\
\boldsymbol{\Gamma}^{\rm{red}} -\gamma\mathbf{1} & \mathbf{A}^{\rm{red}} -\omega \mathbf{1}\end{bmatrix}
\begin{bmatrix}
x^{\Re}   \\
x^{\Im}
\end{bmatrix}
=  -
\begin{bmatrix}
\xi_{\Re}^{x,\rm{red}}(\omega)    \\
\xi_{\Im}^{x,\rm{red}}(\omega)
\end{bmatrix}
\end{split}
\end{equation}
The CPP reduced equation is solved using standard library solvers
to obtain $x_{}^{\Re}$ and $x_{}^{\Im}$.

\item {\it{Construction of the solution and residual vectors in the full (singles) space}}:
The solution vectors at iteration $n$ are linear combinations of the trial basis with the reduced space solution vectors as coefficients:
\begin{equation}
  {t}_{\Re}^{x,(n)} = \sum_{i}^{n_{\rm{red}}} x^{\Re}_{i} \boldsymbol{b}_i,\quad 
  {t}_{\Im}^{x,(n)} = \sum_{i}^{n_{\rm{red}}} x^{\Im}_{i} \boldsymbol{b}_i
\end{equation}
They are formally introduced
in the effective CPP equation to yield the residual vectors
\begin{align}
R_{\Re,S}^{(n)} 
& = \sum_i^{n_{\rm{red}}} x_i^\Re {\sigma}^\Re_{i,S} - \omega \ t_{\Re,S}^{x,(n)} - \sum_i^{n_{\rm{red}}} x_i^\Im {\sigma}^\Im_{i,S} + \gamma \ t_{\Im,S}^{x,(n)} + \xi^{x,\rm{eff}}_{\Re,S}
\\
R_{\Im,S}^{(n)} 
&=\sum_i^{n_{\rm{red}}} x_i^\Im \sigma^\Re_{i,S} - \omega t_{\Im,S}^{x,(n)} + \sum_i^{n_{\rm{red}}} x_i^\Re \sigma^\Im_{i,S} - \gamma \ t_{\Re,S}^{x,(n)} + \xi^{x,\rm{eff}}_{\Im,S}
\end{align}
Solution and residual vectors are, alike the linearly transformed one, stored as vectors of twice the size of a singles amplitude. 

\item {\it{Generation of the new trial vectors from the preconditioned residuals.}}
If the residual vectors of step 5 are larger than a preset threshold, new trial vectors are generated and a new iteration is made. 
In practice, we split the (tentative) trial vector into two vectors:
\begin{align}
  (\tilde{b}_{{2n-1}})_{ai} & = 
    \frac{\epsilon_a-\epsilon_i-\omega}{(\epsilon_a-\epsilon_i-\omega)^2 + \gamma^2} \cdot R^{(n)}_{\Re,ai}
   +\frac{\gamma}{(\epsilon_a-\epsilon_i-\omega)^2 + \gamma^2} \cdot R^{(n)}_{\Im,ai}
\\
  (\tilde{b}_{{2n}})_{ai} & = 
\frac{\epsilon_a-\epsilon_i-\omega}{(\epsilon_a-\epsilon_i-\omega)^2 + \gamma^2} \cdot R^{(n)}_{\Im,ai}
-\frac{\gamma}{(\epsilon_a-\epsilon_i-\omega)^2 + \gamma^2} \cdot R^{(n)}_{\Re,ai}
\end{align}
that are normalised and then orthogonalized onto the previous trial vectors. If after this step their norm is smaller than a linear-dependence threshold they are discarded, otherwise they are normalized once more and added to the set of trial vectors.
\item {\it Extension of the reduced space and iteration
until convergence.}
If the residuals for all equations have decreased below a user-defined threshold, the procedure is stopped, else
the reduced space is extended as in step 3 and steps 4--6 are repeated until convergence.
\end{enumerate}

\subsection{Building blocks: the RHS vectors}
\label{Sec:RHS}

The perturbation operators are in general assumed to be either real or purely imaginary. As a consequence, the (not partitioned) RHS vectors $\xi^x$
 for the first-order amplitude equations are either real or purely imaginary.
In the case of real perturbations (e.g. electric dipole), the effective RHS vectors simplify to
\begin{align}
\xi^{x,\textrm{eff}}_{\Re,S}(\omega,\gamma) ~=~& 
\xi^x_{S} - \mathbf{A}_{SD} \frac{\Delta}{\Delta^2+\gamma^2}  \xi^x_{D},\\
   \xi^{x,\text{eff}}_{\Im,S}(\omega,\gamma)~=~& - \mathbf{A}_{SD} \frac{\gamma}{\Delta^2+\gamma^2} \xi^x_{D}.
\end{align}
{Within RI-CC2,} the doubles elements of the RHS vector $\xi^x_D$ are computed only on the fly and immediately contracted with the elements of the {singles-doubles block of the Jacobian matrix} $\mathbf{A}_{SD}$, either in a loop over pairs of occupied or pairs of virtual orbital {indices}. This entails computing
\begin{align}
    (\tilde{\xi}^{x}_{\Re})^{ij}_{ab}
    = \frac{-\Delta_{aibj}}{\Delta_{aibj}^2 + \gamma^2} \xi^{x,ij}_{ab} 
    ~,
\\
  (\tilde{\xi}^x_{\Im})_{ab}^{ij}
  = \frac{-\gamma}{\Delta_{aibj}^2 + \gamma^2} \xi^{x,ij}_{ab} 
  ~,
\end{align}
where the elements of the unmodified doubles part of the RHS vector are:~\cite{RICC2:Friese:2012}
\begin{align}
    \xi^{x,ij}_{ab} = \hat{P}^{ij}_{ab} \Big( \sum_c t^{ij}_{ac} \hat{h}^x_{cb} - \sum_k t^{ik}_{ab} \hat{h}^x_{kj} \Big)
\end{align}
{In the expression above, 
$t^{ij}_{ab}$
are the zero-order 
double amplitudes
and $\hat{h}^x_{pq}$ 
are the integrals of the one-electron operator 
$x$, similarity transformed
with the exponential function of the single excitation cluster operator (for their definition, see Appendix~\ref{appendix}).
$\hat{P}^{ij}_{ab}$ is a symmetrization operator, defined by $\hat{P}^{pr}_{qs} f_{pq,rs} = f_{pq,rs} + f_{rs,pq}$.~\cite{RICC2:Friese:2012}} 

Then, we contract  $\tilde{\xi}^{x,ij}_{{\Re},ab}$ 
and 
$\tilde{\xi}^{x,ij}_{{\Im},ab}$
with the elements of the singles-doubles matrix $\mathbf{A}_{SD}$. 
In general, the contraction of 
$\mathbf{A}_{SD}$
with a doubles vector $b^{kl}_{cd}$ can be written~\cite{RICC2:Friese:2012}
\begin{align}
   \sum_{ckdl} 
   {A}_{ai,ckdl} b^{kl}_{cd} = 
     + \sum_{cdk} \big( 2 b^{ik}_{cd} - b^{ik}_{dc} \big) (kd\hat{|}ac)
     - \sum_{dkl} \big( 2 b^{kl}_{ad} - b^{kl}_{da} \big) (ld\hat{|}ki)
     + \sum_{ck}  \big( 2 b^{ik}_{ac} - b^{ik}_{ca} \big) \hat{F}_{kc}
\end{align}
where $\hat{F}_{kc}$ is the Fock matrix,~\cite{RICC2TMES} and $(ld\hat{|}ki)$ are the two-electron integrals of the $T_1$-similarity transformed Hamiltonian operator (see their definition in Appendix~\ref{appendix}). 
{Within RI-CC2, the two-electron integrals are approximated as~\cite{RICC2:geoopt,RI:integral:1,RI:integral:2,RI:integral:3}}
\begin{align}
\hat{\big( pq | rs \big )} = 
 \sum_{Q} \hat{B}_{Q,pq}\hat{B}_{Q,rs}~,
\end{align}
where
\begin{align}
\label{hatBQ}
  \hat{B}_{Q,pq}
  = \sum_P (pq\hat{|}P)V^{-\frac{1}{2}}_{PQ}
  =
  \revS{
  \sum_{\mu\nu}\Lambda^{p}_{\mu p} \Lambda^{h}_{\nu q}  \sum_{P} (\mu\nu|P) V^{-\frac{1}{2}}_{PQ}~.
  }
\end{align}
The $\Lambda^p$ and $\Lambda^h$ matrices are the $T_1$-transformed molecular orbital coefficients,
whose definition is given in Appendix~\ref{appendix}.
With this, the first two terms can be rewritten, {\it e.g.}, like
\begin{align}
\label{1termksi}
\sum_{cdk} \big( 2 b^{ik}_{cd} - b^{ik}_{dc} \big) (kd\hat{|}ac) = 
\sum_{Qc}\left(
\sum_{dk}
\big( 2 b^{ik}_{cd} - b^{ik}_{dc} \big) \hat{B}_{Q,kd}\right)\hat{B}_{Q,ac}
= \sum_{Qc}
\bar{Y}_{Q,ci} 
   \hat{B}_{Q,ac}
\end{align}
where we introduced the $\bar{Y}$ intermediate:
\begin{equation}
\label{Ybar}
    \bar{Y}_{Q,ai} =  \sum_{bj} \big( 2 b^{ij}_{ab} - b^{ij}_{ba} \big) \hat{B}_{Q,jb}
\end{equation}
In the case of the CPP RHS vectors, these intermediates become
\begin{align}
\label{YintermediateRe}
    \bar{Y}^{x,\Re}_{Q,ai} = & \sum_{bj} \big( 2 \tilde{\xi}^{x,ij}_{\Re,ab} - \tilde{\xi}^{x,ij}_{\Re,ba} \big) \hat{B}_{Q,jb}
\\
\label{YintermediateIm}
  \bar{Y}^{x,\Im}_{Q,ai} = & \sum_{bj} \big( 2 \tilde{\xi}^{x,ij}_{\Im,ab} - \tilde{\xi}^{x,ij}_{\Im,ba} \big) \hat{B}_{Q,jb}
\end{align}
and the real and imaginary {parts of the effective singles} RHS vectors are computed 
as
\begin{align}
   \xi^{x,\text{eff}}_{\Re, ai} 
 =~& \xi_{ai} + \sum_{ck} \big( 2 \tilde{\xi}^{x,ik}_{\Re,ac}-\tilde{\xi}^{x,ik}_{\Re,ca} \big) \hat{F}_{kc}
   +\sum_{cQ}
   \bar{Y}^{x,\Re}_{Q,ci} 
   \hat{B}_{Q,ac}
   -\sum_{kQ} \bar{Y}^{x,\Re}_{Q,ak}
   \hat{B}_{Q,ki}
\end{align}
\begin{align}
    \xi^{x,\text{eff}}_{\Im,ai} =~&
     \sum_{ck} \big( 2 \tilde{\xi}^{x,ik}_{\Im,ac} -\tilde{\xi}^{x,ik}_{\Im,ca} \big) \hat{F}_{kc}
      + \sum_{cQ} \bar{Y}^{x,\Im}_{Q,ci} \hat{B}_{Q,ac}
      -\sum_{kQ} \bar{Y}^{x,\Im}_{Q,ak}  \hat{B}_{Q,ki}~.
\end{align}


In case of an imaginary perturbation ${\mathcal{X}}$ (e.g., the magnetic dipole moment or the linear momentum), the effective RHS vector reads 
\begin{align}
    \xi^{{\mathcal{X}},\text{eff}}_{\Re,S}(\omega,\gamma)~=~& \mathbf{A}_{SD} \frac{+\gamma}{\Delta^2 +\gamma^2} \xi^{\mathcal{X}}_D
\\
   \xi^{{\mathcal{X}},\text{eff}}_{\Im,S}(\omega,\gamma)~=~& \xi^{\mathcal{X}}_S + \mathbf{A}_{SD}   \frac{-\Delta }{\Delta^2+\gamma^2}  \xi^{\mathcal{X}}_{D}
\end{align}
that is,
\begin{align}
\xi^{{\mathcal{X}},\text{eff}}_{\Re,ai}~=~&
   \phantom{xxx}
     -   \sum_{ck} \big(
     \tilde{\xi}^{{\mathcal{X}},ik}_{\Re,ac} -\tilde{\xi}^{{\mathcal{X}},ik}_{\Re,ca} \big) \hat{F}_{kc}
      - \sum_{cQ}
      \bar{Y}^{{\mathcal{X}},\Re}_{Q,ci} \hat{B}_{Q,ac}
      +\sum_{kQ} \bar{Y}^{{\mathcal{X}},\Re}_{Q,ak}  \hat{B}_{Q,ki}  
\\
    \xi^{{\mathcal{X}},\text{eff}}_{\Im,ai}~=~  &  \xi^\chi_{ai} + 
    \sum_{ck} \big( 2 \tilde{\xi}^{{\mathcal{X}},ik}_{\Im,ac}-
    \tilde{\xi}^{{\mathcal{X}},ik}_{\Im,ca} \big) \hat{F}_{kc}
   +\sum_{cQ} \bar{Y}^{{\mathcal{X}},\Im}_{Q,ci} 
   \hat{B}_{Q,ac}
   -\sum_{kQ} \bar{Y}^{{\mathcal{X}},\Im}_{Q,ak} 
   \hat{B}_{Q,ki}~.  
\end{align}

\subsection{Building blocks: the Jacobian transformation}
To build the reduced-space quantities needed in the CPP solver we need,
for each trial vector $b$, the result of its transformations with the effective matrices
$\mathbf{A}^{\text{eff}}_{SS}(\omega,\gamma)$ and $\boldsymbol{\Gamma}^{\text{eff}}_{SS}(\omega,\gamma)$,
typically referred to as $\sigma$ vectors.
To keep the overhead for CPP small, the transformations with the two matrices are done together.

We express the result of the transformation of a singles trial vector {$b$} with the {doubles-singles Jacobian matrix} $\mathbf{A}_{DS}$ as one-index transformed two-electron integrals 
\begin{align}
 \sum_{ck} A_{aibj,ck} b_{ck} =~& \langle {^{ij}_{ab}} | 
 [\hat{H},\tau_{ck}] 
 \text{HF} \rangle b_{ck} = {(ai\bar{|}bj)}, \\
 (ai\bar{|}bj) =~& 
 {\hat{P}}^{ab}_{ij}
\sum_{\alpha\beta\gamma\delta}
\left(
\bar{\Lambda}^p_{\alpha a} \Lambda^h_{\revS{\beta i}}
+
{\Lambda}^p_{\alpha a} \bar{\Lambda}^h_{\revS{\beta i}}
\right)
{\Lambda}^p_{\gamma b} \Lambda^h_{\delta j}
(\alpha\beta|\gamma\delta) 
\end{align}
where $\bar{\Lambda}^p$ 
and 
$\bar{\Lambda}^h$
are defined as in Appendix~\ref{appendix},
with the singles trial vector $b_1$
in place of the singles response amplitudes $t^x_1$.
These 4-index integrals are evaluated on-the-fly from three-centre intermediates~\cite{RICC2TMES}
($\hat{B}_{Q,ai}$, Eq.~\eqref{hatBQ}, and $\bar{B}_{Q,ai}$, given in Appendix~\ref{appendix})
and combined with the energy denominators from $\boldsymbol{\Delta}$ 
into intermediate doubles amplitudes.
In other words, for the CPP implementation, the following intermediate doubles amplitudes are built:
\begin{align}
   \tilde{b}^{\Re,ij}_{ab} =  & 
   \frac{ -\Delta_{aibj} }{ \Delta_{aibj}^2 + \gamma^2 } {(ai\overline{|}bj)}
\\
   \tilde{b}^{\Im,ij}_{ab} =  & \frac{ -\gamma }{ \Delta_{aibj}^2 + \gamma^2 } {(ai\overline{|}bj)}
\end{align}
With these, the transformations with $\mathbf{A}^{\textrm{eff}}_{SS}$ and $\boldsymbol{\Gamma}^{\textrm{eff}}_{SS}$ can be expressed as:
\begin{align}
  \mathbf{A}_{SS}^{\textrm{eff}}(\omega,\gamma) b_{S} = & \mathbf{A}_{SS} b_{S} + \sum_{ckdl} \mathbf{A}_{S,ckdl} \tilde{b}^{\Re,kl}_{cd}
\\
  \boldsymbol{\Gamma}_{SS}^{\textrm{eff}}(\omega,\gamma) b_{S} = &   \sum_{ckdl} \mathbf{A}_{S,ckdl} \tilde{b}^{\Im,kl}_{cd}
\end{align}
The contribution $\mathbf{A}_{SS} b_S$ is unchanged compared to the standard (non-CPP) solver.\revS{REF!!!}
The other contributions are evaluated in a way similar (and partially using the same routines) to 
the contributions to the effective right-hand-sides discussed in Section~\ref{Sec:RHS}: 
\begin{align}
\nonumber
 \sigma^{{\textrm{eff}}}_{\Re,ai} = 
 & 
\sum_{ck}
{A}^{{\textrm{eff}}}_{ai,ck}(\omega,\gamma) b_{ck}
\\\nonumber
 \\
 = & \sum_{ck} {A}_{ai,ck} b_{ck}  
   + \sum_{ck} \big( 2 \tilde{b}^{\Re,ik}_{ac}-\tilde{b}^{\Re,ik}_{ca} \big) \hat{F}_{kc}
  +\sum_{cQ} \bar{Y}^{\Re}_{Q,ci} \hat{B}_{Q,ac}
  -\sum_{kQ} \bar{Y}^{\Re}_{Q,ak} \hat{B}_{Q,ki} 
  +  \sum_{ck} (2t^{ik}_{ac} - t^{ik}_{ca}) \bar{F}_{kc}
\end{align}
and
\begin{align}
\nonumber
    \sigma^{\text{{\textrm{eff}}}}_{\Im,ai} = & \sum_{ck} \Gamma^{{\textrm{eff}}}_{ai,ck}(\omega,\gamma) b_{ck}
\\ =~&
     \sum_{ck} \big( 2 \tilde{b}^{\Im,ik}_{ac} - \tilde{b}^{\Im,ik}_{ca} \big) \hat{F}_{kc}
      + \sum_{cQ} \bar{Y}^{\Im}_{Q,ci} \hat{B}_{Q,ac}
      -\sum_{kQ} \bar{Y}^{\Im}_{Q,ak}  \hat{B}_{Q,ki}
\end{align}
The real and imaginary $\bar{Y}$ intermediates are, 
as in Eq.~\eqref{Ybar},
using the real and imaginary intermediate doubles amplitude trial vectors defined above.

\subsection{
The first-order perturbed densities}
Once the real and imaginary response amplitudes have been obtained, we can build the real and imaginary linear response functions needed for the properties and spectra discussed in Section~\ref{Properties}.
This entails computing contractions of the (complex) response amplitudes with the $\eta^x$ vectors and with the $\bf{F}$ matrix.

The contributions from the terms of the type $\eta^x \cdot t^y$ are formulated as contractions of densities and one-electron integrals of the perturbation operator~\cite{RICC2:Friese:2012,RICC2TMES}
\begin{align}
\eta^x \cdot t^y = \sum_{pq} D^\eta_{pq}(t^y) \hat{h}^x_{pq} ~.
\end{align}
We do so as, for large systems, we do not want to store the doubles parts of $\eta^x$ and $t^y$. Also, to recalculate the doubles parts of both vectors for every dot product, i.e. for every pair of perturbations $x$ and $y$, would require a number of ${\cal{N}}^5$-scaling steps that increases with the number of operator pairs. 
Via densities, on the other hand, the number of ${\cal{N}}^5$-scaling steps increases only linear with the number of operators.
The explicit density blocks are~\cite{RICC2TMES} 
\begin{align}
D_{ij}^\eta(t^x) &= -\sum_a \bar{t}_{ja} t^{x}_{ai} - X^x_{ij}\\
D_{ia}^\eta(t^x) &= 
C^x_{ai}
- \sum_k t^x_{ak} X_{ik}
- \sum_b Y_{ba} t^{x}_{bi}
\\
D_{ai}^\eta(t^x) & = 0 \\
D_{ab}^\eta(t^x) & = \sum_i \bar{t}_{ia} t^{x}_{bi} + Y^x_{ba}
\end{align}
The real part of $\mathbf{D}^\eta(t^x)$ is computed from the real part of $t^x$ as in the standard response case.~\cite{RICC2TMES} The imaginary part of $\mathbf{D}^\eta(t^x)$ is done in the same way using the imaginary part of $t^x$.
The contributions to the densities from the singles amplitudes are straightforward to compute since the singles are stored on disk and can be read from file when needed.
Complications arise from the doubles response amplitudes $t^{x,ij}_{ab}$, as they should also be implemented with ${\cal O}({\cal N}^2)$-scaling memory demands.
The expression for the doubles part of the response amplitudes is:
\begin{align}
    t^{x,ij}_{ab} =& - \Big\{ \hat{P}^{ij}_{ab} \Big(
         \sum_c t^{ij}_{ac} \hat{h}^x_{cb} - \sum_k t^{ik}_{ab} \hat{h}^x_{kj} \Big) + (ai\bar{|}bj)^x \Big\}/(\epsilon_a - \epsilon_i + \epsilon_b - \epsilon_j - \omega - i\gamma ),
\end{align}
The (complex) $t^x_S$-dressed  4-index integrals are evaluated within the RI approximation as:
\begin{align}
    (ai\bar{|}bj)^x = & \hat{P}^{ij}_{ab}  
\left\{    
    \sum_Q \bar{B}^{x,\Re}_{Q,ai} \hat{B}_{Q,bj}
    +i\sum_Q \bar{B}^{x,\Im}_{Q,ai} \hat{B}_{Q,bj} 
\right\}    
    ~,
\end{align}
where the three-centre intermediates $\bar{B}^{x,\Re}$ and $\bar{B}^{x,\Im}$ are built 
with, respectively, the real and imaginary part of the singles amplitudes $t^x_{ai}$,
\revS{see Appendix~\ref{appendix}}.

As described elsewhere,~\cite{RICC2:Winter:2011,RICC2:Friese:2012}
the ground-state double amplitudes are evaluated on the fly within the RI approximation and with a numerical Laplace transformation of the denominators
\begin{align}
    t^{ij}_{ab} = & \frac{- \sum_Q \hat{B}_{Q,ai} \hat{B}_{Q,bj}}{ (\epsilon_a - \epsilon_i + \epsilon_b - \epsilon_j)}
  \approx  - \sum_{m} \sum_Q \hat{K}^m_{Q,ai} \hat{K}^m_{Q,bj}
\end{align} 
with $\hat{K}^m_{Q,ai} = \hat{B}_{Q,ai}
\sqrt{\omega_m}\exp\{-(\epsilon_a-\epsilon_i)\theta_m\}$, $\theta_m$ being the Laplace sampling
points and $\omega_m$ the weights.~\cite{RICC2:Friese:2012}
This allows to do the transformation with the one-electron integrals for the perturbation operator $x$ at the level of the $\hat{K}$ intermediates:\cite{RICC2:Friese:2012}
\begin{align}
  \bar{K}^{m,x}_{Q,ai} = \sum_{c} \hat{K}^{m}_{Q,ci} \hat{h}^x_{ac} - \sum_k \hat{K}^{m}_{Q,ak} \hat{h}^x_{ki} ~,
\end{align}
(assuming that $x$ is purely real)
so that we can compute 
the real and the imaginary
response double amplitudes on the fly as
\begin{align}
t^{x,ij}_{\Re,ab} = & \hat{P}^{ij}_{ab} \Big\{ -\sum_m \sum_Q \bar{K}^{m,x}_{Q,ai} \hat{K}^m_{Q,bj} + \sum_Q \bar{B}^{x,\Re}_{Q,ai} \hat{B}_{Q,bj}   \Big\}\cdot \frac{-\Delta_{aibj} }{\Delta_{aibj}^2 + \gamma^2 } 
- \hat{P}^{ij}_{ab} \sum_{Q} \bar{B}^{x,\Im}_{Q,ai} \hat{B}_{Q,bj} \cdot \frac{-\gamma}{\Delta_{aibj}^2 + \gamma^2 }
~,
\\
t^{x,ij}_{\Im,ab} = & \hat{P}^{ij}_{ab} \Big\{ -\sum_m \sum_Q \bar{K}^{m,x}_{Q,ai} \hat{K}^m_{Q,bj} + \sum_Q \bar{B}^{x,\Re}_{Q,ai} \hat{B}_{Q,bj}   \Big\}\cdot \frac{-\gamma}{\Delta_{aibj}^2 + \gamma^2 } 
+ \hat{P}^{ij}_{ab} \sum_Q \bar{B}^{x,\Im}_{Q,ai} \hat{B}_{Q,bj} \cdot
\frac{-\Delta_{aibj}}{\Delta_{aibj}^2 + \gamma^2}
~.
\end{align}
%
The doubles of the first-order response amplitudes are constructed in a loop over pairs of occupied orbitals $i$ and $j$.
In the same loop, the doubles of the ground-state Lagrange multipliers $\bar{t}^{ij}_{ab}$ are built.
The response amplitudes $t^{x,ij}_{bc}$ are then contracted with the Lagrange multipliers to the intermediates:
\begin{align}
    Y^x_{ab} = \sum_{cij} \bar{t}^{ij}_{ac} t^{x,ij}_{bc}
\end{align}
and
\begin{align}
    C^x_{ai} = & \sum_{bj} \big( 2t^{x,ij}_{ab} - t^{x,ij}_{ba} \big) \bar{t}_{jb}
\end{align}
Then, the same procedure is repeated 
within a loop over pairs of virtual orbital indices $a$ and $b$ (with occupied and virtual orbitals interchanged) to calculate:
\begin{align}
X^x_{ik} = \sum_{abk} \bar{t}^{jk}_{ab} t^{x,ik}_{ab}~.
\end{align}
The real and imaginary parts for the doubles are computed together to avoid having to compute the doubles multipliers twice, and thus the real and imaginary parts of $C^x$, $Y^x$, and $X^x$ are evaluated together.
{Eventually, the individual blocks of the density $\mathbf{D}^{\eta}(t^x)$ are 
put together 
from these intermediates and the singles parts for the response amplitudes and Lagrange multipliers.}

\subsection{The F-matrix contractions}

Similar to the evaluation of $\eta^x \cdot t^y$, also the F-matrix contractions are organised such that all ${\cal O}({\cal N}^5)$-scaling steps only depend on one perturbation, and only cheap, low-scaling, steps 
depend on both amplitude response vectors.
The F-matrix contraction is first rewritten as:
\begin{equation}
    F t^x t^y = \sigma^x \cdot t^y
\end{equation}
with
\begin{align}
 \sigma^x_\mu = \sum_{\nu=\nu_1,\nu_2} F_{\nu\mu} t^x_\nu ~.
\end{align}
The singles and doubles blocks of $\sigma^x$ are partitioned as summarised in Table~\ref{tab:sigma}.
Scheme~\ref{fig:F} in appendix~\ref{appendix} summarises the main steps in the actual evaluation of the F-matrix contribution to the linear response function.
In difference to standard response theory, in the CPP case all intermediates depending on the response amplitudes, i.e.\ carrying an upper index $x$ or $y$, are complex.
The contributions to the real and imaginary parts of the intermediates are evaluated with the real and imaginary parts of $t^x$,  respectively, 
as described for standard response theory in Ref.\ \citenum{RICC2:Friese:2012}.

%
The explicit evaluation of the doubles blocks is avoided by reformulating the contraction of $\sigma^x_{iajb}$ with $t^{y,ij}_{ab}$ as in the following:
\begin{align}
 \frac{1}{2} \sum_{ijab} \sigma^{I,x}_{iajb} t^{y,ij}_{ab} =  & \frac{1}{2} \sum_{iajb} \hat{P}^{ij}_{ab} \Big[ \bar{t}_{ia}  \Big( 2 t^{y,ij}_{ab} - t^{y,ij}_{ba} \Big) \Big] \bar{F}^x_{jb}
   =  \sum_{ia} C_{ai}^y \bar{F}^x_{ia}
\end{align}
\begin{align}
    \frac{1}{2} \hat{P}^{ij}_{ab} \Big( \sigma^{G,x}_{iajb} + \sigma^{H,x}_{iajb} \Big) t^{y,ij}_{ab} = & \sum_{jbQ} \Big[ - \sum_{ck} \Big( \bar{t}_{jc} t^x_{ck} B_{Q,kb} + t^x_{ck} \bar{t}_{kb} B_{Q,jc} \Big) \Big] Y^y_{Q,bj}
    = \sum_{jbQ} \breve{B}^x_{Q,jb} Y^y_{Q,bj}
    \label{Bupx}
\end{align}

\begin{table}
\caption{The singles and doubles blocks of $\sigma^x$.
We refer to Appendix~\ref{appendix} for further definitions of intermediates.
\label{tab:sigma}}
\begin{tabular}{l}\\
\hline\hline
\multicolumn{1}{c}{
$\sigma^x_{ia} =  \sigma^{0,x}_{ia} + \sigma^{F,x}_{ia} + \sigma^{JG,x}_{ia} + \sigma^{JH,x}_{ia} + \sigma^{I,x}_{ia} + \sigma^{J^\prime,x}_{ia}$}
\\\hline
$
\begin{aligned}
\sigma^{0,x}_{ia}=~ 2\bar{F}^x_{ia}
\end{aligned}
$
\\
$
\begin{aligned}
\sigma^{F,x}_{ia} & =
\sum_{cdk} \bar{t}^{ki}_{cd} (ck\bar{|}da)^x - \sum_{ckl} \bar{t}^{kl}_{ca} (ck\bar{|}il)^x 
= \sum_{dQ} \Big( \breve{Y}^x_{Q,id} \hat{B}_{Q,da} + \breve{Y}_{Q,id} \bar{B}^x_{Q,da} \Big) - \sum_{lQ} \Big( \breve{Y}^{x}_{Q,al} \hat{B}_{Q,il} + \breve{Y}_{Q,al} \bar{B}^x_{Q,il} \Big)
\end{aligned}
$
\\
$
\begin{aligned}
\sigma^{JG,x}_{ia} & =
- \sum_j \bar{t}_{ja} \bar{F}^x_{ij} -\sum_j \bar{t}_{ja} \sum_{cdk} \Big(2t^{x,jk}_{cd} - t^{x,kj}_{cd}\Big) (kd|ic) = 
-\sum_j \bar{E}^{x,2}_{ij} \bar{t}_{ja} 
\end{aligned}
$
\\
$
\begin{aligned}
\sigma^{JH,x}_{ia} = &  \sum_b \bar{t}_{ib} \bar{F}^x_{ba}  +\sum_b \bar{t}_{ib} \sum_{dkl} \Big(2t^{x,kl}_{bd} - t^{x,lk}_{bd}\Big) (ld|ka)
= \sum_b \bar{t}_{ib} \bar{E}^{x,1}_{ba}
\end{aligned}
$
\\
$
\begin{aligned}
\sigma^{I,x}_{ia} & 
 = \sum_{ck} C^x_{ck} \Big[ 2(kc|ia) - (ic|ka) \Big]
= 
\sum_Q \Big( 2 \sum_{ck} B_{Q,ck} C^x_{ck}\Big) B_{Q,ia} - 
\sum_{Qk} \Big(\sum_c B_{Q,ic} C^x_{ck} \Big) B_{Q,ka}
\\
&  
= 
\sum_{Q\beta} \Big\{ 2m^{x}_Q C_{\beta i} - \sum_k M^{x}_{Q,ik} C_{\beta k} \Big\}B_{Q,\beta a}
\end{aligned}
$
\\
$
\begin{aligned}
\sigma^{J^\prime,x}_{ia} & =   \sum_{bj} \bar{t}_{jb} \Big[ 2(bj\bar{|}ia)^x - (ij\bar{|}ba)^x \Big]
\\
 & 
 = 
 2\sum_Q \Big(\sum_{jb} \bar{B}^x_{Q,bj} \bar{t}_{jb} \Big) B_{Q,ia} - \sum_{Qb} \Big(\sum_{j} \bar{B}^x_{Q,ij}\bar{t}_{jb} \Big) \sum_\beta \Lambda^p_{\beta b}  \hat{B}_{Q,\beta a} 
- \sum_{Qb} \Big( \sum_{j} \hat{B}_{Q,ij} \bar{t}_{jb} \Big)  \sum_\beta
\revS{
\bar{\Lambda}^{p,x}_{\beta b} 
}
\hat{B}_{Q,\beta a}
\\
& = 
2 \sum_Q \breve{i}^x_Q B_{Q,ia}
     - \sum_{Qb} \Big(\sum_{j} \bar{B}^x_{Q,ij}\bar{t}_{jb} \Big) \sum_\beta \Lambda^p_{\beta b}  \hat{B}_{Q,\beta a} - \sum_{Qb} \Big( \sum_{j} \hat{B}_{Q,ij} \bar{t}_{jb} \Big)  \sum_\beta 
     \revS{
     \bar{
     \Lambda}^{p,x}_{\beta b} 
     } \hat{B}_{Q,\beta a}
\end{aligned}
$
\\\hline
\multicolumn{1}{c}{
$
\sigma^x_{iajb} =  
\sigma^{I,x}_{iajb} + \sigma^{G,x}_{iajb} + \sigma^{H,x}_{iajb}
$
}
\\\hline
$
\sigma^{I,x}_{iajb} =  2 \bar{t}_{ia} \bar{F}^x_{jb} - \bar{t}_{ja} \bar{F}^x_{ib}
$
\\
$ 
\sigma^{G,x}_{iajb} =  -\sum_{ck} \bar{t}_{jc} t^x_{ck} \Big[ 2 (kb|ia) - (ka|ib) \Big]
$ 
\\
$ 
\sigma^{H,x}_{iajb} =  - \sum_{ck}  t^x_{ck} \bar{t}_{ka} \Big[ 2(jb|ic) - (jc|ib) \Big]
$
\\
\hline\hline
\end{tabular}
\end{table}
For the definition of the intermediates we refer to Appendix~\ref{appendix}.
Scheme~\ref{fig:F} in appendix~\ref{appendix} summarises the main steps in the evaluation of the F-matrix contribution to the linear response function.
\clearpage
\section{Results and discussion}
\subsection{Computational details}

The CPP solver for RI-CC2 has been implemented in a development version of the Turbomole program package.~\cite{TURBOMOLE,Turbomole2020}
The stick spectra calculations were  performed using previously implemented RI-CC2 functionalities in TURBOMOLE.\cite{RICC2ENE,RICC2TMES}

The structure of C$_{60}$ used in the OPA calculations 
was taken from Ref.~\citenum{Haettig:C60:str}.
It originates from a 
geometry optimization at the level of second-order M{\o}ller-Plesset Perturbation Theory (MP2)~\cite{MP_original} theory with Dunning's cc-pVTZ {basis} set.~\cite{Dunning_basis}
The structures of the molecules considered for the $C_6$ coefficients (alkanes, unsaturated hydrocarbons, aldehydes, and ketones) 
are also MP2/cc-pVTZ optimized structures from the literature.~\cite{CPP:ADC}
{The Cartesian coordinates of all molecular systems considered are reported in the SI file.} 
The structures of the helicenes~\cite{RICC2_OR_Friese} in the ECD calculations are 
MP2/cc-pVTZ optimized ones.
According to the standard convention for 
helicoidal systems, we used ($-$)-5-helicene (M), ($-$)-6-helicene (M), and ($+$)-7-helicene (P)
structures.
The structures of the fullerenes in the $C_6$ calculations are the same B3LYP/cc-pVDZ optimized ones used in Ref.~\citenum{CPP:Joanna:C6}.
In the calculations of the OPA spectra of C$_{60}$ we adopted the aug-cc-pVDZ basis set.
The calculations of $C_6$ dispersion coefficients of the fullerenes 
were carried out using the cc-pVDZ basis set. 
For all other molecules, the aug-cc-pVTZ basis set was used. This also applies to the 
ECD calculations on the helicenes.
The frozen-core approximation was 
used for the helicenes and in the calculations of the $C_6$ coefficients of the selected set of fullerenes.
An optimized auxiliary basis set matching the chosen atomic orbital basis 
was employed in all calculations.~\cite{bas:Weigend01a}

For comparison with the CPP spectra, the stick spectra were broadened using the Lorentzian function 
\begin{equation}
g_{j}(\omega) = \frac{\gamma}{(\omega-\omega_j)^2 + \gamma^2}
\label{LorBrod}
\end{equation}
with half width at half maximum (HWHM) value
$\gamma=0.004556$ a.u.
The frequency steps in the CPP calculations varied between 0.0025 a.u. and 0.01 a.u.
Cubic spline was used for the interpolation between the computed points to obtain the CPP spectrum. 

The $C_6$ coefficients were obtained according to Eq.~\ref{C6eq}, with $A = B$. 
The integral was evaluated using a Gauss-Legendre integration scheme, with a transformation of variables as suggested in Ref.~\citenum{Amos:C6}
and followed by a Gauss-Legendre quadrature in the interval $-1 \leq t \leq +1$.
A 12-point scheme was adopted.

\subsection{One-photon absorption: C$_{60}$}
\begin{figure}[hbt!]
\includegraphics[scale=0.8]{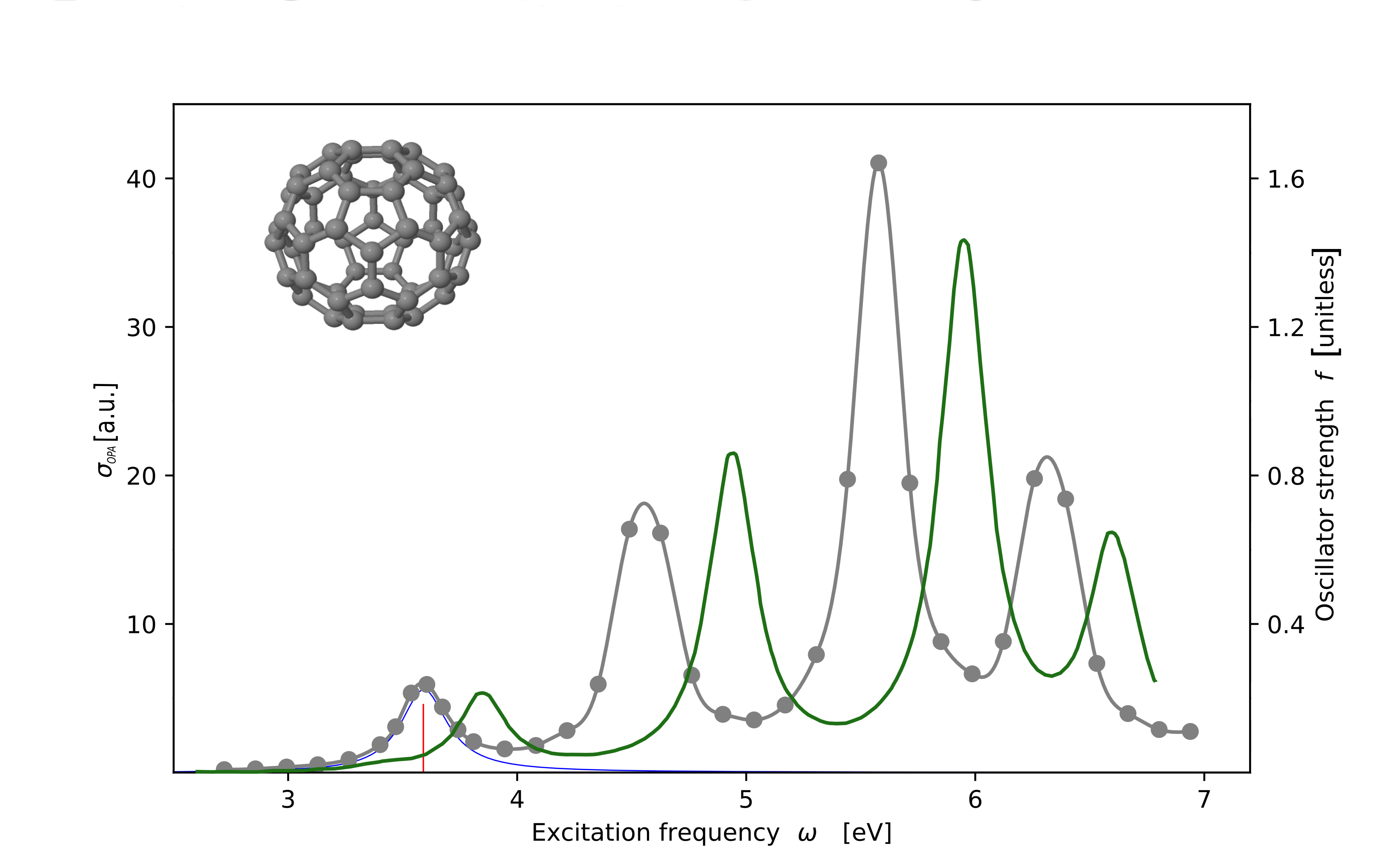}
\caption{\label{fig:labels} C$_{60}$. 
RI-CC2/aug-cc-pVDZ (all electrons correlated)
UV-vis  OPA spectra from standard linear response (blue) and CPP (grey) calculations.
The red stick is the only excitation with non zero intensity obtained converging 15 roots; in blue is its corresponding 
Lorentzian broadened spectrum, using  $\gamma=0.004556$ a.u. 
{The CPP spectrum shown as grey libe is a cubic spline of the computed CPP grid points.}
The spectrum in green is the CPP-B3LYP/pol-Sadlej one from Ref.~\onlinecite{CPP:Joanna:C6}. 
}
\label{C60_UV}
\end{figure}

The UV spectrum of C$_{60}$ obtained at the CPP-RI-CC2/aug-cc-pVDZ level (all electrons correlated) is shown in Figure \ref{C60_UV}. The spectrum is compared with the CPP-KS-TDDFT result of Ref.~\onlinecite{CPP:Joanna:solver2}.
C$_{60}$ is a prototypical case where the application of the CPP algorithm is particularly advantageous. When running in D$_{2h}$ symmetry (as done with the large majority of quantum chemistry codes),  straightforward application of regular solvers to obtain the stick spectrum results in an exceedingly large number of roots with no intensity to be converged.

In the example below, we converged 15 states, which covered an energy range up to 4.23 eV, and only obtained 
one state with non-zero intensity
%
at $\sim$3.6 eV, shown in Fig.~\ref{C60_UV} as a red stick. 
On the other hand the spectra computed with CPP-RI-CC2 and with a previously reported CPP-B3LYP approach (using the pol-Sadley [10s6p4d|5s3p2d] basis),~\cite{CPP:Joanna:solver2} cover the frequency range up to 7 eV and show very similar profiles, with 4 peaks of different intensities. The intensity of the bands is slightly larger in RI-CC2 compared to CPP-B3LYP, except for the second peak. The CPP-B3LYP spectrum is blue-shifted by $\sim$0.3 eV) with respect to the one obtained with CPP-RI-CC2.
%

\clearpage

\subsection{Electronic circular dichroism: Helicenes} 


Helicenes are prototypical systems that 
show chiro-optical activity not because of the presence of chiral centers ({\it e.g.}, asymmetric carbons), but because of the handedness of their helical structure, also known as axial chirality,
as the clockwise and counterclockwise helices are non-superposable. 
By convention, a left-handed helix is minus and labelled M, whereas a right-handed helix is plus and labelled P. 
The $n$-helicenes are also a prototypical example of overcrowded aromatic chromophores, and
the enantiomers possess a strong optical activity,\cite{Brown:1971:5:5-helicene,Newman:1967:6-helicene,Brickell:1971:7-helicene,Weigang:Emission:CD:1966,GOEDICKE1970937} which makes them ideal test systems for our CPP-RI-CC2 computational scheme. 
\revS{ECD spectra of helicenes have been theoretically studied before,~\cite{Furche2000,Koehn-PhD,Helicenes:Sergio,BUSS1996309,Helicenes:2012} \textit{e.g.} in 2000 by \citeauthor{Furche2000} at TDDFT level\cite{Furche2000} 
and, for 5- and 6-helicene, in 2003 by \citeauthor{Koehn-PhD}\cite{Koehn-PhD}
at CC2 level using the aug-cc-pVDZ basis supplemented with center of mass functions. The stick spectra in this later study included the lowest 24 and 20 states, respectively. 
In 2012 a combined theoretical and experimental study on several helicenes was also presented by 
\citeauthor{Helicenes:2012},\cite{Helicenes:2012} where the computed ECD spectra were obtained at the RI-CC2 level using the TZVPP basis set and 40 excited states.  
To illustrate the CPP approach,
we here extend the RI-CC2 studies of Refs.~\citenum{Koehn-PhD} and \citenum{Helicenes:2012} by investigating the penta-,  hexa-, and hepta-helicenes
using the larger aug-cc-pVTZ basis set.}
{Experimental spectra were re-digitized from the original references and are shown together with the calculated ones.}

\begin{figure}[hbt!]
\includegraphics[scale=0.8]{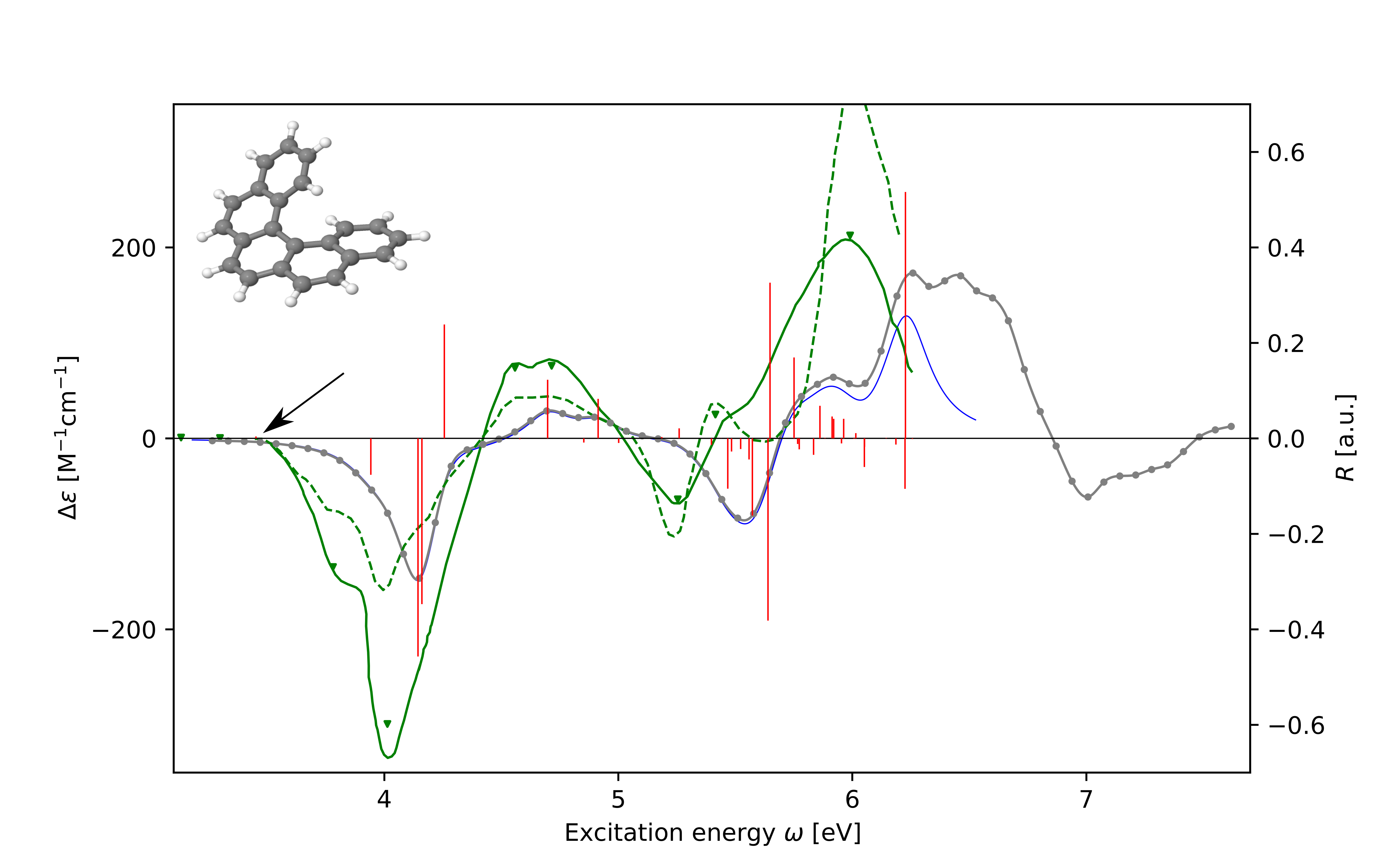}
\caption{\label{fig:ECD:H5} ($-$)-5-helicene (M):
(frozen-core) RI-CC2/aug-cc-pVTZ  ECD spectra from resonant linear response and CPP  calculations. The stick spectrum is reported in red. The spectrum in blue is a Lorentzian broadening of the stick spectrum. The grey circles are the CPP points, and the grey line is a cubic spline of CPP points. The solid green line is the experimental spectrum re-digitized from Ref.~\citenum{Brown:1971:5:5-helicene},
which is (supposedly) derived from the 
measurement in {\it iso}-octane  in Ref.~\citenum{GOEDICKE1970937} (green triangles). {The dashed green line is  the experimental spectrum re-digitized from Ref.~\citenum{Helicenes:2012}, recorded in 98:2 $n$-hexane/2-propanol.}}
\end{figure}

The ECD spectra of ($-$)-5-helicene are shown in Fig.~\ref{fig:ECD:H5}. By converging 40 excited states, we could obtain the (broadened) stick spectrum up to approximately {6.3 eV.}
Experimental~\cite{GOEDICKE1970937,Brown:1971:5:5-helicene,Helicenes:2012} and CPP spectra cover the frequency range up to 6.2 eV and 7.7 eV,
respectively.
One~\cite{GOEDICKE1970937,Brown:1971:5:5-helicene} of the shown experimental spectra was recorded in  \textit{iso}-octane. Note that we re-digitised the experimental spectrum reported in Figure 2 of Ref.\citenum{Brown:1971:5:5-helicene}. According to the authors,~\cite{Brown:1971:5:5-helicene}
this experimental spectrum was 
taken from the work of \citeauthor{GOEDICKE1970937},\cite{GOEDICKE1970937}
even though no image of the spectrum is actually given by \citeauthor{GOEDICKE1970937},
who only report individual values of 
$\Delta\epsilon$ at given wavelengths.
Spectral data from both articles is presented 
as green continuum line and triangles in Fig.~\ref{fig:ECD:H5}. 
We observe small inconsistencies at around 4 eV and 5.5 eV between the spectrum re-digitized from Ref.\citenum{Brown:1971:5:5-helicene} and the spectral points taken from Ref.\citenum{GOEDICKE1970937} (green triangles).
\revS{The experimental spectrum from Ref.~\citenum{Helicenes:2012},
recorded in 98:2 $n$-hexane/2-propanol, 
is shown as dashed green line.}

The stick spectrum starts with one positive peak of symmetry A and very low intensity (marked by an arrow). Roughly in the same region, experiment~\cite{GOEDICKE1970937,Brown:1971:5:5-helicene} shows two low intensity positive features (ca.\ 50 times weaker than the rest of the spectrum).~\cite{GOEDICKE1970937,Brown:1971:5:5-helicene} 
The CPP and {the} Lorentzian-broadened spectra are practically indistinguishable up to around {5.85 eV},
where differences start to emerge, as peaks may be missing.
The computed and experimental spectra have similar features: two negative bands, one at around {4 eV} 
and one just above {5 eV},
two positive overlapping bands 
at {around 4.5$-$4.7 eV}
and a feature-rich positive band, starting 
in between 5 and 6 eV, clearly due to a large number of transitions. The computed spectra (in vacuo) are slightly blue-shifted and of lower intensity 
compared to the experimental data in \textit{iso}-octane.~\cite{GOEDICKE1970937}
%

\begin{figure}[hbt!]
\includegraphics[scale=0.8]{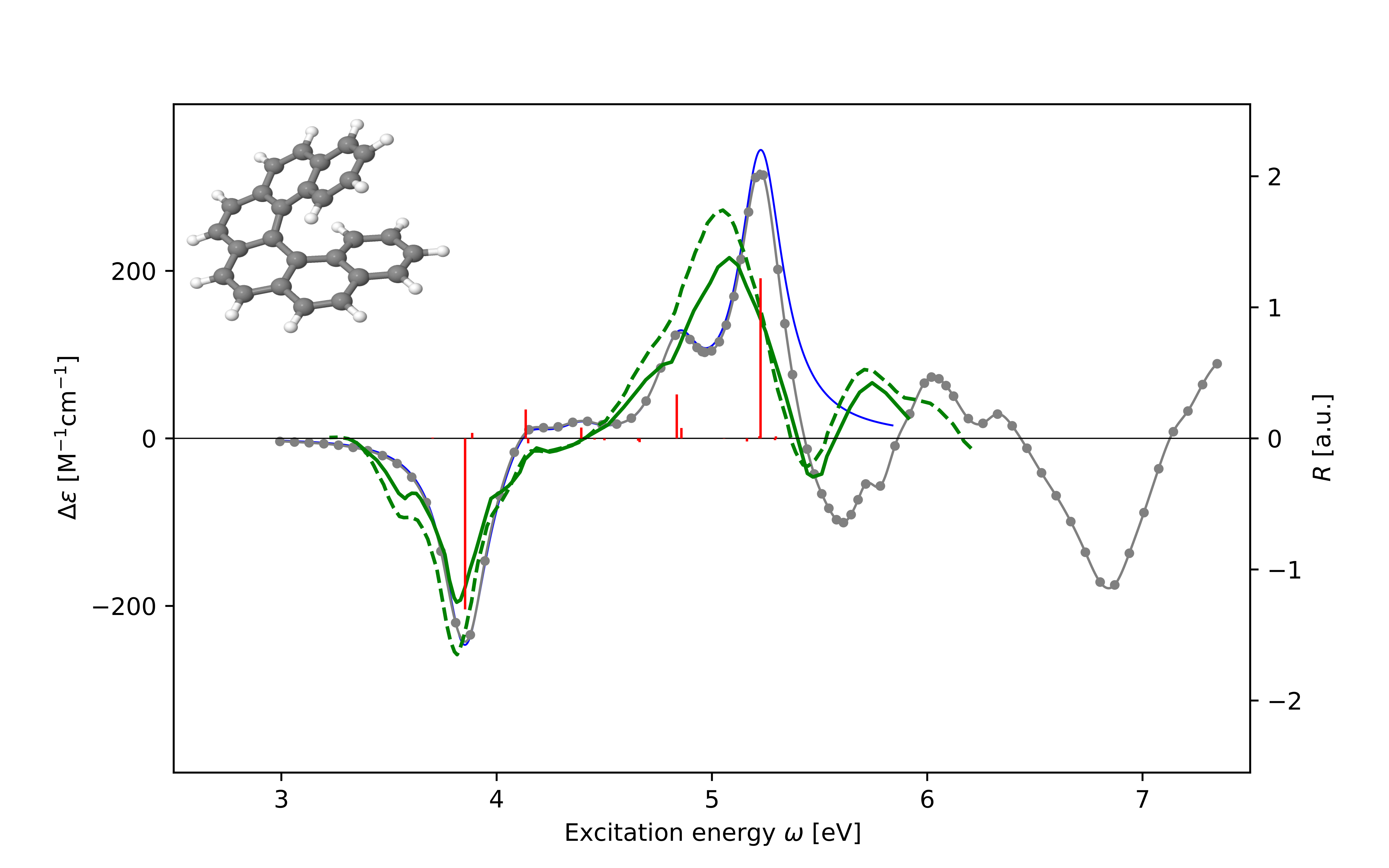}
\caption{\label{fig:ECD:H6} ($-$)-6-helicene (M): (frozen-core) RI-CC2/aug-cc-pVTZ ECD spectra from standard linear response and CPP  calculations. The stick spectrum is reported in red. The spectrum in blue is a Lorentzian broadening of the red stick spectrum. 
The grey circles are the CPP points, and the grey line is a cubic spline {through} the CPP points. A {mirror} image of the experimental spectrum of ($+$)-6-helicene (P) from Ref.~\onlinecite{Newman:1967:6-helicene}, recorded in methanol, 
is shown as green solid line. A dashed green line shows the experimental spectrum
recorded in acetonitrile, re-digitised from Ref.~\citenum{Helicenes:2012}.
} 
\end{figure}

The ECD spectra for ($-$)-6-helicene are presented in Fig.~\ref{fig:ECD:H6}. Note that the experimental measurement from Ref.~\citenum{Newman:1967:6-helicene} was carried out in methanol on the P structure, so we have reversed its sign when comparing it in Fig.~\ref{fig:ECD:H6} with the spectra computed for the M enantiomer (solid green line).
\revS{The experimental CD spectrum recorded in acetonitrile from Ref.~\citenum{Helicenes:2012} is also shown as dashed green line.}

As the system size increases, it becomes progressively more challenging to converge the stick spectra. For the 6-helicene, our stick spectrum contains the first 20 excited states. This, however, only covers the region up to 
{5.2 eV}.
The CPP spectrum was computed up to {7.3 eV}.
The CPP and broadened stick spectra start to differ around {5.2 eV}. 
Indeed, the intensity of the strongest positive peak predicted by the CPP spectrum is slightly lower than the one obtained from broadening the stick spectrum, probably the effect of the broad negative band, located in between 5.5 eV and 5.7 eV,
clearly not present in the broadened spectrum as the corresponding sticks were not computed. 

All in all, as for 5-helicene, the computed and experimental spectra of 6-helicene have rather similar features: a relatively strong negative peak at around {3.8 eV};
two (partly overlapping) positive peaks in between {4.8 eV and 5.3 eV}, 
followed by a bisignate band in between 5.5 eV and 6.3 eV.
The computed first negative peak at {3.9 eV} 
is {marginally} {blue}-shifted 
with respect to the experimental band. 
%
The band intensities in the simulated spectrum are only slightly {larger} than the corresponding ones in the experimental spectrum recorded in methanol. The lowest energy band is practically overlapping with the same band from the experimental measurement in acetonitrile.~\cite{Helicenes:2012}
%

\begin{figure}[hbt!]
\includegraphics[width=16cm]{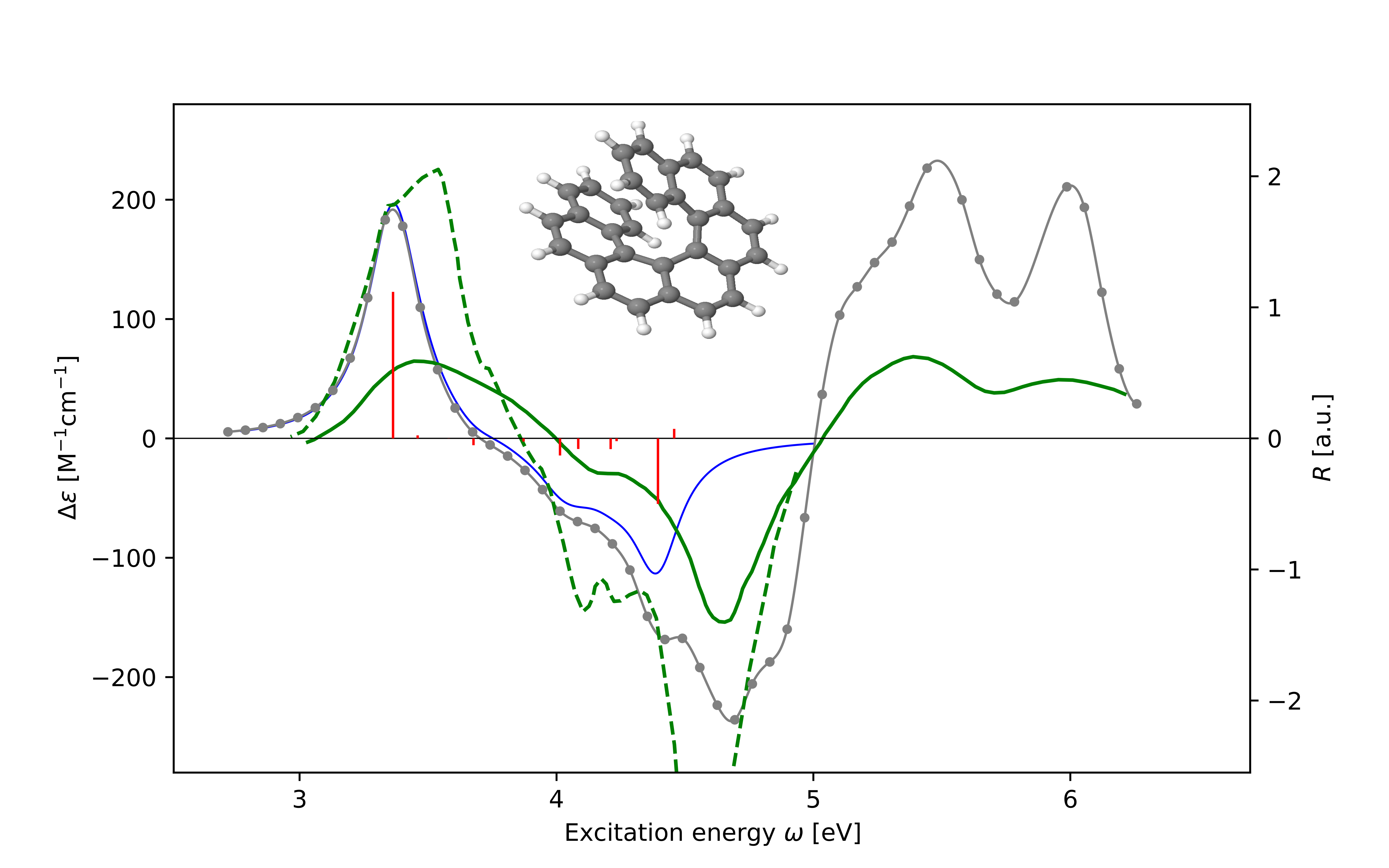}
\caption{\label{fig:ECD:H7}  (+)-7-helicene (P): (frozen-core) RI-CC2/aug-cc-pVTZ  ECD spectra from resonant linear response and CPP calculations. The stick spectrum is reported in red. The spectrum in blue is a Lorentzian broadening of the red stick spectrum. The grey circles are the CPP points, and the grey line is a cubic spline through the CPP points. A mirror image of the experimental spectrum of ($-$)-7-helicene (M) from Ref.~\onlinecite{Brickell:1971:7-helicene}, recorded in ethanol, is shown in solid green.
The experimental spectrum in chloroform~\cite{MARTIN1974343,Helicenes:2012}
is given as dashed green line.}
\end{figure}

The ECD spectra for (+)-7-helicene are presented in Fig.~\ref{fig:ECD:H7}. The mirror image of the experimental spectrum  of ($-$)-7-helicene, recorded in ethanol by 
~\citeauthor{Brickell:1971:7-helicene}~\cite{Brickell:1971:7-helicene}, is also shown as green solid line on Fig. \ref{fig:ECD:H7}.
\revS{The experimental spectrum in chloroform
reported by \citeauthor{Helicenes:2012},
originally taken from \citeauthor{MARTIN1974343},
is also shown as dashed line}.
The CPP spectrum was obtained up to {6.25 eV} 
and shows a well-separated positive peak 
{between 3 and 3.7 eV;}
a feature-rich negative band between {3.7 and 5 eV,} 
clearly with contributions from several transitions of different intensity,
and with maximum at {4.7 eV;}
and two positive peaks at 
{5.44 eV and 5.99 eV,} 
respectively. 
All peaks are of comparable intensity. 
Not surprising, with the increase in complexity of the system, our ability to compute the stick spectrum deteriorates. 
Indeed, for 7-helicene, we only succeeded in converging 12 states, which covers the region up to 4.5 eV,
thus only reproducing the first (positive) peak and half of the negative broad band. 

Despite the different environments, the experimental spectral profiles are,
as in the previous two cases, quite similar to the computed one, with a broad positive band at lower energy,  a structured negative one in the intermediate region and two positive 
bands in the upper frequency region. 
The experimental intensity of the spectrum in ethanol is, on the other hand, roughly {two times} {lower}, whereas the one in chloroform is a more intense, in particular in the intermediate frequency region.

\subsection{The $C_6$ dispersion coefficients}

In Table~\ref{tab:C6:orgmol} we present the 
 $C_{6}$~dispersion coefficients for the dimers of a set of ten organic molecules and in Table \ref{tab:C6:fullerenes} the results for six different fullerenes are collected.

\begin{table}[htp!]
\caption{\label{tab:C6:orgmol}RI-CC2/aug-cc-pVTZ $C_{6}$ dispersion coefficients (a.u.) of the dimers of ten organic molecules, and comparison with previous theoretical results,  obtained at the ADC(2)/Sadlej-pVTZ\cite{CPP:ADC} and CCSD/Sadlej-pVTZ\cite{CPP:ADC} levels of theory. 
}
\begin{tabular}{lccccr}
\hline
\hline
Molecules 	  &	{RI-CC2}  &  ADC(2)\cite{CPP:ADC} &{(ADC(2)~$-$~CC2)}& CCSD\cite{CPP:ADC}
\\
\hline
Acetaldehyde	&	432.8	&	434.3	&		1.5	    &	407.2		
\\
Acetone	        &	834.0	&	832.0	&		$-$2.0	    &	787.4	
\\
Benzene	        &	1874	&	1926	&		52.0	&	1786		
\\
Butane	        &	1285	&	1263	&		$-$21.8	&	1224		
\\
Ethane	        &	374.5	&	365.9	&	$-$8.6	    &	357.3		
\\
Ethene	        &	305.9	&	299.8	&		$-$6.1	    &	287.3	
\\
Formaldehyde	&	154.0	&	157.6	&		3.6	    &	144.8		
\\
Methane	        &	126.3	&	122.7	&		$-$3.6	    &	120.7	
\\
Pentane	        &	1950	&	1918	&		$-$31.6	&	1855		
\\
Propane	        &	759.9	&	745.1	&		$-$14.8	&	724.2		
\\
\hline
\hline
\end{tabular}
\end{table}


The RI-CC2 values for the $C_6$ coefficients of the organic molecules are in line with the results of a previous theoretical study  at the ADC(2)/Sadlej-pVTZ level.\cite{CPP:ADC} 
The difference (in absolute value) between CC2 and ADC(2) becomes larger as the size of the molecular system increases.
Both RI-CC2 and ADC(2) results are systematically larger than corresponding CCSD/Sadlej-pVTZ results from the literature,~\cite{CPP:ADC}
which were obtained using a Lanczos-based 
implementation of the polarizability at imaginary frequencies~\cite{Lanczos2} and a moderate
chain length.

The CC2/cc-pVDZ {results for the} $C_6$ coefficients for the fullerenes, see  Table~\ref{tab:C6:fullerenes}, are compared to literature results at the CAM-B3LYP, B3LYP, and TD-HF levels of theory,~\cite{CPP:Joanna:C6} 
obtained with the pol-Sadlej basis set. 
We note that our basis set is on the small side, so our results are probably not fully converged. Indeed, adding one set of augmented functions increased the coefficient for C$_{60}$ to 117.7$\times10^3$. At the MP2 geometry used in the OPA calculations, 
the $C_{6}$ coefficient of C$_{60}$ changes from 96.00$\times10^3$ (cc-pVDZ) to 116.3$\times10^3$ (aug-cc-pVDZ) to 115.4$\times10^3$ (aug-cc-pVTZ).
A reference value, obtained by the Differential Dipole Oscillator Strength Distribution (DOSD) approach is available for C$_{60}$. In the DOSD approach, the $C_{6}$ coefficients are derived from
the  dipole  oscillator  strength  distributions constructed from theoretical and experimental photoabsorption  cross  sections,  combined  with  constraints  provided  by  the  Kuhn–Reiche–Thomas sum rule, and molar refractivity data.~\cite{DOSD} 
As already commented upon in Ref.~\citenum{CPP:Joanna:C6}, the TD-HF/pol-Sadley result is the closest to the DOSD value, but the good agreement is probably fortuitous.

\begin{figure}[hbt!]
\includegraphics[width=16cm]{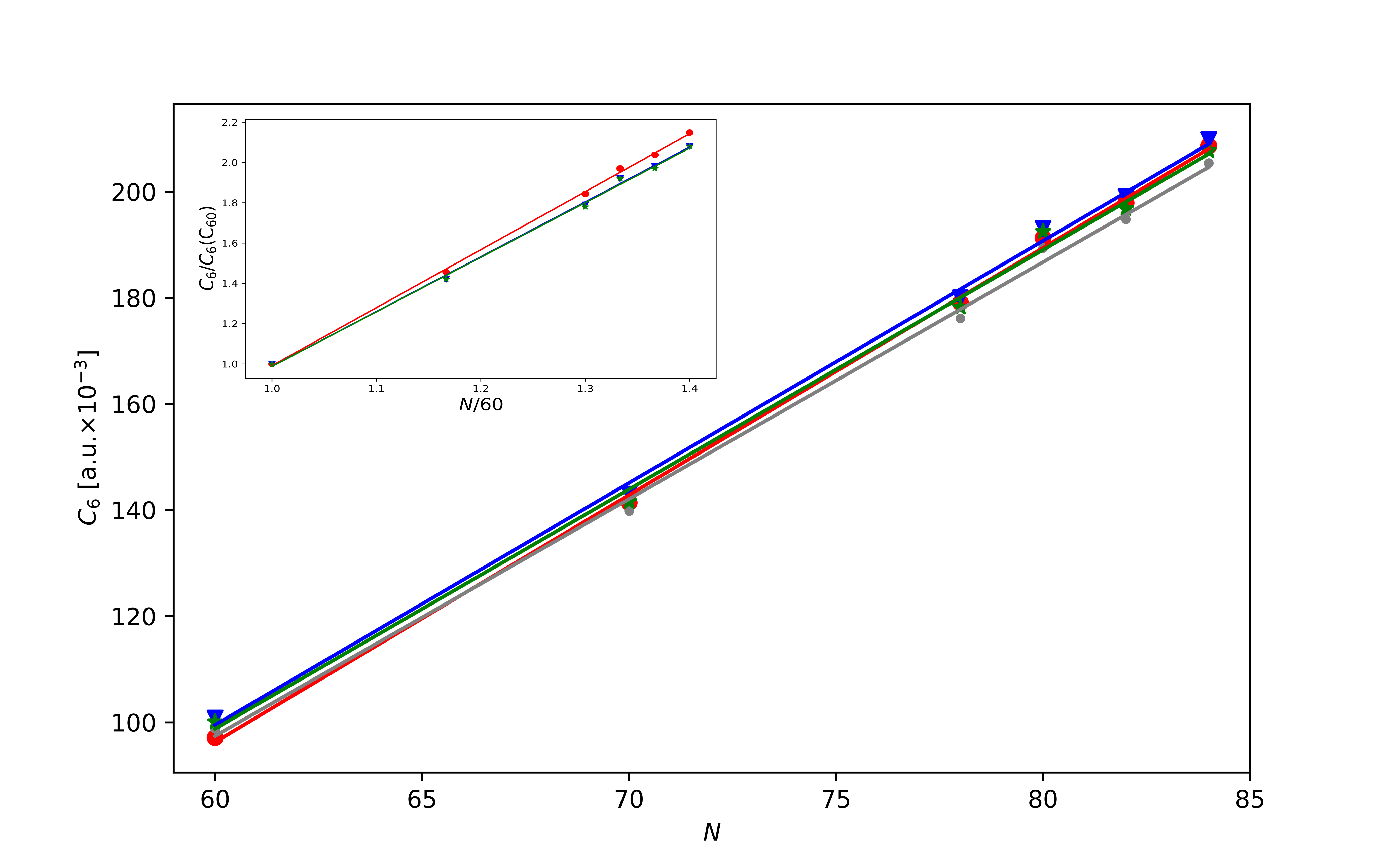}
\caption{\label{fig:C6:linreg} $C_{6}$ coefficients [a.u.$\times 10^{-3}$] of six fullerenes as a function of the number of carbon atoms $N$, plotted for four different electronic structure methods. The inset shows the ratios $C_6$(C$_N$)/$C_{6}$(C$_{60}$) versus $N/60$. Red: RI-CC2; blue: B3LYP; grey: CAMB3LYP; green: TD-HF results. The lines are linear regressions of the $C_6$ points.
The regression coefficients for the lines in the main panel are:
$r=0.9995$ (RI-CC2); $r=0.9986$ (HF); 
$r=0.9991$ (B3LYP) and $r=0.9989$ (CAM-B3LYP).
Those of the lines in the inset are:
$r=0.9995$ (RI-CC2); $r=0.9988$ (HF); 
$r=0.9991$ (B3LYP) and $r=0.9988$ (CAM-B3LYP).}
\end{figure}

\begin{figure}[hbt!]
\includegraphics[width=16cm]{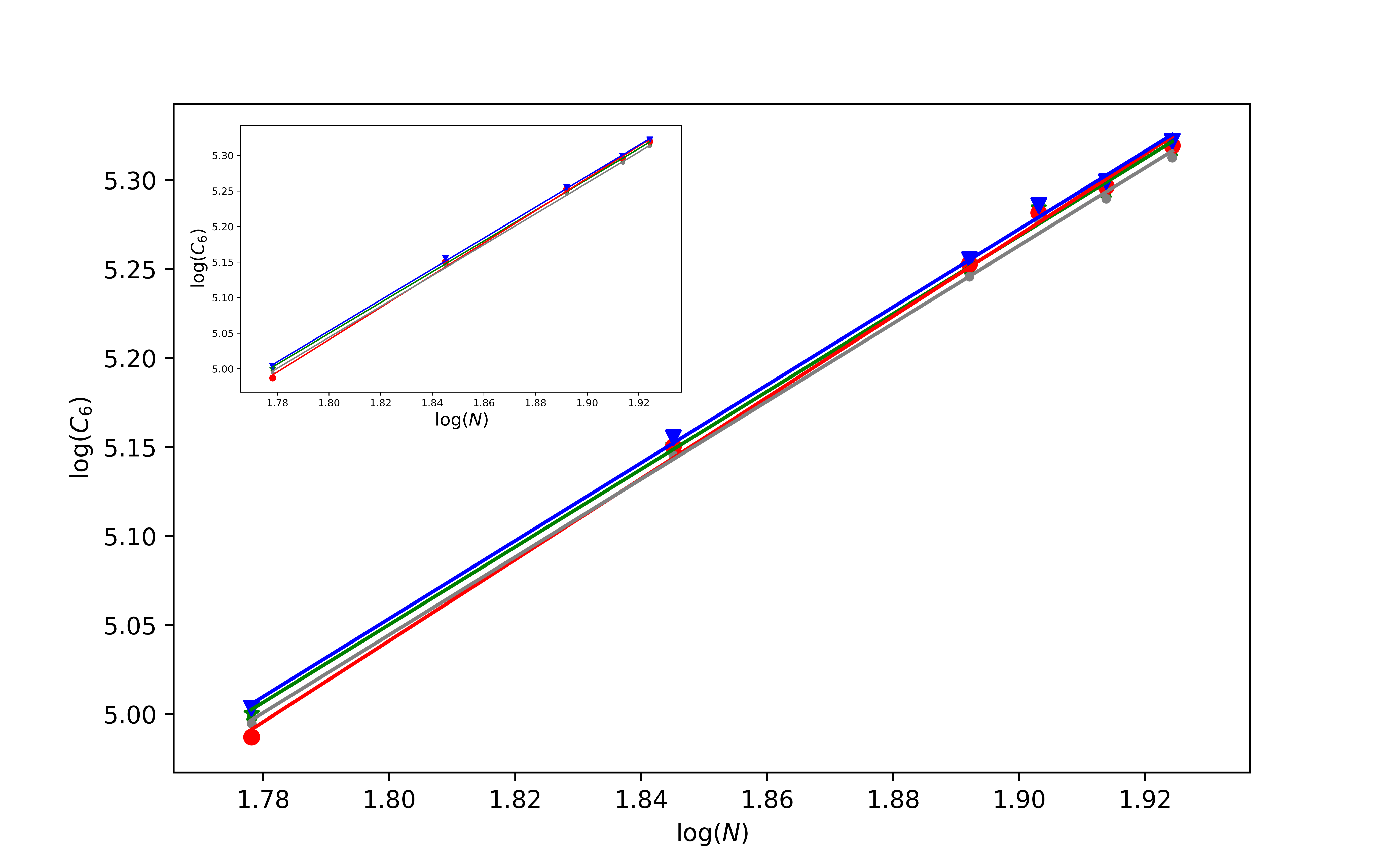}
\caption{\label{fig:C6:log} Base-10 logarithm of the $C_{6}$ coefficients of six fullerenes as a function of $\log{N}$, for four different electronic structure methods. Red: RI-CC2 (this work, $r=0.9992$); blue: B3LYP ($r=0.9994)$; grey: CAMB3LYP ($r=0.9994$); green: TD-HF  ($r=0.9991$) results. The inset shows the data without consideration of C$_{80}$, which slightly improves the linear regression coefficient: $r = 0.9998$ for HF, $r = 0.9998$ for B3LYP,
$r=0.9998$  for CAM-B3LYP, and $r=0.9994$ for RI-CC2.}
\end{figure}

In Figure~\ref{fig:C6:linreg}, the $C_6$ coefficients are plotted as a function of the number $N$ of carbon atoms in the considered fullerenes.
\revS{In the inset, a plot of the ratios $C_6$(C$_N$)/$C_{6}$(C$_{60}$) 
versus the $N/60$ is given.
Figure~\ref{fig:C6:log} reports the base-10 logarithm of $C_6$ coefficients as a function of $log{N}$. The latter figure is used to determine the exponent $\eta$ at the RI-CC2 level of the ansatz $C_6 \propto N^\eta$, as also reported by ~\citeauthor{CPP:Joanna:C6}.~\cite{CPP:Joanna:C6}
\citeauthor{CPP:Joanna:C6}~\cite{CPP:Joanna:C6} found that the $C_6$ coefficients were non-additive, and scaled roughly as $N^{2.2}$ for the three methods they considered. The exponent for the $C_6$ power-dependence on $N$ was therefore much smaller than the values predicted based on a classical-metallic spherical-shell approximation of the fullerenes ($\approx$2.75).~\cite{PhysRevLett.109.233203}
In a later study, some of the same authors\cite{Saidi:2016} proposed a model based on classical electrodynamics
that yielded $C_6 \propto N^{2.8}$.
Our results at the CC2 level, based on small-size fullerenes, give $\eta = 2.3$, {\it i.e.} only marginally larger than the HF/DFT estimates of \citeauthor{CPP:Joanna:C6}.
Removing C$_{80}$ from the series slightly improves the linear regression coefficients, but does not significantly change the value
of $\eta$.


}
\clearpage
\begin{table}[htp!]
\caption{\label{tab:C6:fullerenes}RI-CC2 $C_{6}$ dispersion coefficients 
[a.u.$\times$10$^{-3}$] for a set of fullerene dimers, and comparison with previous literature results.~\cite{CPP:Joanna:C6,DOSD} 
}
\begin{tabular}{lccccc}
\hline
\hline
 {Molecule}	  &	{RI-CC2/}  & B3LYP/  
 & CAM-B3LYP/  &TD-HF/ & DOSD~\cite{DOSD} 
 \\
 & cc-pVDZ & pol-Sadley~\cite{CPP:Joanna:C6} & 
 pol-Sadley~\cite{CPP:Joanna:C6}& pol-Sadley~\cite{CPP:Joanna:C6} & \\
\hline
C$_{60}$               
&   97.08$^a$  & 100.8 & 98.8  & 100.1 & 100.3  \\
C$_{70}$               &  141.4  & 143.0 & 139.8 & 141.6 &        \\
C$_{78}$               &  179.1  & 180.0 & 176.1 & 178.2 &        \\
C$_{80}$               &  191.3  & 193.1 & 189.4 & 192.5 &        \\
C$_{82}$               &  197.9  & 199.1 & 194.8 & 196.8 &        \\
C$_{84}$               &  208.6  & 209.8 & 205.4 & 207.7 &        \\
\hline
\hline
\end{tabular}\\
$^a$117.7$\times 10^3$ (aug-cc-pVDZ); \\
At MP2 geometry, 
96.00$\times 10^3$ (cc-pVDZ); 116.3$\times 10^3$ (aug-cc-pVDZ);
115.4$\times 10^3$ (aug-cc-pVTZ)
\end{table}

\clearpage
\section{Conclusions}
We have presented an implementation of
a damped linear response solver and of the damped linear response function within 
the resolution-of-identity CC2 method in TURBOMOLE.~\cite{Turbomole2020}
The LR-CPP-RI-CC2 approach allows to directly compute, e.g.,  ECD and OPA spectra of systems with a high density of excited states, where a standard (stick-spectrum) response approach is hardly or not applicable.
The combination of the RI approximation with a partitioned formulation which avoids the storage and I/O of four-index two-electron integrals and double excitation amplitudes (employing a Laplace transformation of orbital energy denominators) together with an OpenMP parallelization makes the LR-CPP-RI-CC2 approach applicable to molecular systems as large as fullerenes and helicenes.

Examples of application of the approach included
the OPA spectra of C$_{60}$,
the ECD spectra of $n$-helicenes ($n= 5, 6, 7$) and the $C_6$ dispersion coefficients for a sample of organic molecules and fullerenes.

The CPP solver for RI-CC2 is also a fundamental step-stone for the implementation of higher-order response properties in a RI-CC2
CPP framework, like, e.g. RIXS and MCD, as well as for the extension to excited state properties.
\clearpage
\section{Appendix}
\label{appendix}
\subsection*{Additional definitions}
\begin{itemize}
\item $T_1$-similarity-transformed MO coefficients: $$\boldsymbol{\Lambda}^p = {\bf{C}}({\bf{1}}-{\bf{t}}^T_1)~; \quad
\boldsymbol{\Lambda}^h = {\bf{C}}({\bf{1}}+{\bf{t}}_1)$$
\item $T_1$-similarity-transformed one-electron integrals: $${\hat{h}}^x_{pq} = \sum_{\alpha\beta}\Lambda^p_{\alpha p}\Lambda^h_{\beta q}
h^x_{\alpha \beta}
$$
\item $T_1$-similarity-transformed two-electron integrals: $$(pq\hat{|}rq) = \sum_{\alpha\beta\gamma\delta}\Lambda^p_{\alpha p}\Lambda^h_{\beta q}\Lambda^p_{\gamma r}\Lambda^h_{\delta s}(\alpha\beta|\gamma\delta)$$
\item the elements of $\hat{\mathbf{F}}$ are defined like the elements of the usual Fock matrix, but evaluated with $T_1$-similarity-transformed one- and two-electron integrals
\item 
One-index transformed
$\breve{\boldsymbol{\Lambda}}^p$
and $\breve{\boldsymbol{\Lambda}}^h$ matrices: 
$$\breve{\Lambda}^p_{\beta i} = \sum_{a} {\Lambda}^p_{\beta a}{\bar{t}}_{ia} \quad\text{and}\quad
\breve{\Lambda}^h_{\beta a} = -\sum_{i} {\Lambda}^h_{\beta i}{\bar{t}}_{ia}$$
\item  
$t^x$-dressed one-index transformed ${\bar{\boldsymbol{\Lambda}}}^{p,x}$ 
and $\bar{\boldsymbol{\Lambda}}^{h,x}$ matrices {(*)}: 
$$\bar{\Lambda}^{p,x}_{\beta a} = -\sum_{i} C_{\beta i} t^x_{ai}
\quad\text{and}\quad
{\bar{\Lambda}^{h,x}_{\beta i} = \sum_{a} C_{\beta a}{t}^x_{ai}}
$$
\item barred one-electron and three- and four-centre two-electron integrals {(*)}:
\begin{align*}
   \bar{h}^x_{pq} = & \sum_{\alpha\beta}
    \left(
\bar{\Lambda}^{p,x}_{\alpha p} \Lambda^h_{\beta q}
+
{\Lambda}^p_{\alpha p} \bar{\Lambda}^{h,x}_{\beta q}
\right) h_{\alpha\beta}
\\
  \bar{B}^x_{Q,pq}
  = & \sum_P (pq\overline{|}P)V^{-\frac{1}{2}}_{PQ}
  =
  \sum_{\alpha\beta}
  \left(
\bar{\Lambda}^{p,x}_{\alpha p} \Lambda^h_{\beta q}
+
{\Lambda}^p_{\alpha p} \bar{\Lambda}^{h,x}_{\beta q}
\right)
  \sum_{P} (\alpha\beta|P) V^{-\frac{1}{2}}_{PQ}~.
\\
  \bar{(pq|rs)}^x = & 
   \hat{P}^{pr}_{qs} \sum_{\alpha\beta\gamma\delta}
    \left(
\bar{\Lambda}^{p,x}_{\alpha p} \Lambda^h_{\beta q}
+
{\Lambda}^p_{\alpha p} \bar{\Lambda}^{h,x}_{\beta q}
\right) 
 \Lambda^p_{\gamma r} \Lambda^h_{\delta s} (\alpha\beta|\gamma\delta)
\end{align*}
Here it is understood that $\bar{\Lambda}^{p,x}_{\alpha p}$ vanishes if $p$ is an occupied and $\bar{\Lambda}^{h,x}_{\beta q}$ vanishes if $q$ is a virtual index.
%
\item barred Fock matrices and $E$ intermediates {(*)}:
   \begin{align*}
       \bar{F}^x_{ia} = & \sum_{ck} \left[ 2(ia|kc) - (ic|ka) \right] t^x_{ck}
      \\
        \bar{F}^x_{ab} = & -\sum_j t^x_{aj} \hat{F}_{jb} + \sum_{ck} \left[ 2(ab|kc) - (ac|kb) \right] t^x_{ck}
      \\
        \bar{F}^x_{ij} = & + \sum_b \hat{F}_{jb} t^x_{bi} + \sum_{ck} \left[ 2(ij|kc) - (ic|kj) \right] t^x_{ck}
      \\
         \bar{E}^{x,2}_{ij} = & \bar{F}^x_{ij} + \sum_{cdk} \left[ 2 t^{x,jk}_{cd} - t^{x,jk}_{dc} \right] (kd|ic)
      \\
         \bar{E}^{x,2}_{ba} = & \bar{F}^x_{ba} + \sum_{dkl} \left[ 2 t^{x,kl}_{bd} - t^{x,kl}_{db} \right] (ld|ka)
   \end{align*}
\end{itemize}
Above an asterisk (*) indicates that the respective intermediates depend linearly on the complex response amplitude $t^x$ and have been generalised for the CPP case such that their real and imaginary parts are evaluated, respectively, with the real and imaginary parts of $t^x$.

\begin{figure}
\centering
\caption{Main steps in the evaluation of the F-matrix contraction terms. (\text{*}) indicates the terms that have been generalised in the CPP case (complex). \label{fig:F}}
\begin{minipage}[r]{1\textwidth}
\begin{framed}
\begin{itemize}
\footnotesize
\item Compute $\breve{Y}_{Q,ck} =
 \sum_{dj} \bar{t}_{dc}^{jk} \hat{B}_{Q,dj}$ \quad \quad (standard code)
\item Compute $\bar{F}^x_{ia}$, $\bar{E}^{x,1}_{ab}$, $\bar{E}^{x,2}_{ij}$, $\bar{B}^x_{Q,bj}$, $\bar{B}^x_{Q,ij}$, $C^x_{ck}$, $Y^x_{Q,ai}$ (*)
    %
    \item Compute dressed integrals $\breve{B}^x_{Q,jb} 
    {= - \sum_{ck} \Big( \bar{t}_{jc} t^x_{ck} B_{Q,kb} + t^x_{ck} \bar{t}_{kb} B_{Q,jc} \Big)} $ (*) (Eq.~\ref{Bupx})
    \item Compute $\sigma^{0,x}_{ia} = 2 \bar{F}^x_{ia}$, \ $\sigma^{JG,x}_{ia}$
    and $\sigma^{JH,x}_{ia}$ (*)
    \item Compute  intermediates: 
     \begin{itemize}
\item 
$\breve{\Lambda}^p_{\beta i}$, $\breve{\Lambda}^h_{\beta a}$ 
and  $\breve{B}_{Q,ia} = \sum_{\alpha\beta{P}} \Lambda^h_{\alpha a} \breve{\Lambda}^p_{\beta i} (\alpha\beta|P) V^{-1/2}_{PQ} - {\sum_k}\hat{B}_{Q,ik} \bar{t}_{ka}$ \quad \quad (standard code)
\item $\bar{t}^{ij}_{ab} = (2-\hat{P}_{ij}) \hat{P}^{ij}_{ab} \Big(\sum_Q \breve{B}_{Q,ai} B_{Q,jb} + \bar{t}_{ia} \hat{F}_{jb} \Big) / \Big( \epsilon_i - \epsilon_a + \epsilon_j - \epsilon_b\Big)$ \quad \quad (standard code)
    \item 
    ${\breve{Y}^x_{Q,ia}} = \sum_{bj} \bar{t}^{ij}_{ab} \bar{B}^x_{Q,jb}$ (*)
    \item $m^x_Q = \sum_{ck} B_{Q,ck} C^x_{ck}$ and $M^x_{Q,ik} = \sum_c B_{Q,ic}C^x_{ck}$ (*)
    \item $\breve{i}^x_Q = \sum_{jb} \bar{B}^x_{Q,bj} \bar{t}_{jb}$  (*)
    \item
    $\breve{\Gamma}^x_{Q,\beta i} = \sum_{Pa} \big(\breve{Y}^{x}_{P,ia} - \sum_j \bar{t}_{ja} \bar{B}^x_{P,ij} \big) V^{-1/2}_{PQ} \Lambda^p_{\beta a} - \sum_{Pk} M^x_{P,ik} V^{-1/2}_{PQ} \Lambda^p_{\beta k} + 2\sum_P (m^x_P + \breve{i}^x_P) V^{-1/2}_{PQ} \Lambda^p_{\beta i}$~~(*) 
        \item  $\sigma^{IJ^\prime_{12}F_{1},x}_{i\alpha} = \sum_{Q\beta} \breve{\Gamma}^x_{Q,\beta i} (Q|\beta\alpha)$ 
        \end{itemize}
        \item Compute $\sigma^{IJ^\prime_{12}F_1,x}_{ia} = \sum_\alpha \sigma^{IJ^\prime_{12}F_1}_{i\alpha} C_{\alpha a}$
        \quad ($\sigma^{I,x}_{ia}$ + 1st term of $\sigma^{F,x}_{ia}$
        + 1st \& 2nd term of 
        $\sigma^{J^{\prime},x}_{ia})$
        ~~(*)
        \item 
        Compute $\sigma^{F_{3},x}_{ia} = -\sum_{lQ} \breve{Y}^{x}_{Q,al} \hat{B}_{Q,il}$ \quad (3rd term of $\sigma^{F,x}_{ia}$)~~(*) 
    \item Compute intermediate $\breve{\Gamma}^{x''}_{Q,i\beta} = \sum_{Pd} \Big( \breve{Y}_{P,id} - \sum_j \hat{B}_{P,ij} \bar{t}_{jd} \Big) \bar{\Lambda}^{p,x}_{\beta d} V^{-1/2}_{PQ}$~~(*) 
    \item Compute $\sigma^{F_2J^{\prime}_3,x}_{i\alpha} = \sum_{Q\beta} \breve{\Gamma}^{x''}_{Q,i\beta} (Q|\beta\alpha)$~~(*) 
    \item 
    Compute $\sigma^{F_2J^\prime_3,x}_{ia} = \sum_\alpha \sigma^{F_2J_3,x}_{i\alpha} C_{\alpha a}$ (2nd term of $\sigma^{F,x}_{ia}$+
    3rd term of $\sigma^{J^{\prime},x}_{ia}$) ~~(*) 
    \item Calculate $\sigma^{F_4,x}_{ia} = -\sum_{lQ} \breve{Y}_{Q,al} \bar{B}^x_{Q,il}$~~(*) 
    \item Contract with single amplitudes and add the doubles contributions \\
    $ Ft^x t^y = \sum_{ai} \sigma^x_{ia} t^{y}_{ai} + \sum_{ai} \bar{F}^x_{ia} C^y_{ai} 
    + \sum_{Q,bj}
    \breve{B}^x_{Q,jb} \cdot Y^y_{Q,bj}$~~(*) 
\end{itemize}
\end{framed}
\end{minipage}
\end{figure}
\clearpage

\section*{Data availability}
The data that support the findings of this study are available
within the article and its supplementary material.

\begin{acknowledgments}
We thank Dr. Thomas Fransson (University of Heidelberg) and Prof. Wissam Saidi (University of Pittsburgh) 
for sending us the Cartesian coordinates of the molecular systems considered for the $C_6$ coefficients.
D. A. F. thanks Dr. Rasmus Faber (DTU) and Dr. Alireza Marefat Khah (RUB) for valuable discussions.
D. A. F. and S. C. acknowledge financial support from the Marie Sk{\l}odowska-Curie European
Training Network ``COSINE-COmputational Spectroscopy In Natural sciences and Engineering'', Grant Agreement
No. 765739.
S. C. acknowledges 
the Independent Research
Fund Denmark – Natural Sciences, Research Project 2, grant no. 7014-00258B.
C. H. acknowledges financial support by the DFG through grant no. HA 2588/8.
\end{acknowledgments}

\bibliography{daniil}

\begin{thebibliography}{69}%
\makeatletter
\providecommand \@ifxundefined [1]{%
 \@ifx{#1\undefined}
}%
\providecommand \@ifnum [1]{%
 \ifnum #1\expandafter \@firstoftwo
 \else \expandafter \@secondoftwo
 \fi
}%
\providecommand \@ifx [1]{%
 \ifx #1\expandafter \@firstoftwo
 \else \expandafter \@secondoftwo
 \fi
}%
\providecommand \natexlab [1]{#1}%
\providecommand \enquote  [1]{``#1''}%
\providecommand \bibnamefont  [1]{#1}%
\providecommand \bibfnamefont [1]{#1}%
\providecommand \citenamefont [1]{#1}%
\providecommand \href@noop [0]{\@secondoftwo}%
\providecommand \href [0]{\begingroup \@sanitize@url \@href}%
\providecommand \@href[1]{\@@startlink{#1}\@@href}%
\providecommand \@@href[1]{\endgroup#1\@@endlink}%
\providecommand \@sanitize@url [0]{\catcode `\\12\catcode `\$12\catcode
  `\&12\catcode `\#12\catcode `\^12\catcode `\_12\catcode `\%12\relax}%
\providecommand \@@startlink[1]{}%
\providecommand \@@endlink[0]{}%
\providecommand \url  [0]{\begingroup\@sanitize@url \@url }%
\providecommand \@url [1]{\endgroup\@href {#1}{\urlprefix }}%
\providecommand \urlprefix  [0]{URL }%
\providecommand \Eprint [0]{\href }%
\providecommand \doibase [0]{http://dx.doi.org/}%
\providecommand \selectlanguage [0]{\@gobble}%
\providecommand \bibinfo  [0]{\@secondoftwo}%
\providecommand \bibfield  [0]{\@secondoftwo}%
\providecommand \translation [1]{[#1]}%
\providecommand \BibitemOpen [0]{}%
\providecommand \bibitemStop [0]{}%
\providecommand \bibitemNoStop [0]{.\EOS\space}%
\providecommand \EOS [0]{\spacefactor3000\relax}%
\providecommand \BibitemShut  [1]{\csname bibitem#1\endcsname}%
\let\auto@bib@innerbib\@empty
\bibitem [{\citenamefont {Kristensen}\ \emph {et~al.}(2009)\citenamefont
  {Kristensen}, \citenamefont {Kauczor}, \citenamefont {Kjaergaard},\ and\
  \citenamefont {J{\o}rgensen}}]{DampedResponse}%
  \BibitemOpen
  \bibfield  {author} {\bibinfo {author} {\bibfnamefont {K.}~\bibnamefont
  {Kristensen}}, \bibinfo {author} {\bibfnamefont {J.}~\bibnamefont {Kauczor}},
  \bibinfo {author} {\bibfnamefont {T.}~\bibnamefont {Kjaergaard}}, \ and\
  \bibinfo {author} {\bibfnamefont {P.}~\bibnamefont {J{\o}rgensen}},\
  }\bibfield  {title} {\enquote {\bibinfo {title} {Quasienergy formulation of
  damped response theory},}\ }\href {\doibase 10.1063/1.3173828} {\bibfield
  {journal} {\bibinfo  {journal} {J. Chem. Phys.}\ }\textbf {\bibinfo {volume}
  {131}},\ \bibinfo {pages} {044112} (\bibinfo {year} {2009})}\BibitemShut
  {NoStop}%
\bibitem [{\citenamefont {Norman}\ \emph {et~al.}(2001)\citenamefont {Norman},
  \citenamefont {Bishop}, \citenamefont {Jensen},\ and\ \citenamefont
  {Oddershede}}]{Norman:CPP:2001}%
  \BibitemOpen
  \bibfield  {author} {\bibinfo {author} {\bibfnamefont {P.}~\bibnamefont
  {Norman}}, \bibinfo {author} {\bibfnamefont {D.~M.}\ \bibnamefont {Bishop}},
  \bibinfo {author} {\bibfnamefont {H.~J.~A.}\ \bibnamefont {Jensen}}, \ and\
  \bibinfo {author} {\bibfnamefont {J.}~\bibnamefont {Oddershede}},\ }\bibfield
   {title} {\enquote {\bibinfo {title} {Near-resonant absorption in the
  time-dependent self-consistent field and multiconfigurational self-consistent
  field approximations},}\ }\href {\doibase 10.1063/1.1415081} {\bibfield
  {journal} {\bibinfo  {journal} {J. Chem. Phys.}\ }\textbf {\bibinfo {volume}
  {115}},\ \bibinfo {pages} {10323--10334} (\bibinfo {year}
  {2001})}\BibitemShut {NoStop}%
\bibitem [{\citenamefont {Norman}\ \emph {et~al.}(2005)\citenamefont {Norman},
  \citenamefont {Bishop}, \citenamefont {Jensen},\ and\ \citenamefont
  {Oddershede}}]{Norman:CPP:2005}%
  \BibitemOpen
  \bibfield  {author} {\bibinfo {author} {\bibfnamefont {P.}~\bibnamefont
  {Norman}}, \bibinfo {author} {\bibfnamefont {D.~M.}\ \bibnamefont {Bishop}},
  \bibinfo {author} {\bibfnamefont {H.~J.~A.}\ \bibnamefont {Jensen}}, \ and\
  \bibinfo {author} {\bibfnamefont {J.}~\bibnamefont {Oddershede}},\ }\bibfield
   {title} {\enquote {\bibinfo {title} {Nonlinear response theory with
  relaxation: The first-order hyperpolarizability},}\ }\href@noop {} {\bibfield
   {journal} {\bibinfo  {journal} {J. Chem. Phys.}\ }\textbf {\bibinfo {volume}
  {123}},\ \bibinfo {pages} {194103} (\bibinfo {year} {2005})}\BibitemShut
  {NoStop}%
\bibitem [{\citenamefont {Jensen}, \citenamefont {Autschbach},\ and\
  \citenamefont {Schatz}(2005)}]{Jensen:CPP:2005}%
  \BibitemOpen
  \bibfield  {author} {\bibinfo {author} {\bibfnamefont {L.}~\bibnamefont
  {Jensen}}, \bibinfo {author} {\bibfnamefont {J.}~\bibnamefont {Autschbach}},
  \ and\ \bibinfo {author} {\bibfnamefont {G.~C.}\ \bibnamefont {Schatz}},\
  }\bibfield  {title} {\enquote {\bibinfo {title} {Finite lifetime effects on
  the polarizability within time-dependent density-functional theory},}\
  }\href@noop {} {\bibfield  {journal} {\bibinfo  {journal} {J. Chem. Phys.}\
  }\textbf {\bibinfo {volume} {122}},\ \bibinfo {eid} {224115} (\bibinfo {year}
  {2005})}\BibitemShut {NoStop}%
\bibitem [{\citenamefont {Ekstr\"om}\ \emph {et~al.}(2006)\citenamefont
  {Ekstr\"om}, \citenamefont {Norman}, \citenamefont {Carravetta},\ and\
  \citenamefont {\AA{}gren}}]{Norman:CPP:Xray:2006}%
  \BibitemOpen
  \bibfield  {author} {\bibinfo {author} {\bibfnamefont {U.}~\bibnamefont
  {Ekstr\"om}}, \bibinfo {author} {\bibfnamefont {P.}~\bibnamefont {Norman}},
  \bibinfo {author} {\bibfnamefont {V.}~\bibnamefont {Carravetta}}, \ and\
  \bibinfo {author} {\bibfnamefont {H.}~\bibnamefont {\AA{}gren}},\ }\bibfield
  {title} {\enquote {\bibinfo {title} {{Polarization Propagator for X-Ray
  Spectra}},}\ }\href {\doibase 10.1103/PhysRevLett.97.143001} {\bibfield
  {journal} {\bibinfo  {journal} {Phys. Rev. Lett.}\ }\textbf {\bibinfo
  {volume} {97}},\ \bibinfo {pages} {143001} (\bibinfo {year}
  {2006})}\BibitemShut {NoStop}%
\bibitem [{\citenamefont {Ekstr{\"o}m}\ and\ \citenamefont
  {Norman}(2006)}]{Norman:CPP:Xray:2006:1}%
  \BibitemOpen
  \bibfield  {author} {\bibinfo {author} {\bibfnamefont {U.}~\bibnamefont
  {Ekstr{\"o}m}}\ and\ \bibinfo {author} {\bibfnamefont {P.}~\bibnamefont
  {Norman}},\ }\bibfield  {title} {\enquote {\bibinfo {title} {X-ray absorption
  spectra from the resonant-convergent first-order polarization propagator
  approach},}\ }\href {\doibase 10.1103/PhysRevA.74.042722} {\bibfield
  {journal} {\bibinfo  {journal} {Phys. Rev. A}\ }\textbf {\bibinfo {volume}
  {74}},\ \bibinfo {pages} {042722} (\bibinfo {year} {2006})}\BibitemShut
  {NoStop}%
\bibitem [{\citenamefont {Fahleson}, \citenamefont {{\AA}gren},\ and\
  \citenamefont {Norman}(2016)}]{CPP:XTPA:Norman2016}%
  \BibitemOpen
  \bibfield  {author} {\bibinfo {author} {\bibfnamefont {T.}~\bibnamefont
  {Fahleson}}, \bibinfo {author} {\bibfnamefont {H.}~\bibnamefont {{\AA}gren}},
  \ and\ \bibinfo {author} {\bibfnamefont {P.}~\bibnamefont {Norman}},\
  }\bibfield  {title} {\enquote {\bibinfo {title} {{A Polarization Propagator
  for Nonlinear X-ray Spectroscopies}},}\ }\href {\doibase
  10.1021/acs.jpclett.6b00750} {\bibfield  {journal} {\bibinfo  {journal} {J.
  Phys. Chem. Lett.}\ }\textbf {\bibinfo {volume} {7}},\ \bibinfo {pages}
  {1991--1995} (\bibinfo {year} {2016})}\BibitemShut {NoStop}%
\bibitem [{\citenamefont {Coriani}\ \emph
  {et~al.}(2012{\natexlab{a}})\citenamefont {Coriani}, \citenamefont
  {Fransson}, \citenamefont {Christiansen},\ and\ \citenamefont
  {Norman}}]{Lanczos1}%
  \BibitemOpen
  \bibfield  {author} {\bibinfo {author} {\bibfnamefont {S.}~\bibnamefont
  {Coriani}}, \bibinfo {author} {\bibfnamefont {T.}~\bibnamefont {Fransson}},
  \bibinfo {author} {\bibfnamefont {O.}~\bibnamefont {Christiansen}}, \ and\
  \bibinfo {author} {\bibfnamefont {P.}~\bibnamefont {Norman}},\ }\bibfield
  {title} {\enquote {\bibinfo {title} {{Asymmetric-Lanczos-Chain-Driven
  Implementation of Electronic Resonance Convergent Coupled-Cluster Linear
  Response Theory}},}\ }\href@noop {} {\bibfield  {journal} {\bibinfo
  {journal} {J. Chem. Theory Comput.}\ }\textbf {\bibinfo {volume} {8}},\
  \bibinfo {pages} {1616--1628} (\bibinfo {year}
  {2012}{\natexlab{a}})}\BibitemShut {NoStop}%
\bibitem [{\citenamefont {Coriani}\ \emph
  {et~al.}(2012{\natexlab{b}})\citenamefont {Coriani}, \citenamefont
  {Christiansen}, \citenamefont {Fransson},\ and\ \citenamefont
  {Norman}}]{Lanczos2}%
  \BibitemOpen
  \bibfield  {author} {\bibinfo {author} {\bibfnamefont {S.}~\bibnamefont
  {Coriani}}, \bibinfo {author} {\bibfnamefont {O.}~\bibnamefont
  {Christiansen}}, \bibinfo {author} {\bibfnamefont {T.}~\bibnamefont
  {Fransson}}, \ and\ \bibinfo {author} {\bibfnamefont {P.}~\bibnamefont
  {Norman}},\ }\bibfield  {title} {\enquote {\bibinfo {title} {Coupled-cluster
  response theory for near-edge x-ray-absorption fine structure of atoms and
  molecules},}\ }\href@noop {} {\bibfield  {journal} {\bibinfo  {journal}
  {Phys. Rev. A}\ }\textbf {\bibinfo {volume} {85}},\ \bibinfo {pages} {022507}
  (\bibinfo {year} {2012}{\natexlab{b}})}\BibitemShut {NoStop}%
\bibitem [{\citenamefont {Faber}\ and\ \citenamefont
  {Coriani}(2019)}]{CPP:RIXS:CC}%
  \BibitemOpen
  \bibfield  {author} {\bibinfo {author} {\bibfnamefont {R.}~\bibnamefont
  {Faber}}\ and\ \bibinfo {author} {\bibfnamefont {S.}~\bibnamefont
  {Coriani}},\ }\bibfield  {title} {\enquote {\bibinfo {title} {Resonant
  inelastic x-ray scattering and nonresonant x-ray emission spectra from
  coupled-cluster (damped) response theory},}\ }\href {\doibase
  10.1021/acs.jctc.8b01020} {\bibfield  {journal} {\bibinfo  {journal} {J.
  Chem. Theory Comput.}\ }\textbf {\bibinfo {volume} {15}},\ \bibinfo {pages}
  {520--528} (\bibinfo {year} {2019})}\BibitemShut {NoStop}%
\bibitem [{\citenamefont {Norman}(2011)}]{Norman:PCCP:2011}%
  \BibitemOpen
  \bibfield  {author} {\bibinfo {author} {\bibfnamefont {P.}~\bibnamefont
  {Norman}},\ }\bibfield  {title} {\enquote {\bibinfo {title} {A perspective on
  nonresonant and resonant electronic response theory for time-dependent
  molecular properties},}\ }\href {\doibase 10.1039/C1CP21951K} {\bibfield
  {journal} {\bibinfo  {journal} {Phys. Chem. Chem. Phys.}\ }\textbf {\bibinfo
  {volume} {13}},\ \bibinfo {pages} {20519--20535} (\bibinfo {year}
  {2011})}\BibitemShut {NoStop}%
\bibitem [{\citenamefont {Kristensen}\ \emph {et~al.}(2011)\citenamefont
  {Kristensen}, \citenamefont {Kauczor}, \citenamefont {Thorvaldsen},
  \citenamefont {J{\o}rgensen}, \citenamefont {Kj{\ae}rgaard},\ and\
  \citenamefont {Rizzo}}]{CPP:TPA}%
  \BibitemOpen
  \bibfield  {author} {\bibinfo {author} {\bibfnamefont {K.}~\bibnamefont
  {Kristensen}}, \bibinfo {author} {\bibfnamefont {J.}~\bibnamefont {Kauczor}},
  \bibinfo {author} {\bibfnamefont {A.~J.}\ \bibnamefont {Thorvaldsen}},
  \bibinfo {author} {\bibfnamefont {P.}~\bibnamefont {J{\o}rgensen}}, \bibinfo
  {author} {\bibfnamefont {T.}~\bibnamefont {Kj{\ae}rgaard}}, \ and\ \bibinfo
  {author} {\bibfnamefont {A.}~\bibnamefont {Rizzo}},\ }\bibfield  {title}
  {\enquote {\bibinfo {title} {Damped response theory description of two-photon
  absorption},}\ }\href@noop {} {\bibfield  {journal} {\bibinfo  {journal} {J.
  Chem. Phys.}\ }\textbf {\bibinfo {volume} {134}},\ \bibinfo {pages} {214104}
  (\bibinfo {year} {2011})}\BibitemShut {NoStop}%
\bibitem [{\citenamefont {Rehn}, \citenamefont {Dreuw},\ and\ \citenamefont
  {Norman}(2017)}]{CPP:RIXS:ADC}%
  \BibitemOpen
  \bibfield  {author} {\bibinfo {author} {\bibfnamefont {D.~R.}\ \bibnamefont
  {Rehn}}, \bibinfo {author} {\bibfnamefont {A.}~\bibnamefont {Dreuw}}, \ and\
  \bibinfo {author} {\bibfnamefont {P.}~\bibnamefont {Norman}},\ }\bibfield
  {title} {\enquote {\bibinfo {title} {{Resonant Inelastic X-ray Scattering
  Amplitudes and Cross Sections in the Algebraic Diagrammatic
  Construction/Intermediate State Representation (ADC/ISR) Approach}},}\
  }\href@noop {} {\bibfield  {journal} {\bibinfo  {journal} {J. Chem. Theory
  Comput.}\ }\textbf {\bibinfo {volume} {13}},\ \bibinfo {pages} {5552--5559}
  (\bibinfo {year} {2017})}\BibitemShut {NoStop}%
\bibitem [{\citenamefont {Jiemchooroj}, \citenamefont {Sernelius},\ and\
  \citenamefont {Norman}(2004)}]{CPP:C6:Jiemchooroj}%
  \BibitemOpen
  \bibfield  {author} {\bibinfo {author} {\bibfnamefont {A.}~\bibnamefont
  {Jiemchooroj}}, \bibinfo {author} {\bibfnamefont {B.~E.}\ \bibnamefont
  {Sernelius}}, \ and\ \bibinfo {author} {\bibfnamefont {P.}~\bibnamefont
  {Norman}},\ }\bibfield  {title} {\enquote {\bibinfo {title} {$c_6$
  dipole-dipole dispersion coefficients for the n-alkanes: Test of an
  additivity procedure},}\ }\href@noop {} {\bibfield  {journal} {\bibinfo
  {journal} {Phys. Rev. A}\ }\textbf {\bibinfo {volume} {69}},\ \bibinfo
  {pages} {044701} (\bibinfo {year} {2004})}\BibitemShut {NoStop}%
\bibitem [{\citenamefont {Fransson}\ \emph {et~al.}(2017)\citenamefont
  {Fransson}, \citenamefont {Rehn}, \citenamefont {Dreuw},\ and\ \citenamefont
  {Norman}}]{CPP:ADC}%
  \BibitemOpen
  \bibfield  {author} {\bibinfo {author} {\bibfnamefont {T.}~\bibnamefont
  {Fransson}}, \bibinfo {author} {\bibfnamefont {D.~R.}\ \bibnamefont {Rehn}},
  \bibinfo {author} {\bibfnamefont {A.}~\bibnamefont {Dreuw}}, \ and\ \bibinfo
  {author} {\bibfnamefont {P.}~\bibnamefont {Norman}},\ }\bibfield  {title}
  {\enquote {\bibinfo {title} {Static polarizabilities and c$_6$ dispersion
  coefficients using the algebraic-diagrammatic construction scheme for the
  complex polarization propagator},}\ }\href@noop {} {\bibfield  {journal}
  {\bibinfo  {journal} {J. Chem. Phys.}\ }\textbf {\bibinfo {volume} {146}},\
  \bibinfo {pages} {094301} (\bibinfo {year} {2017})}\BibitemShut {NoStop}%
\bibitem [{\citenamefont {Kauczor}\ and\ \citenamefont
  {Norman}(2014)}]{CPP:Joanna:solver2}%
  \BibitemOpen
  \bibfield  {author} {\bibinfo {author} {\bibfnamefont {J.}~\bibnamefont
  {Kauczor}}\ and\ \bibinfo {author} {\bibfnamefont {P.}~\bibnamefont
  {Norman}},\ }\bibfield  {title} {\enquote {\bibinfo {title} {Efficient
  calculations of molecular linear response properties for spectral regions},}\
  }\href@noop {} {\bibfield  {journal} {\bibinfo  {journal} {J. Chem. Theory
  Comput.}\ }\textbf {\bibinfo {volume} {10}},\ \bibinfo {pages} {2449--2455}
  (\bibinfo {year} {2014})}\BibitemShut {NoStop}%
\bibitem [{\citenamefont {Scheurer}\ \emph {et~al.}(2020)\citenamefont
  {Scheurer}, \citenamefont {Fransson}, \citenamefont {Norman}, \citenamefont
  {Dreuw},\ and\ \citenamefont {Rehn}}]{MaxADC}%
  \BibitemOpen
  \bibfield  {author} {\bibinfo {author} {\bibfnamefont {M.}~\bibnamefont
  {Scheurer}}, \bibinfo {author} {\bibfnamefont {T.}~\bibnamefont {Fransson}},
  \bibinfo {author} {\bibfnamefont {P.}~\bibnamefont {Norman}}, \bibinfo
  {author} {\bibfnamefont {A.}~\bibnamefont {Dreuw}}, \ and\ \bibinfo {author}
  {\bibfnamefont {D.~R.}\ \bibnamefont {Rehn}},\ }\bibfield  {title} {\enquote
  {\bibinfo {title} {{Complex excited state polarizabilities in the ADC/ISR
  framework}},}\ }\href@noop {} {\bibfield  {journal} {\bibinfo  {journal} {J.
  Chem. Phys.}\ }\textbf {\bibinfo {volume} {153}},\ \bibinfo {pages} {074112}
  (\bibinfo {year} {2020})}\BibitemShut {NoStop}%
\bibitem [{\citenamefont {Kauczor}\ \emph {et~al.}(2013)\citenamefont
  {Kauczor}, \citenamefont {Norman}, \citenamefont {Christiansen},\ and\
  \citenamefont {Coriani}}]{CPP:Kauczor:CC}%
  \BibitemOpen
  \bibfield  {author} {\bibinfo {author} {\bibfnamefont {J.}~\bibnamefont
  {Kauczor}}, \bibinfo {author} {\bibfnamefont {P.}~\bibnamefont {Norman}},
  \bibinfo {author} {\bibfnamefont {O.}~\bibnamefont {Christiansen}}, \ and\
  \bibinfo {author} {\bibfnamefont {S.}~\bibnamefont {Coriani}},\ }\bibfield
  {title} {\enquote {\bibinfo {title} {Communication: A reduced-space algorithm
  for the solution of the complex linear response equations used in coupled
  cluster damped response theory},}\ }\href {\doibase 10.1063/1.4840275}
  {\bibfield  {journal} {\bibinfo  {journal} {J. Chem. Phys.}\ }\textbf
  {\bibinfo {volume} {139}},\ \bibinfo {pages} {211102} (\bibinfo {year}
  {2013})}\BibitemShut {NoStop}%
\bibitem [{\citenamefont {Reinholdt}, \citenamefont {N{\o}rby},\ and\
  \citenamefont {Kongsted}(2018)}]{CPP:MCD:PE}%
  \BibitemOpen
  \bibfield  {author} {\bibinfo {author} {\bibfnamefont {P.}~\bibnamefont
  {Reinholdt}}, \bibinfo {author} {\bibfnamefont {M.~S.}\ \bibnamefont
  {N{\o}rby}}, \ and\ \bibinfo {author} {\bibfnamefont {J.}~\bibnamefont
  {Kongsted}},\ }\bibfield  {title} {\enquote {\bibinfo {title} {{Modeling of
  Magnetic Circular Dichroism and UV/Vis Absorption Spectra Using Fluctuating
  Charges or Polarizable Embedding within a Resonant-Convergent Response Theory
  Formalism}},}\ }\href@noop {} {\bibfield  {journal} {\bibinfo  {journal} {J.
  Chem. Theory Comput.}\ }\textbf {\bibinfo {volume} {14}},\ \bibinfo {pages}
  {6391--6404} (\bibinfo {year} {2018})}\BibitemShut {NoStop}%
\bibitem [{\citenamefont {N{\o}rby}, \citenamefont {Coriani},\ and\
  \citenamefont {Kongsted}(2018)}]{CPP:PE:MCD}%
  \BibitemOpen
  \bibfield  {author} {\bibinfo {author} {\bibfnamefont {M.}~\bibnamefont
  {N{\o}rby}}, \bibinfo {author} {\bibfnamefont {S.}~\bibnamefont {Coriani}}, \
  and\ \bibinfo {author} {\bibfnamefont {J.}~\bibnamefont {Kongsted}},\
  }\bibfield  {title} {\enquote {\bibinfo {title} {Modeling magnetic circular
  dichroism within the polarizable embedding approach},}\ }\href@noop {}
  {\bibfield  {journal} {\bibinfo  {journal} {Theor. Chim. Acta}\ }\textbf
  {\bibinfo {volume} {137}} (\bibinfo {year} {2018})}\BibitemShut {NoStop}%
\bibitem [{\citenamefont {Fransson}, \citenamefont {Burdakova},\ and\
  \citenamefont {Norman}(2016)}]{CPP:4components}%
  \BibitemOpen
  \bibfield  {author} {\bibinfo {author} {\bibfnamefont {T.}~\bibnamefont
  {Fransson}}, \bibinfo {author} {\bibfnamefont {D.}~\bibnamefont {Burdakova}},
  \ and\ \bibinfo {author} {\bibfnamefont {P.}~\bibnamefont {Norman}},\
  }\bibfield  {title} {\enquote {\bibinfo {title} {{K}- and {L}-edge x-ray
  absorption spectrum calculations of closed-shell carbon{,} silicon{,}
  germanium{,} and sulfur compounds using damped four-component density
  functional response theory},}\ }\href {\doibase 10.1039/C6CP00561F}
  {\bibfield  {journal} {\bibinfo  {journal} {Phys. Chem. Chem. Phys.}\
  }\textbf {\bibinfo {volume} {18}},\ \bibinfo {pages} {13591--13603} (\bibinfo
  {year} {2016})}\BibitemShut {NoStop}%
\bibitem [{\citenamefont {Konecny}\ \emph {et~al.}(2019)\citenamefont
  {Konecny}, \citenamefont {Repisky}, \citenamefont {Ruud},\ and\ \citenamefont
  {Komorovsky}}]{damped_4components}%
  \BibitemOpen
  \bibfield  {author} {\bibinfo {author} {\bibfnamefont {L.}~\bibnamefont
  {Konecny}}, \bibinfo {author} {\bibfnamefont {M.}~\bibnamefont {Repisky}},
  \bibinfo {author} {\bibfnamefont {K.}~\bibnamefont {Ruud}}, \ and\ \bibinfo
  {author} {\bibfnamefont {S.}~\bibnamefont {Komorovsky}},\ }\bibfield  {title}
  {\enquote {\bibinfo {title} {{Relativistic four-component linear damped
  response TDDFT for electronic absorption and circular dichroism
  calculations}},}\ }\href@noop {} {\bibfield  {journal} {\bibinfo  {journal}
  {J. Chem. Phys.}\ }\textbf {\bibinfo {volume} {151}},\ \bibinfo {pages}
  {194112} (\bibinfo {year} {2019})}\BibitemShut {NoStop}%
\bibitem [{\citenamefont {Jiemchooroj}\ and\ \citenamefont
  {Norman}(2007)}]{CPP:ECD}%
  \BibitemOpen
  \bibfield  {author} {\bibinfo {author} {\bibfnamefont {A.}~\bibnamefont
  {Jiemchooroj}}\ and\ \bibinfo {author} {\bibfnamefont {P.}~\bibnamefont
  {Norman}},\ }\bibfield  {title} {\enquote {\bibinfo {title} {Electronic
  circular dichroism spectra from the complex polarization propagator},}\
  }\href@noop {} {\bibfield  {journal} {\bibinfo  {journal} {J. Chem. Phys.}\
  }\textbf {\bibinfo {volume} {126}},\ \bibinfo {pages} {134102} (\bibinfo
  {year} {2007})}\BibitemShut {NoStop}%
\bibitem [{\citenamefont {Faber}\ and\ \citenamefont
  {Coriani}(2020)}]{CVS-CPP:1}%
  \BibitemOpen
  \bibfield  {author} {\bibinfo {author} {\bibfnamefont {R.}~\bibnamefont
  {Faber}}\ and\ \bibinfo {author} {\bibfnamefont {S.}~\bibnamefont
  {Coriani}},\ }\bibfield  {title} {\enquote {\bibinfo {title}
  {{Core–valence-separated coupled-cluster-singles-and-doubles
  complex-polarization-propagator approach to X-ray spectroscopies}},}\ }\href
  {\doibase 10.1039/C9CP03696B} {\bibfield  {journal} {\bibinfo  {journal}
  {Phys. Chem. Chem. Phys.}\ }\textbf {\bibinfo {volume} {22}},\ \bibinfo
  {pages} {2642} (\bibinfo {year} {2020})}\BibitemShut {NoStop}%
\bibitem [{\citenamefont {Solheim}\ \emph {et~al.}(2008)\citenamefont
  {Solheim}, \citenamefont {Ruud}, \citenamefont {Coriani},\ and\ \citenamefont
  {Norman}}]{CPP:MCD}%
  \BibitemOpen
  \bibfield  {author} {\bibinfo {author} {\bibfnamefont {H.}~\bibnamefont
  {Solheim}}, \bibinfo {author} {\bibfnamefont {K.}~\bibnamefont {Ruud}},
  \bibinfo {author} {\bibfnamefont {S.}~\bibnamefont {Coriani}}, \ and\
  \bibinfo {author} {\bibfnamefont {P.}~\bibnamefont {Norman}},\ }\bibfield
  {title} {\enquote {\bibinfo {title} {Complex polarization propagator
  calculations of magnetic circular dichroism spectra},}\ }\href {\doibase
  10.1063/1.2834924} {\bibfield  {journal} {\bibinfo  {journal} {J. Chem.
  Phys.}\ }\textbf {\bibinfo {volume} {128}},\ \bibinfo {pages} {094103}
  (\bibinfo {year} {2008})}\BibitemShut {NoStop}%
\bibitem [{\citenamefont {Fahleson}\ \emph {et~al.}(2015)\citenamefont
  {Fahleson}, \citenamefont {Kauczor}, \citenamefont {Norman}, \citenamefont
  {Santoro}, \citenamefont {Improta},\ and\ \citenamefont
  {Coriani}}]{CPP:MCD:Nucleic}%
  \BibitemOpen
  \bibfield  {author} {\bibinfo {author} {\bibfnamefont {T.}~\bibnamefont
  {Fahleson}}, \bibinfo {author} {\bibfnamefont {J.}~\bibnamefont {Kauczor}},
  \bibinfo {author} {\bibfnamefont {P.}~\bibnamefont {Norman}}, \bibinfo
  {author} {\bibfnamefont {F.}~\bibnamefont {Santoro}}, \bibinfo {author}
  {\bibfnamefont {R.}~\bibnamefont {Improta}}, \ and\ \bibinfo {author}
  {\bibfnamefont {S.}~\bibnamefont {Coriani}},\ }\bibfield  {title} {\enquote
  {\bibinfo {title} {{TD-DFT Investigation of the Magnetic Circular Dichroism
  Spectra of Some Purine and Pyrimidine Bases of Nucleic Acids}},}\ }\href
  {\doibase 10.1021/jp512468k} {\bibfield  {journal} {\bibinfo  {journal} {J.
  Chem. Phys.}\ }\textbf {\bibinfo {volume} {119}},\ \bibinfo {pages}
  {5476--5489} (\bibinfo {year} {2015})}\BibitemShut {NoStop}%
\bibitem [{\citenamefont {Vaara}\ \emph {et~al.}(2014)\citenamefont {Vaara},
  \citenamefont {Rizzo}, \citenamefont {Kauczor}, \citenamefont {Norman},\ and\
  \citenamefont {Coriani}}]{CPP:NSCD}%
  \BibitemOpen
  \bibfield  {author} {\bibinfo {author} {\bibfnamefont {J.}~\bibnamefont
  {Vaara}}, \bibinfo {author} {\bibfnamefont {A.}~\bibnamefont {Rizzo}},
  \bibinfo {author} {\bibfnamefont {J.}~\bibnamefont {Kauczor}}, \bibinfo
  {author} {\bibfnamefont {P.}~\bibnamefont {Norman}}, \ and\ \bibinfo {author}
  {\bibfnamefont {S.}~\bibnamefont {Coriani}},\ }\bibfield  {title} {\enquote
  {\bibinfo {title} {Nuclear spin circular dichroism},}\ }\href@noop {}
  {\bibfield  {journal} {\bibinfo  {journal} {J. Chem. Phys.}\ }\textbf
  {\bibinfo {volume} {140}},\ \bibinfo {pages} {134103} (\bibinfo {year}
  {2014})}\BibitemShut {NoStop}%
\bibitem [{\citenamefont {Cukras}\ \emph {et~al.}(2016)\citenamefont {Cukras},
  \citenamefont {Kauczor}, \citenamefont {Norman}, \citenamefont {Rizzo},
  \citenamefont {Rikken},\ and\ \citenamefont {Coriani}}]{CPP:MCHD}%
  \BibitemOpen
  \bibfield  {author} {\bibinfo {author} {\bibfnamefont {J.}~\bibnamefont
  {Cukras}}, \bibinfo {author} {\bibfnamefont {J.}~\bibnamefont {Kauczor}},
  \bibinfo {author} {\bibfnamefont {P.}~\bibnamefont {Norman}}, \bibinfo
  {author} {\bibfnamefont {A.}~\bibnamefont {Rizzo}}, \bibinfo {author}
  {\bibfnamefont {G.~L. J.~A.}\ \bibnamefont {Rikken}}, \ and\ \bibinfo
  {author} {\bibfnamefont {S.}~\bibnamefont {Coriani}},\ }\bibfield  {title}
  {\enquote {\bibinfo {title} {A complex-polarization-propagator protocol for
  magneto-chiral axial dichroism and birefringence dispersion},}\ }\href
  {\doibase 10.1039/C6CP01465H} {\bibfield  {journal} {\bibinfo  {journal}
  {Phys. Chem. Chem. Phys.}\ }\textbf {\bibinfo {volume} {18}},\ \bibinfo
  {pages} {13267--13279} (\bibinfo {year} {2016})}\BibitemShut {NoStop}%
\bibitem [{\citenamefont {Nanda}\ \emph {et~al.}(2020)\citenamefont {Nanda},
  \citenamefont {Vidal}, \citenamefont {Faber}, \citenamefont {Coriani},\ and\
  \citenamefont {Krylov}}]{NandaPCCP}%
  \BibitemOpen
  \bibfield  {author} {\bibinfo {author} {\bibfnamefont {K.}~\bibnamefont
  {Nanda}}, \bibinfo {author} {\bibfnamefont {M.~L.}\ \bibnamefont {Vidal}},
  \bibinfo {author} {\bibfnamefont {R.}~\bibnamefont {Faber}}, \bibinfo
  {author} {\bibfnamefont {S.}~\bibnamefont {Coriani}}, \ and\ \bibinfo
  {author} {\bibfnamefont {A.~I.}\ \bibnamefont {Krylov}},\ }\bibfield  {title}
  {\enquote {\bibinfo {title} {{How to stay out of trouble in RIXS calculations
  within the equation-of-motion coupled-cluster damped response theory
  framework? Safe hitchhiking in the excitation manifold by means of
  core-valence separation}},}\ }\href {\doibase doi: 10.1039/c9cp03688a}
  {\bibfield  {journal} {\bibinfo  {journal} {Phys. Chem. Chem. Phys.}\
  }\textbf {\bibinfo {volume} {22}},\ \bibinfo {pages} {2629} (\bibinfo {year}
  {2020})}\BibitemShut {NoStop}%
\bibitem [{\citenamefont {Furche}\ \emph {et~al.}(2014)\citenamefont {Furche},
  \citenamefont {Ahlrichs}, \citenamefont {H{\"a}ttig}, \citenamefont
  {Klopper}, \citenamefont {Sierka},\ and\ \citenamefont
  {Weigend}}]{turbomole2014}%
  \BibitemOpen
  \bibfield  {author} {\bibinfo {author} {\bibfnamefont {F.}~\bibnamefont
  {Furche}}, \bibinfo {author} {\bibfnamefont {R.}~\bibnamefont {Ahlrichs}},
  \bibinfo {author} {\bibfnamefont {C.}~\bibnamefont {H{\"a}ttig}}, \bibinfo
  {author} {\bibfnamefont {W.}~\bibnamefont {Klopper}}, \bibinfo {author}
  {\bibfnamefont {M.}~\bibnamefont {Sierka}}, \ and\ \bibinfo {author}
  {\bibfnamefont {F.}~\bibnamefont {Weigend}},\ }\bibfield  {title} {\enquote
  {\bibinfo {title} {Turbomole},}\ }\href@noop {} {\bibfield  {journal}
  {\bibinfo  {journal} {WIREs Comput Mol. Sci.}\ }\textbf {\bibinfo {volume}
  {4}},\ \bibinfo {pages} {91--100} (\bibinfo {year} {2014})}\BibitemShut
  {NoStop}%
\bibitem [{\citenamefont {Balasubramani}\ \emph {et~al.}(2020)\citenamefont
  {Balasubramani}, \citenamefont {Chen}, \citenamefont {Coriani}, \citenamefont
  {Diedenhofen}, \citenamefont {Frank}, \citenamefont {Franzke}, \citenamefont
  {Furche}, \citenamefont {Grotjahn}, \citenamefont {Harding}, \citenamefont
  {Hättig}, \citenamefont {Hellweg}, \citenamefont {Helmich-Paris},
  \citenamefont {Holzer}, \citenamefont {Huniar}, \citenamefont {Kaupp},
  \citenamefont {Marefat~Khah}, \citenamefont {Karbalaei~Khani}, \citenamefont
  {Müller}, \citenamefont {Mack}, \citenamefont {Nguyen}, \citenamefont
  {Parker}, \citenamefont {Perlt}, \citenamefont {Rappoport}, \citenamefont
  {Reiter}, \citenamefont {Roy}, \citenamefont {Rückert}, \citenamefont
  {Schmitz}, \citenamefont {Sierka}, \citenamefont {Tapavicza}, \citenamefont
  {Tew}, \citenamefont {van Wüllen}, \citenamefont {Voora}, \citenamefont
  {Weigend}, \citenamefont {Wodyński},\ and\ \citenamefont
  {Yu}}]{Turbomole2020}%
  \BibitemOpen
  \bibfield  {author} {\bibinfo {author} {\bibfnamefont {S.~G.}\ \bibnamefont
  {Balasubramani}}, \bibinfo {author} {\bibfnamefont {G.~P.}\ \bibnamefont
  {Chen}}, \bibinfo {author} {\bibfnamefont {S.}~\bibnamefont {Coriani}},
  \bibinfo {author} {\bibfnamefont {M.}~\bibnamefont {Diedenhofen}}, \bibinfo
  {author} {\bibfnamefont {M.~S.}\ \bibnamefont {Frank}}, \bibinfo {author}
  {\bibfnamefont {Y.~J.}\ \bibnamefont {Franzke}}, \bibinfo {author}
  {\bibfnamefont {F.}~\bibnamefont {Furche}}, \bibinfo {author} {\bibfnamefont
  {R.}~\bibnamefont {Grotjahn}}, \bibinfo {author} {\bibfnamefont {M.~E.}\
  \bibnamefont {Harding}}, \bibinfo {author} {\bibfnamefont {C.}~\bibnamefont
  {Hättig}}, \bibinfo {author} {\bibfnamefont {A.}~\bibnamefont {Hellweg}},
  \bibinfo {author} {\bibfnamefont {B.}~\bibnamefont {Helmich-Paris}}, \bibinfo
  {author} {\bibfnamefont {C.}~\bibnamefont {Holzer}}, \bibinfo {author}
  {\bibfnamefont {U.}~\bibnamefont {Huniar}}, \bibinfo {author} {\bibfnamefont
  {M.}~\bibnamefont {Kaupp}}, \bibinfo {author} {\bibfnamefont
  {A.}~\bibnamefont {Marefat~Khah}}, \bibinfo {author} {\bibfnamefont
  {S.}~\bibnamefont {Karbalaei~Khani}}, \bibinfo {author} {\bibfnamefont
  {T.}~\bibnamefont {Müller}}, \bibinfo {author} {\bibfnamefont
  {F.}~\bibnamefont {Mack}}, \bibinfo {author} {\bibfnamefont {B.~D.}\
  \bibnamefont {Nguyen}}, \bibinfo {author} {\bibfnamefont {S.~M.}\
  \bibnamefont {Parker}}, \bibinfo {author} {\bibfnamefont {E.}~\bibnamefont
  {Perlt}}, \bibinfo {author} {\bibfnamefont {D.}~\bibnamefont {Rappoport}},
  \bibinfo {author} {\bibfnamefont {K.}~\bibnamefont {Reiter}}, \bibinfo
  {author} {\bibfnamefont {S.}~\bibnamefont {Roy}}, \bibinfo {author}
  {\bibfnamefont {M.}~\bibnamefont {Rückert}}, \bibinfo {author}
  {\bibfnamefont {G.}~\bibnamefont {Schmitz}}, \bibinfo {author} {\bibfnamefont
  {M.}~\bibnamefont {Sierka}}, \bibinfo {author} {\bibfnamefont
  {E.}~\bibnamefont {Tapavicza}}, \bibinfo {author} {\bibfnamefont {D.~P.}\
  \bibnamefont {Tew}}, \bibinfo {author} {\bibfnamefont {C.}~\bibnamefont {van
  Wüllen}}, \bibinfo {author} {\bibfnamefont {V.~K.}\ \bibnamefont {Voora}},
  \bibinfo {author} {\bibfnamefont {F.}~\bibnamefont {Weigend}}, \bibinfo
  {author} {\bibfnamefont {A.}~\bibnamefont {Wodyński}}, \ and\ \bibinfo
  {author} {\bibfnamefont {J.~M.}\ \bibnamefont {Yu}},\ }\bibfield  {title}
  {\enquote {\bibinfo {title} {{TURBOMOLE: Modular program suite for {\it ab
  initio} quantum-chemical and condensed-matter simulations}},}\ }\href@noop {}
  {\bibfield  {journal} {\bibinfo  {journal} {J. Chem. Phys.}\ }\textbf
  {\bibinfo {volume} {152}},\ \bibinfo {pages} {184107} (\bibinfo {year}
  {2020})}\BibitemShut {NoStop}%
\bibitem [{\citenamefont {Christiansen}, \citenamefont {J{\o}rgensen},\ and\
  \citenamefont {H{\"a}ttig}(1998)}]{Christiansen:1998}%
  \BibitemOpen
  \bibfield  {author} {\bibinfo {author} {\bibfnamefont {O.}~\bibnamefont
  {Christiansen}}, \bibinfo {author} {\bibfnamefont {P.}~\bibnamefont
  {J{\o}rgensen}}, \ and\ \bibinfo {author} {\bibfnamefont {C.}~\bibnamefont
  {H{\"a}ttig}},\ }\bibfield  {title} {\enquote {\bibinfo {title} {Response
  functions from fourier component variational perturbation theory applied to a
  time-averaged quasienergy},}\ }\href@noop {} {\bibfield  {journal} {\bibinfo
  {journal} {Int. J. Quantum Chem.}\ }\textbf {\bibinfo {volume} {68}},\
  \bibinfo {pages} {1--52} (\bibinfo {year} {1998})}\BibitemShut {NoStop}%
\bibitem [{\citenamefont {Faber}\ \emph {et~al.}(2020)\citenamefont {Faber},
  \citenamefont {Ghidinelli}, \citenamefont {H{\"a}ttig},\ and\ \citenamefont
  {Coriani}}]{CPP:MCD:CC}%
  \BibitemOpen
  \bibfield  {author} {\bibinfo {author} {\bibfnamefont {R.}~\bibnamefont
  {Faber}}, \bibinfo {author} {\bibfnamefont {S.}~\bibnamefont {Ghidinelli}},
  \bibinfo {author} {\bibfnamefont {C.}~\bibnamefont {H{\"a}ttig}}, \ and\
  \bibinfo {author} {\bibfnamefont {S.}~\bibnamefont {Coriani}},\ }\bibfield
  {title} {\enquote {\bibinfo {title} {Magnetic circular dichroism spectra from
  resonant and damped coupled cluster response theory},}\ }\href@noop {}
  {\bibfield  {journal} {\bibinfo  {journal} {J. Chem. Phys.}\ }\textbf
  {\bibinfo {volume} {153}},\ \bibinfo {pages} {114105} (\bibinfo {year}
  {2020})}\BibitemShut {NoStop}%
\bibitem [{\citenamefont {Warnke}\ and\ \citenamefont
  {Furche}(2012)}]{warnke_furche:ECD}%
  \BibitemOpen
  \bibfield  {author} {\bibinfo {author} {\bibfnamefont {I.}~\bibnamefont
  {Warnke}}\ and\ \bibinfo {author} {\bibfnamefont {F.}~\bibnamefont
  {Furche}},\ }\bibfield  {title} {\enquote {\bibinfo {title} {Circular
  dichroism: electronic},}\ }\href@noop {} {\bibfield  {journal} {\bibinfo
  {journal} {WIREs Comput Mol. Sci.}\ }\textbf {\bibinfo {volume} {2}},\
  \bibinfo {pages} {150--166} (\bibinfo {year} {2012})}\BibitemShut {NoStop}%
\bibitem [{\citenamefont {Pedersen}\ \emph {et~al.}(2004)\citenamefont
  {Pedersen}, \citenamefont {Koch}, \citenamefont {Boman},\ and\ \citenamefont
  {de~Mer{\'a}s}}]{PEDERSEN:2004:modvel}%
  \BibitemOpen
  \bibfield  {author} {\bibinfo {author} {\bibfnamefont {T.~B.}\ \bibnamefont
  {Pedersen}}, \bibinfo {author} {\bibfnamefont {H.}~\bibnamefont {Koch}},
  \bibinfo {author} {\bibfnamefont {L.}~\bibnamefont {Boman}}, \ and\ \bibinfo
  {author} {\bibfnamefont {A.~M.~S.}\ \bibnamefont {de~Mer{\'a}s}},\ }\bibfield
   {title} {\enquote {\bibinfo {title} {{Origin invariant calculation of
  optical rotation without recourse to London orbitals}},}\ }\href@noop {}
  {\bibfield  {journal} {\bibinfo  {journal} {Chem. Phys. Lett.}\ }\textbf
  {\bibinfo {volume} {393}},\ \bibinfo {pages} {319} (\bibinfo {year}
  {2004})}\BibitemShut {NoStop}%
\bibitem [{\citenamefont {Friese}\ and\ \citenamefont
  {H{\"a}ttig}(2014)}]{RICC2_OR_Friese}%
  \BibitemOpen
  \bibfield  {author} {\bibinfo {author} {\bibfnamefont {D.~H.}\ \bibnamefont
  {Friese}}\ and\ \bibinfo {author} {\bibfnamefont {C.}~\bibnamefont
  {H{\"a}ttig}},\ }\bibfield  {title} {\enquote {\bibinfo {title} {Optical
  rotation calculations on large molecules using the approximate coupled
  cluster model {CC2} and the resolution-of-the-identity approximation},}\
  }\href {\doibase 10.1039/C3CP54338B} {\bibfield  {journal} {\bibinfo
  {journal} {Phys. Chem. Chem. Phys.}\ }\textbf {\bibinfo {volume} {16}},\
  \bibinfo {pages} {5942--5951} (\bibinfo {year} {2014})}\BibitemShut {NoStop}%
\bibitem [{\citenamefont {Grimme}(2006)}]{Grimme:2006}%
  \BibitemOpen
  \bibfield  {author} {\bibinfo {author} {\bibfnamefont {S.}~\bibnamefont
  {Grimme}},\ }\bibfield  {title} {\enquote {\bibinfo {title} {Semiempirical
  {GGA}-type density functional constructed with a long-range dispersion
  correction},}\ }\href {\doibase https://doi.org/10.1002/jcc.20495} {\bibfield
   {journal} {\bibinfo  {journal} {J. Comput. Chem.}\ }\textbf {\bibinfo
  {volume} {27}},\ \bibinfo {pages} {1787--1799} (\bibinfo {year}
  {2006})}\BibitemShut {NoStop}%
\bibitem [{\citenamefont {Tkatchenko}\ and\ \citenamefont
  {Scheffler}(2009)}]{Tkatchenko2009}%
  \BibitemOpen
  \bibfield  {author} {\bibinfo {author} {\bibfnamefont {A.}~\bibnamefont
  {Tkatchenko}}\ and\ \bibinfo {author} {\bibfnamefont {M.}~\bibnamefont
  {Scheffler}},\ }\bibfield  {title} {\enquote {\bibinfo {title} {{Accurate
  Molecular Van Der Waals Interactions from Ground-State Electron Density and
  Free-Atom Reference Data}},}\ }\href {\doibase
  10.1103/PhysRevLett.102.073005} {\bibfield  {journal} {\bibinfo  {journal}
  {Phys. Rev. Lett.}\ }\textbf {\bibinfo {volume} {102}},\ \bibinfo {pages}
  {073005} (\bibinfo {year} {2009})}\BibitemShut {NoStop}%
\bibitem [{\citenamefont {Grimme}\ \emph {et~al.}(2010)\citenamefont {Grimme},
  \citenamefont {Antony}, \citenamefont {Ehrlich},\ and\ \citenamefont
  {Krieg}}]{Grimme:2010}%
  \BibitemOpen
  \bibfield  {author} {\bibinfo {author} {\bibfnamefont {S.}~\bibnamefont
  {Grimme}}, \bibinfo {author} {\bibfnamefont {J.}~\bibnamefont {Antony}},
  \bibinfo {author} {\bibfnamefont {S.}~\bibnamefont {Ehrlich}}, \ and\
  \bibinfo {author} {\bibfnamefont {H.}~\bibnamefont {Krieg}},\ }\bibfield
  {title} {\enquote {\bibinfo {title} {A consistent and accurate ab initio
  parametrization of density functional dispersion correction ({DFT-D}) for the
  94 elements {H-Pu}},}\ }\href {\doibase 10.1063/1.3382344} {\bibfield
  {journal} {\bibinfo  {journal} {J. Chem. Phys.}\ }\textbf {\bibinfo {volume}
  {132}},\ \bibinfo {pages} {154104} (\bibinfo {year} {2010})}\BibitemShut
  {NoStop}%
\bibitem [{\citenamefont {Christiansen}, \citenamefont {Koch},\ and\
  \citenamefont {J{\o}rgensen}(1995)}]{CC2}%
  \BibitemOpen
  \bibfield  {author} {\bibinfo {author} {\bibfnamefont {O.}~\bibnamefont
  {Christiansen}}, \bibinfo {author} {\bibfnamefont {H.}~\bibnamefont {Koch}},
  \ and\ \bibinfo {author} {\bibfnamefont {P.}~\bibnamefont {J{\o}rgensen}},\
  }\bibfield  {title} {\enquote {\bibinfo {title} {{The second-order
  approximate coupled cluster singles and doubles model CC2}},}\ }\href@noop {}
  {\bibfield  {journal} {\bibinfo  {journal} {Chem. Phys. Lett.}\ }\textbf
  {\bibinfo {volume} {243}},\ \bibinfo {pages} {409--418} (\bibinfo {year}
  {1995})}\BibitemShut {NoStop}%
\bibitem [{\citenamefont {H{\"a}ttig}\ and\ \citenamefont
  {Weigend}(2000)}]{RICC2ENE}%
  \BibitemOpen
  \bibfield  {author} {\bibinfo {author} {\bibfnamefont {C.}~\bibnamefont
  {H{\"a}ttig}}\ and\ \bibinfo {author} {\bibfnamefont {F.}~\bibnamefont
  {Weigend}},\ }\bibfield  {title} {\enquote {\bibinfo {title} {{CC2}
  excitation energy calculations on large molecules using the resolution of the
  identity approximation},}\ }\href@noop {} {\bibfield  {journal} {\bibinfo
  {journal} {J. Chem. Phys.}\ }\textbf {\bibinfo {volume} {113}},\ \bibinfo
  {pages} {5154--5161} (\bibinfo {year} {2000})}\BibitemShut {NoStop}%
\bibitem [{\citenamefont {Friese}\ \emph {et~al.}(2012)\citenamefont {Friese},
  \citenamefont {Winter}, \citenamefont {Balzerowski}, \citenamefont {Schwan},\
  and\ \citenamefont {H{\"a}ttig}}]{RICC2:Friese:2012}%
  \BibitemOpen
  \bibfield  {author} {\bibinfo {author} {\bibfnamefont {D.~H.}\ \bibnamefont
  {Friese}}, \bibinfo {author} {\bibfnamefont {N.~O.}\ \bibnamefont {Winter}},
  \bibinfo {author} {\bibfnamefont {P.}~\bibnamefont {Balzerowski}}, \bibinfo
  {author} {\bibfnamefont {R.}~\bibnamefont {Schwan}}, \ and\ \bibinfo {author}
  {\bibfnamefont {C.}~\bibnamefont {H{\"a}ttig}},\ }\bibfield  {title}
  {\enquote {\bibinfo {title} {Large scale polarizability calculations using
  the approximate coupled cluster model {CC2} and {MP2} combined with the
  resolution-of-the-identity approximation.}}\ }\href@noop {} {\bibfield
  {journal} {\bibinfo  {journal} {J. Chem. Phys.}\ }\textbf {\bibinfo {volume}
  {136}},\ \bibinfo {pages} {174106--211120} (\bibinfo {year}
  {2012})}\BibitemShut {NoStop}%
\bibitem [{\citenamefont {H{\"a}ttig}\ and\ \citenamefont
  {K{\"o}hn}(2002)}]{RICC2TMES}%
  \BibitemOpen
  \bibfield  {author} {\bibinfo {author} {\bibfnamefont {C.}~\bibnamefont
  {H{\"a}ttig}}\ and\ \bibinfo {author} {\bibfnamefont {A.}~\bibnamefont
  {K{\"o}hn}},\ }\bibfield  {title} {\enquote {\bibinfo {title} {{Transition
  moments and excited-state first-order properties in the coupled-cluster model
  CC2 using the resolution-of-the-identity approximation}},}\ }\href@noop {}
  {\bibfield  {journal} {\bibinfo  {journal} {J. Chem. Phys.}\ }\textbf
  {\bibinfo {volume} {117}},\ \bibinfo {pages} {6939--6951} (\bibinfo {year}
  {2002})}\BibitemShut {NoStop}%
\bibitem [{\citenamefont {H{\"a}ttig}(2003)}]{RICC2:geoopt}%
  \BibitemOpen
  \bibfield  {author} {\bibinfo {author} {\bibfnamefont {C.}~\bibnamefont
  {H{\"a}ttig}},\ }\bibfield  {title} {\enquote {\bibinfo {title} {{Geometry
  optimizations with the coupled-cluster model CC2 using the
  resolution-of-the-identity approximation}},}\ }\href@noop {} {\bibfield
  {journal} {\bibinfo  {journal} {J. Chem. Phys.}\ }\textbf {\bibinfo {volume}
  {118}},\ \bibinfo {pages} {7751--7761} (\bibinfo {year} {2003})}\BibitemShut
  {NoStop}%
\bibitem [{\citenamefont {Vahtras}, \citenamefont {Alml{\"o}f},\ and\
  \citenamefont {Feyereisen}(1993)}]{RI:integral:1}%
  \BibitemOpen
  \bibfield  {author} {\bibinfo {author} {\bibfnamefont {O.}~\bibnamefont
  {Vahtras}}, \bibinfo {author} {\bibfnamefont {J.}~\bibnamefont {Alml{\"o}f}},
  \ and\ \bibinfo {author} {\bibfnamefont {M.~W.}\ \bibnamefont {Feyereisen}},\
  }\bibfield  {title} {\enquote {\bibinfo {title} {{Integral approximations for
  LCAO-SCF calculations}},}\ }\href@noop {} {\bibfield  {journal} {\bibinfo
  {journal} {Chem. Phys. Lett.}\ }\textbf {\bibinfo {volume} {213}},\ \bibinfo
  {pages} {514--518} (\bibinfo {year} {1993})}\BibitemShut {NoStop}%
\bibitem [{\citenamefont {Dunlap}, \citenamefont {Connolly},\ and\
  \citenamefont {Sabin}(1979)}]{RI:integral:2}%
  \BibitemOpen
  \bibfield  {author} {\bibinfo {author} {\bibfnamefont {B.~I.}\ \bibnamefont
  {Dunlap}}, \bibinfo {author} {\bibfnamefont {J.~W.~D.}\ \bibnamefont
  {Connolly}}, \ and\ \bibinfo {author} {\bibfnamefont {J.~R.}\ \bibnamefont
  {Sabin}},\ }\bibfield  {title} {\enquote {\bibinfo {title} {{On some
  approximations in applications of X$\alpha$ theory}},}\ }\href@noop {}
  {\bibfield  {journal} {\bibinfo  {journal} {J. Chem. Phys.}\ }\textbf
  {\bibinfo {volume} {71}},\ \bibinfo {pages} {3396--3402} (\bibinfo {year}
  {1979})}\BibitemShut {NoStop}%
\bibitem [{\citenamefont {Whitten}(1973)}]{RI:integral:3}%
  \BibitemOpen
  \bibfield  {author} {\bibinfo {author} {\bibfnamefont {J.~L.}\ \bibnamefont
  {Whitten}},\ }\bibfield  {title} {\enquote {\bibinfo {title} {Coulombic
  potential energy integrals and approximations},}\ }\href@noop {} {\bibfield
  {journal} {\bibinfo  {journal} {J. Chem. Phys.}\ }\textbf {\bibinfo {volume}
  {58}},\ \bibinfo {pages} {4496--4501} (\bibinfo {year} {1973})}\BibitemShut
  {NoStop}%
\bibitem [{\citenamefont {Winter}\ and\ \citenamefont
  {H{\"a}ttig}(2011)}]{RICC2:Winter:2011}%
  \BibitemOpen
  \bibfield  {author} {\bibinfo {author} {\bibfnamefont {N.~O.~C.}\
  \bibnamefont {Winter}}\ and\ \bibinfo {author} {\bibfnamefont
  {C.}~\bibnamefont {H{\"a}ttig}},\ }\bibfield  {title} {\enquote {\bibinfo
  {title} {Scaled opposite-spin {CC2} for ground and excited states with fourth
  order scaling computational costs},}\ }\href@noop {} {\bibfield  {journal}
  {\bibinfo  {journal} {J. Chem. Phys.}\ }\textbf {\bibinfo {volume} {134}},\
  \bibinfo {pages} {184101} (\bibinfo {year} {2011})}\BibitemShut {NoStop}%
\bibitem [{TUR()}]{TURBOMOLE}%
  \BibitemOpen
  \href@noop {} {\enquote {\bibinfo {title} {{TURBOMOLE V7.5 2020}, a
  development of {University of Karlsruhe} and {Forschungszentrum Karlsruhe
  GmbH}, 1989--2007, {TURBOMOLE GmbH}, since 2007; available from
  https://www.turbomole.org},}\ }\BibitemShut {NoStop}%
\bibitem [{\citenamefont {H{\"a}ttig}, \citenamefont {Hellweg},\ and\
  \citenamefont {K{\"o}hn}(2006)}]{Haettig:C60:str}%
  \BibitemOpen
  \bibfield  {author} {\bibinfo {author} {\bibfnamefont {C.}~\bibnamefont
  {H{\"a}ttig}}, \bibinfo {author} {\bibfnamefont {A.}~\bibnamefont {Hellweg}},
  \ and\ \bibinfo {author} {\bibfnamefont {A.}~\bibnamefont {K{\"o}hn}},\
  }\bibfield  {title} {\enquote {\bibinfo {title} {{Distributed memory parallel
  implementation of energies and gradients for second-order M{\o}ller--Plesset
  perturbation theory with the resolution-of-the-identity approximation}},}\
  }\href@noop {} {\bibfield  {journal} {\bibinfo  {journal} {Phys. Chem. Chem.
  Phys.}\ }\textbf {\bibinfo {volume} {8}},\ \bibinfo {pages} {1159--1169}
  (\bibinfo {year} {2006})}\BibitemShut {NoStop}%
\bibitem [{\citenamefont {C}\ and\ \citenamefont
  {Plesset}(1934)}]{MP_original}%
  \BibitemOpen
  \bibfield  {author} {\bibinfo {author} {\bibfnamefont {C.~M.}\ \bibnamefont
  {C}}\ and\ \bibinfo {author} {\bibfnamefont {M.~S.}\ \bibnamefont
  {Plesset}},\ }\bibfield  {title} {\enquote {\bibinfo {title} {Note on an
  approximation treatment for many-electron systems},}\ }\href@noop {}
  {\bibfield  {journal} {\bibinfo  {journal} {Phys. Rev.}\ }\textbf {\bibinfo
  {volume} {46}},\ \bibinfo {pages} {618} (\bibinfo {year} {1934})}\BibitemShut
  {NoStop}%
\bibitem [{\citenamefont {Dunning~Jr}(1989)}]{Dunning_basis}%
  \BibitemOpen
  \bibfield  {author} {\bibinfo {author} {\bibfnamefont {T.~H.}\ \bibnamefont
  {Dunning~Jr}},\ }\bibfield  {title} {\enquote {\bibinfo {title} {{Gaussian
  basis sets for use in correlated molecular calculations. I. The atoms boron
  through neon and hydrogen}},}\ }\href@noop {} {\bibfield  {journal} {\bibinfo
   {journal} {J. Chem. Phys.}\ }\textbf {\bibinfo {volume} {90}},\ \bibinfo
  {pages} {1007--1023} (\bibinfo {year} {1989})}\BibitemShut {NoStop}%
\bibitem [{\citenamefont {Kauczor}, \citenamefont {Norman},\ and\ \citenamefont
  {Saidi}(2013)}]{CPP:Joanna:C6}%
  \BibitemOpen
  \bibfield  {author} {\bibinfo {author} {\bibfnamefont {J.}~\bibnamefont
  {Kauczor}}, \bibinfo {author} {\bibfnamefont {P.}~\bibnamefont {Norman}}, \
  and\ \bibinfo {author} {\bibfnamefont {W.~A.}\ \bibnamefont {Saidi}},\
  }\bibfield  {title} {\enquote {\bibinfo {title} {{Non-additivity of
  polarizabilities and van der Waals $C_6$ coefficients of fullerenes}},}\
  }\href@noop {} {\bibfield  {journal} {\bibinfo  {journal} {J. Chem. Phys.}\
  }\textbf {\bibinfo {volume} {138}},\ \bibinfo {pages} {114107} (\bibinfo
  {year} {2013})}\BibitemShut {NoStop}%
\bibitem [{\citenamefont {Weigend}, \citenamefont {K{\"o}hn},\ and\
  \citenamefont {H{\"a}ttig}(2001)}]{bas:Weigend01a}%
  \BibitemOpen
  \bibfield  {author} {\bibinfo {author} {\bibfnamefont {F.}~\bibnamefont
  {Weigend}}, \bibinfo {author} {\bibfnamefont {A.}~\bibnamefont {K{\"o}hn}}, \
  and\ \bibinfo {author} {\bibfnamefont {C.}~\bibnamefont {H{\"a}ttig}},\
  }\bibfield  {title} {\enquote {\bibinfo {title} {Efficient use of the
  correlation consistent basis sets in resolution of the identity {MP2}
  calculations},}\ }\href@noop {} {\bibfield  {journal} {\bibinfo  {journal}
  {J. Chem. Phys.}\ }\textbf {\bibinfo {volume} {116}},\ \bibinfo {pages}
  {3175--3183} (\bibinfo {year} {2001})}\BibitemShut {NoStop}%
\bibitem [{\citenamefont {Amos}\ \emph {et~al.}(1985)\citenamefont {Amos},
  \citenamefont {Handy}, \citenamefont {Knowles}, \citenamefont {Rice},\ and\
  \citenamefont {Stone}}]{Amos:C6}%
  \BibitemOpen
  \bibfield  {author} {\bibinfo {author} {\bibfnamefont {R.~D.}\ \bibnamefont
  {Amos}}, \bibinfo {author} {\bibfnamefont {N.~C.}\ \bibnamefont {Handy}},
  \bibinfo {author} {\bibfnamefont {P.~J.}\ \bibnamefont {Knowles}}, \bibinfo
  {author} {\bibfnamefont {J.~E.}\ \bibnamefont {Rice}}, \ and\ \bibinfo
  {author} {\bibfnamefont {A.~J.}\ \bibnamefont {Stone}},\ }\bibfield  {title}
  {\enquote {\bibinfo {title} {Ab-initio prediction of properties of {CO}$_2$,
  {NH}$_3$, and {CO}$_2\cdots${NH}$_3$},}\ }\href@noop {} {\bibfield  {journal}
  {\bibinfo  {journal} {J. Chem. Phys.}\ }\textbf {\bibinfo {volume} {89}},\
  \bibinfo {pages} {2186--2192} (\bibinfo {year} {1985})}\BibitemShut {NoStop}%
\bibitem [{\citenamefont {Brown}, \citenamefont {Kemp},\ and\ \citenamefont
  {Mason}(1971)}]{Brown:1971:5:5-helicene}%
  \BibitemOpen
  \bibfield  {author} {\bibinfo {author} {\bibfnamefont {A.}~\bibnamefont
  {Brown}}, \bibinfo {author} {\bibfnamefont {C.~M.}\ \bibnamefont {Kemp}}, \
  and\ \bibinfo {author} {\bibfnamefont {S.~F.}\ \bibnamefont {Mason}},\
  }\bibfield  {title} {\enquote {\bibinfo {title} {Electronic absorption,
  polarised excitation, and circular dichroism spectra of [5]-helicene (dibenzo
  [c, g] phenanthrene)},}\ }\href@noop {} {\bibfield  {journal} {\bibinfo
  {journal} {J. Chem. Soc. A}\ ,\ \bibinfo {pages} {751--755}} (\bibinfo {year}
  {1971})}\BibitemShut {NoStop}%
\bibitem [{\citenamefont {Newman}, \citenamefont {Darlak},\ and\ \citenamefont
  {Tsai}(1967)}]{Newman:1967:6-helicene}%
  \BibitemOpen
  \bibfield  {author} {\bibinfo {author} {\bibfnamefont {M.~S.}\ \bibnamefont
  {Newman}}, \bibinfo {author} {\bibfnamefont {R.~S.}\ \bibnamefont {Darlak}},
  \ and\ \bibinfo {author} {\bibfnamefont {L.~L.}\ \bibnamefont {Tsai}},\
  }\bibfield  {title} {\enquote {\bibinfo {title} {Optical properties of
  hexahelicene},}\ }\href@noop {} {\bibfield  {journal} {\bibinfo  {journal}
  {J. Am. Chem. Soc.}\ }\textbf {\bibinfo {volume} {89}},\ \bibinfo {pages}
  {6191--6193} (\bibinfo {year} {1967})}\BibitemShut {NoStop}%
\bibitem [{\citenamefont {Brickell}\ \emph {et~al.}(1971)\citenamefont
  {Brickell}, \citenamefont {Brown}, \citenamefont {Kemp},\ and\ \citenamefont
  {Mason}}]{Brickell:1971:7-helicene}%
  \BibitemOpen
  \bibfield  {author} {\bibinfo {author} {\bibfnamefont {W.~S.}\ \bibnamefont
  {Brickell}}, \bibinfo {author} {\bibfnamefont {A.}~\bibnamefont {Brown}},
  \bibinfo {author} {\bibfnamefont {C.~M.}\ \bibnamefont {Kemp}}, \ and\
  \bibinfo {author} {\bibfnamefont {S.~F.}\ \bibnamefont {Mason}},\ }\bibfield
  {title} {\enquote {\bibinfo {title} {$\pi$-electron absorption and circular
  dichroism spectra of [6]- and [7]-helicene},}\ }\href {\doibase
  10.1039/J19710000756} {\bibfield  {journal} {\bibinfo  {journal} {J. Chem.
  Soc. A}\ ,\ \bibinfo {pages} {756--760}} (\bibinfo {year}
  {1971})}\BibitemShut {NoStop}%
\bibitem [{\citenamefont {Weigang}, \citenamefont {Turner},\ and\ \citenamefont
  {Trouard}(1966)}]{Weigang:Emission:CD:1966}%
  \BibitemOpen
  \bibfield  {author} {\bibinfo {author} {\bibfnamefont {O.~E.}\ \bibnamefont
  {Weigang}}, \bibinfo {author} {\bibfnamefont {J.~A.}\ \bibnamefont {Turner}},
  \ and\ \bibinfo {author} {\bibfnamefont {P.~A.}\ \bibnamefont {Trouard}},\
  }\bibfield  {title} {\enquote {\bibinfo {title} {Emission polarization and
  circular dichroism of hexahelicene},}\ }\href@noop {} {\bibfield  {journal}
  {\bibinfo  {journal} {J. Chem. Phys.}\ }\textbf {\bibinfo {volume} {45}},\
  \bibinfo {pages} {1126--1134} (\bibinfo {year} {1966})}\BibitemShut {NoStop}%
\bibitem [{\citenamefont {Goedicke}\ and\ \citenamefont
  {Stegemeyer}(1970)}]{GOEDICKE1970937}%
  \BibitemOpen
  \bibfield  {author} {\bibinfo {author} {\bibfnamefont {C.}~\bibnamefont
  {Goedicke}}\ and\ \bibinfo {author} {\bibfnamefont {H.}~\bibnamefont
  {Stegemeyer}},\ }\bibfield  {title} {\enquote {\bibinfo {title} {Resolution
  and racemization of pentahelicene},}\ }\href {\doibase
  https://doi.org/10.1016/S0040-4039(01)97871-2} {\bibfield  {journal}
  {\bibinfo  {journal} {Tetrahedron Letters}\ }\textbf {\bibinfo {volume}
  {11}},\ \bibinfo {pages} {937--940} (\bibinfo {year} {1970})}\BibitemShut
  {NoStop}%
\bibitem [{\citenamefont {Furche}\ \emph {et~al.}(2000)\citenamefont {Furche},
  \citenamefont {Ahlrichs}, \citenamefont {Wachsmann}, \citenamefont {Weber},
  \citenamefont {Sobanski}, \citenamefont {V{\"o}gtle},\ and\ \citenamefont
  {Grimme}}]{Furche2000}%
  \BibitemOpen
  \bibfield  {author} {\bibinfo {author} {\bibfnamefont {F.}~\bibnamefont
  {Furche}}, \bibinfo {author} {\bibfnamefont {R.}~\bibnamefont {Ahlrichs}},
  \bibinfo {author} {\bibfnamefont {C.}~\bibnamefont {Wachsmann}}, \bibinfo
  {author} {\bibfnamefont {E.}~\bibnamefont {Weber}}, \bibinfo {author}
  {\bibfnamefont {A.}~\bibnamefont {Sobanski}}, \bibinfo {author}
  {\bibfnamefont {F.}~\bibnamefont {V{\"o}gtle}}, \ and\ \bibinfo {author}
  {\bibfnamefont {S.}~\bibnamefont {Grimme}},\ }\bibfield  {title} {\enquote
  {\bibinfo {title} {{Circular Dichroism of Helicenes Investigated by
  Time-Dependent Density Functional Theory}},}\ }\href {\doibase
  10.1021/ja991960s} {\bibfield  {journal} {\bibinfo  {journal} {J. Am. Chem.
  Soc.}\ }\textbf {\bibinfo {volume} {122}},\ \bibinfo {pages} {1717--1724}
  (\bibinfo {year} {2000})}\BibitemShut {NoStop}%
\bibitem [{\citenamefont {K{\"o}hn}(2003)}]{Koehn-PhD}%
  \BibitemOpen
  \bibfield  {author} {\bibinfo {author} {\bibfnamefont {A.}~\bibnamefont
  {K{\"o}hn}},\ }\href@noop {} {\enquote {\bibinfo {title} {{Analytische
  Gradienten elektronisch angeregter Zust{\"a}nde und Behandlung offenschaliger
  Systeme im Rahmen der Coupled-Cluster-Methode RI-CC2}},}\ } (\bibinfo {year}
  {2003}),\ \bibinfo {note} {{Ph.D.} thesis. Fakult{\"a}t f{\"u}r Chemie und
  Biowissenshaften der Universit{\"a}t Karlsruhe}\BibitemShut {NoStop}%
\bibitem [{\citenamefont {Abbate}\ \emph {et~al.}(2014)\citenamefont {Abbate},
  \citenamefont {Longhi}, \citenamefont {Lebon}, \citenamefont {Castiglioni},
  \citenamefont {Superchi}, \citenamefont {Pisani}, \citenamefont {Fontana},
  \citenamefont {Torricelli}, \citenamefont {Caronna}, \citenamefont {Villani},
  \citenamefont {Sabia}, \citenamefont {Tommasini}, \citenamefont {Lucotti},
  \citenamefont {Mendola}, \citenamefont {Mele},\ and\ \citenamefont
  {Lightner}}]{Helicenes:Sergio}%
  \BibitemOpen
  \bibfield  {author} {\bibinfo {author} {\bibfnamefont {S.}~\bibnamefont
  {Abbate}}, \bibinfo {author} {\bibfnamefont {G.}~\bibnamefont {Longhi}},
  \bibinfo {author} {\bibfnamefont {F.}~\bibnamefont {Lebon}}, \bibinfo
  {author} {\bibfnamefont {E.}~\bibnamefont {Castiglioni}}, \bibinfo {author}
  {\bibfnamefont {S.}~\bibnamefont {Superchi}}, \bibinfo {author}
  {\bibfnamefont {L.}~\bibnamefont {Pisani}}, \bibinfo {author} {\bibfnamefont
  {F.}~\bibnamefont {Fontana}}, \bibinfo {author} {\bibfnamefont
  {F.}~\bibnamefont {Torricelli}}, \bibinfo {author} {\bibfnamefont
  {T.}~\bibnamefont {Caronna}}, \bibinfo {author} {\bibfnamefont
  {C.}~\bibnamefont {Villani}}, \bibinfo {author} {\bibfnamefont
  {R.}~\bibnamefont {Sabia}}, \bibinfo {author} {\bibfnamefont
  {M.}~\bibnamefont {Tommasini}}, \bibinfo {author} {\bibfnamefont
  {A.}~\bibnamefont {Lucotti}}, \bibinfo {author} {\bibfnamefont
  {D.}~\bibnamefont {Mendola}}, \bibinfo {author} {\bibfnamefont
  {A.}~\bibnamefont {Mele}}, \ and\ \bibinfo {author} {\bibfnamefont {D.~A.}\
  \bibnamefont {Lightner}},\ }\bibfield  {title} {\enquote {\bibinfo {title}
  {{Helical Sense-Responsive and Substituent-Sensitive Features in Vibrational
  and Electronic Circular Dichroism, in Circularly Polarized Luminescence, and
  in Raman Spectra of Some Simple Optically Active Hexahelicenes}},}\
  }\href@noop {} {\bibfield  {journal} {\bibinfo  {journal} {J. Phys. Chem. C}\
  }\textbf {\bibinfo {volume} {118}},\ \bibinfo {pages} {1682--1695} (\bibinfo
  {year} {2014})}\BibitemShut {NoStop}%
\bibitem [{\citenamefont {Buss}\ and\ \citenamefont
  {Kolster}(1996)}]{BUSS1996309}%
  \BibitemOpen
  \bibfield  {author} {\bibinfo {author} {\bibfnamefont {V.}~\bibnamefont
  {Buss}}\ and\ \bibinfo {author} {\bibfnamefont {K.}~\bibnamefont {Kolster}},\
  }\bibfield  {title} {\enquote {\bibinfo {title} {Electronic structure
  calculations on helicenes. {C}oncerning the chirality of helically twisted
  aromatic systems},}\ }\href {\doibase
  https://doi.org/10.1016/0301-0104(95)00406-8} {\bibfield  {journal} {\bibinfo
   {journal} {Chem. Phys.}\ }\textbf {\bibinfo {volume} {203}},\ \bibinfo
  {pages} {309 -- 316} (\bibinfo {year} {1996})}\BibitemShut {NoStop}%
\bibitem [{\citenamefont {Nakai}, \citenamefont {Mori},\ and\ \citenamefont
  {Inoue}(2012)}]{Helicenes:2012}%
  \BibitemOpen
  \bibfield  {author} {\bibinfo {author} {\bibfnamefont {Y.}~\bibnamefont
  {Nakai}}, \bibinfo {author} {\bibfnamefont {T.}~\bibnamefont {Mori}}, \ and\
  \bibinfo {author} {\bibfnamefont {Y.}~\bibnamefont {Inoue}},\ }\bibfield
  {title} {\enquote {\bibinfo {title} {{Theoretical and Experimental Studies on
  Circular Dichroism of Carbo[n]helicenes}},}\ }\href@noop {} {\bibfield
  {journal} {\bibinfo  {journal} {J. Phys. Chem. A}\ }\textbf {\bibinfo
  {volume} {116}},\ \bibinfo {pages} {7372--7385} (\bibinfo {year}
  {2012})}\BibitemShut {NoStop}%
\bibitem [{\citenamefont {Martin}\ and\ \citenamefont
  {Marchant}(1974)}]{MARTIN1974343}%
  \BibitemOpen
  \bibfield  {author} {\bibinfo {author} {\bibfnamefont {R.}~\bibnamefont
  {Martin}}\ and\ \bibinfo {author} {\bibfnamefont {M.}~\bibnamefont
  {Marchant}},\ }\bibfield  {title} {\enquote {\bibinfo {title} {Resolution and
  optical properties ([$\alpha$]$_{\rm max}$, ord and cd) of hepta-, octa- and
  nonahelicene},}\ }\href {\doibase
  https://doi.org/10.1016/S0040-4020(01)91468-1} {\bibfield  {journal}
  {\bibinfo  {journal} {Tetrahedron}\ }\textbf {\bibinfo {volume} {30}},\
  \bibinfo {pages} {343 -- 345} (\bibinfo {year} {1974})}\BibitemShut {NoStop}%
\bibitem [{\citenamefont {Kumar}\ and\ \citenamefont {Thakkar}(2011)}]{DOSD}%
  \BibitemOpen
  \bibfield  {author} {\bibinfo {author} {\bibfnamefont {A.}~\bibnamefont
  {Kumar}}\ and\ \bibinfo {author} {\bibfnamefont {A.~J.}\ \bibnamefont
  {Thakkar}},\ }\bibfield  {title} {\enquote {\bibinfo {title} {{Dipole
  polarizability, sum rules, mean excitation energies, and long-range
  dispersion coefficients for buckminsterfullerene C$_{60}$}},}\ }\href@noop {}
  {\bibfield  {journal} {\bibinfo  {journal} {Chem. Phys. Lett.}\ }\textbf
  {\bibinfo {volume} {516}},\ \bibinfo {pages} {208--211} (\bibinfo {year}
  {2011})}\BibitemShut {NoStop}%
\bibitem [{\citenamefont {Ruzsinszky}\ \emph {et~al.}(2012)\citenamefont
  {Ruzsinszky}, \citenamefont {Perdew}, \citenamefont {Tao}, \citenamefont
  {Csonka},\ and\ \citenamefont {Pitarke}}]{PhysRevLett.109.233203}%
  \BibitemOpen
  \bibfield  {author} {\bibinfo {author} {\bibfnamefont {A.}~\bibnamefont
  {Ruzsinszky}}, \bibinfo {author} {\bibfnamefont {J.~P.}\ \bibnamefont
  {Perdew}}, \bibinfo {author} {\bibfnamefont {J.}~\bibnamefont {Tao}},
  \bibinfo {author} {\bibfnamefont {G.~I.}\ \bibnamefont {Csonka}}, \ and\
  \bibinfo {author} {\bibfnamefont {J.~M.}\ \bibnamefont {Pitarke}},\
  }\bibfield  {title} {\enquote {\bibinfo {title} {{Van der Waals Coefficients
  for Nanostructures: Fullerenes Defy Conventional Wisdom}},}\ }\href {\doibase
  10.1103/PhysRevLett.109.233203} {\bibfield  {journal} {\bibinfo  {journal}
  {Phys. Rev. Lett.}\ }\textbf {\bibinfo {volume} {109}},\ \bibinfo {pages}
  {233203} (\bibinfo {year} {2012})}\BibitemShut {NoStop}%
\bibitem [{\citenamefont {Saidi}\ and\ \citenamefont
  {Norman}(2016)}]{Saidi:2016}%
  \BibitemOpen
  \bibfield  {author} {\bibinfo {author} {\bibfnamefont {W.~A.}\ \bibnamefont
  {Saidi}}\ and\ \bibinfo {author} {\bibfnamefont {P.}~\bibnamefont {Norman}},\
  }\bibfield  {title} {\enquote {\bibinfo {title} {{Polarizabilities and van
  der Waals $C_6$ coefficients of fullerenes from an atomistic electrodynamics
  model: Anomalous scaling with number of carbon atoms}},}\ }\href@noop {}
  {\bibfield  {journal} {\bibinfo  {journal} {J. Chem. Phys.}\ }\textbf
  {\bibinfo {volume} {145}},\ \bibinfo {pages} {024311} (\bibinfo {year}
  {2016})}\BibitemShut {NoStop}%
\end{thebibliography}%
\end{document}


\preprint{CPP-RICC2}

\title{Damped (linear) response theory within the resolution-of-identity coupled cluster singles and approximate doubles (RI-CC2) method.
\\
Supplementary Information.}

\author{Daniil A. Fedotov}
\affiliation{DTU Chemistry, Technical University of Denmark, Kemitorvet Bldg 207, DK-2800 Kongens Lyngby, Denmark}%
\author{Sonia Coriani}
\email{soco@kemi.dtu.dk}
\affiliation{DTU Chemistry, Technical University of Denmark, Kemitorvet Bldg 207, DK-2800 Kongens Lyngby, Denmark}
\author{Christof H{\"a}ttig}
\email{christof.haettig@rub.de}
\affiliation{Arbeitsgruppe Quantenchemie, Ruhr-Universit{\"a}t, Bochum D-44780, Germany}

\date{\today}%

\maketitle

\section{Geometries}
\begin{spacing}{.7}
\scriptsize
\begin{longtable}{l*{4}{S[round-mode=places, round-precision=6, table-format=1.8, table-number-alignment=center]}
}\\
\caption{5-helicene. Cartesian coordinates (Bohr) from a MP2/cc-pVTZ optimization in gas phase.}\\
\hline
\hline
Atom  &  $X$ & $Y$ & $Z$  \\
\hline
C	&	 0.007045134994	&	 7.086729926	&	-2.584696133	\\
C	&	-0.3926554651	&	 1.231321039	&	 6.283326197	\\
C	&	 1.21132854200	&	 6.170954495	&	-4.70601801	    \\
C	&	 2.13011765700	&	 3.677612544	&	-4.711638889	\\
C	&	 1.70622528900	&	 2.10258983	    &	-2.674778961	\\
C	&	 0.34616060430	&	 2.938818576	&	-0.5317630611	\\
C	&	-0.1038338008	&	 1.353804225	&	 1.645138276	\\
C	&	-0.3733032127	&	 5.537386162	&	-0.4453290602	\\
C	&	-0.6175298462	&	 2.581898169	&	 3.977660685	\\
C	&	-1.301515398	&	 6.588810254	&	 1.847390565	\\
C	&	-1.288804637	&	 5.185970158	&	 4.012684039	\\
C	&	 3.198869196	&	 2.985578226	&	-6.310947435	\\
C	&	 1.515397812	&	 7.383633034	&	-6.322894843	\\
C	&	-0.7410665718	&	 2.221373258	&	 8.040471967	\\
C	&	-1.868737906	&	 8.555316179	&	 1.876704068	\\
C	&	-1.801067851	&	 6.032324286	&	 5.804833402	\\
C	&	 2.464707084	&	 0.2078951807	&	-2.701026052	\\
C	&	-0.6051160037	&	 9.038190173	&	-2.499096735	\\
C	&	-0.007045135	&	-7.086729926	&	-2.584696133	\\
C	&	 0.3926554651	&	-1.231321039	&	 6.283326197	\\
C	&	-1.211328542	&	-6.170954495	&	-4.70601801	    \\
C	&	-2.130117657	&	-3.677612544	&	-4.711638889	\\
C	&	-1.706225289	&	-2.10258983	    &	-2.674778961	\\
C	&	-0.3461606043	&	-2.938818576	&	-0.5317630611	\\
C	&	 0.1038338008	&	-1.353804225	&	 1.645138276	\\
C	&	 0.3733032127	&	-5.537386162	&	-0.4453290602	\\
C	&	 0.6175298462	&	-2.581898169	&	 3.977660685	\\
C	&	 1.301515398	&	-6.588810254	&	 1.847390565	\\
C	&	 1.288804637	&	-5.185970158	&	 4.012684039	\\
H	&	-3.198869196	&	-2.985578226	&	-6.310947435	\\
H	&	-1.515397812	&	-7.383633034	&	-6.322894843	\\
H	&	 0.7410665718	&	-2.221373258	&	 8.040471967	\\
H	&	 1.868737906	&	-8.555316179	&	 1.876704068	\\
H	&	 1.801067851	&	-6.032324286	&	 5.804833402	\\
H	&	-2.464707084	&	-0.2078951807	&	-2.701026052	\\
H	&	 0.6051160037	&	-9.038190173	&	-2.499096735	\\
\hline
\hline
\end{longtable}
\end{spacing}
\clearpage

\begin{spacing}{.7}
\begin{longtable}{l*{4}{S[round-mode=places, round-precision=6, table-format=1.8, table-number-alignment=center]}
}\\
\scriptsize
\caption{6-helicene. Cartesian coordinates (Bohr) from a MP2/cc-pVTZ optimization in gas phase.}\\
\hline
\hline
Atom  &  $X$ & $Y$ & $Z$  \\
\hline
C	&	2.285540929	&	3.957621293	&	-5.795034471	\\
C	&	-0.8941046691	&	2.147518213	&	6.976780092	\\
C	&	3.263093502	&	1.874425594	&	-4.468084684	\\
C	&	0.0000000000	&	0.0000000000	&	5.643948815	\\
C	&	2.474276279	&	1.367362459	&	-2.032549256	\\
C	&	0.0000000000	&	0.0000000000	&	2.948625249	\\
C	&	0.6703623547	&	2.91111389	&	-0.812394832	\\
C	&	-0.2728339483	&	2.393609979	&	1.691910409	\\
C	&	-0.1427227222	&	5.142819095	&	-2.085792953	\\
C	&	-1.398311873	&	4.419362182	&	3.044711795	\\
C	&	-1.626902611	&	6.960469034	&	-0.7702206265	\\
C	&	-2.118423073	&	6.663656693	&	1.745723566	\\
H	&	4.653624389	&	0.6611385254	&	-5.347459292	\\
H	&	2.878166836	&	4.322974619	&	-7.716774194	\\
H	&	-0.956887733	&	2.077144965	&	9.021362436	\\
H	&	-0.03600633224	&	7.266921592	&	-5.551210884	\\
H	&	-2.577463444	&	5.80636838	&	6.698429404	\\
H	&	-2.244410636	&	8.639717875	&	-1.7647323	\\
H	&	-3.085689852	&	8.128821454	&	2.798430768	\\
H	&	3.280737174	&	-0.217482203	&	-1.028652786	\\
C	&	-0.6402981982	&	-5.586815472	&	-4.596981049	\\
C	&	1.749033131	&	-4.225799867	&	5.696091824	\\
C	&	0.6402981982	&	5.586815472	&	-4.596981049	\\
C	&	-1.749033131	&	4.225799867	&	5.696091824	\\
C	&	-2.285540929	&	-3.957621293	&	-5.795034471	\\
C	&	0.8941046691	&	-2.147518213	&	6.976780092	\\
C	&	-3.263093502	&	-1.874425594	&	-4.468084684	\\
C	&	-2.474276279	&	-1.367362459	&	-2.032549256	\\
C	&	-0.6703623547	&	-2.91111389	&	-0.812394832	\\
C	&	0.2728339483	&	-2.393609979	&	1.691910409	\\
C	&	0.1427227222	&	-5.142819095	&	-2.085792953	\\
C	&	1.398311873	&	-4.419362182	&	3.044711795	\\
C	&	1.626902611	&	-6.960469034	&	-0.7702206265	\\
C	&	2.118423073	&	-6.663656693	&	1.745723566	\\
H	&	-4.653624389	&	-0.6611385254	&	-5.347459292	\\
H	&	-2.878166836	&	-4.322974619	&	-7.716774194	\\
H	&	0.956887733	&	-2.077144965	&	9.021362436	\\
H	&	0.03600633224	&	-7.266921592	&	-5.551210884	\\
H	&	2.577463444	&	-5.80636838	&	6.698429404	\\
H	&	2.244410636	&	-8.639717875	&	-1.7647323	\\
H	&	3.085689852	&	-8.128821454	&	2.798430768	\\
H	&	-3.280737174	&	0.217482203	&	-1.028652786	\\
\hline
\hline
\end{longtable}
\end{spacing}
\clearpage

\begin{spacing}{.7}
\scriptsize
\begin{longtable}{c*{4}{S[round-mode=places, round-precision=6, table-format=1.8, table-number-alignment=center]}}\\
\caption{7-helicene. Cartesian coordinates (Bohr) from a MP2/cc-pVTZ optimization in gas phase.}
\endfirsthead
\endhead
\hline
\hline
Atom  &  $X$ & $Y$ & $Z$  \\
\hline
C	&	0.8604097181	&	5.27993735	&	5.568988053	\\
C	&	0.3402395809	&	2.644899599	&	5.61277473	\\
C	&	-0.08205785243	&	1.35447424	&	3.294087573	\\
C	&	-0.6011759467	&	2.826207332	&	1.07968918	\\
C	&	0.230143736	&	5.371221272	&	1.028907633	\\
C	&	0.9855838793	&	6.557231587	&	3.318960111	\\
C	&	0.3129473107	&	6.686378386	&	-1.315755367	\\
C	&	-0.4482505146	&	5.541900612	&	-3.507215193	\\
C	&	-1.657246356	&	3.139705915	&	-3.443430348	\\
C	&	-1.912786143	&	1.8546871	&	-1.090874615	\\
C	&	0.2572736614	&	1.270108412	&	7.915901156	\\
C	&	-0.2572736614	&	-1.270108412	&	7.915901156	\\
C	&	-0.3402395809	&	-2.644899599	&	5.61277473	\\
C	&	0.08205785243	&	-1.35447424	&	3.294087573	\\
C	&	-0.8604097181	&	-5.27993735	&	5.568988053	\\
C	&	-0.9855838793	&	-6.557231587	&	3.318960111	\\
C	&	-0.230143736	&	-5.371221272	&	1.028907633	\\
C	&	0.6011759467	&	-2.826207332	&	1.07968918	\\
C	&	-0.3129473107	&	-6.686378386	&	-1.315755367	\\
C	&	0.4482505146	&	-5.541900612	&	-3.507215193	\\
C	&	1.657246356	&	-3.139705915	&	-3.443430348	\\
C	&	1.912786143	&	-1.8546871	&	-1.090874615	\\
C	&	2.723330704	&	-2.066773188	&	-5.64113728	\\
H	&	0.2701747857	&	-6.509507363	&	-5.302986072	\\
H	&	-0.5129607659	&	-2.281881135	&	9.67695413	\\
H	&	-1.557963363	&	-8.522123277	&	3.267942529	\\
H	&	-1.24782343	&	-6.237534133	&	7.336250639	\\
C	&	3.5154614	&	0.2771954184	&	-0.9794885285	\\
H	&	-1.050877223	&	-8.596126192	&	-1.341838332	\\
C	&	-2.723330704	&	2.066773188	&	-5.64113728	\\
H	&	-0.2701747857	&	6.509507363	&	-5.302986072	\\
H	&	0.5129607659	&	2.281881135	&	9.67695413	\\
H	&	1.557963363	&	8.522123277	&	3.267942529	\\
H	&	1.24782343	&	6.237534133	&	7.336250639	\\
C	&	-3.5154614	&	-0.2771954184	&	-0.9794885285	\\
H	&	1.050877223	&	8.596126192	&	-1.341838332	\\
C	&	-4.129380202	&	-0.1274899149	&	-5.508977898	\\
C	&	-4.592176923	&	-1.257524033	&	-3.148411045	\\
C	&	4.592176923	&	1.257524033	&	-3.148411045	\\
C	&	4.129380202	&	0.1274899149	&	-5.508977898	\\
H	&	2.464873592	&	-3.028847587	&	-7.429797071	\\
H	&	4.943277421	&	0.9280312344	&	-7.205120773	\\
H	&	5.815077393	&	2.89094877	&	-3.01758107	\\
H	&	3.935260741	&	1.122649434	&	0.8326921985	\\
H	&	-2.464873592	&	3.028847587	&	-7.429797071	\\
H	&	-4.943277421	&	-0.9280312344	&	-7.205120773	\\
H	&	-5.815077393	&	-2.89094877	&	-3.01758107	\\
H	&	-3.935260741	&	-1.122649434	&	0.8326921985	\\
\hline
\hline
\end{longtable}
\end{spacing}

\clearpage

\begin{spacing}{0.6}
\tiny
\begin{longtable}{c*{4}{S[round-mode=places, round-precision=6, table-format=1.8, table-number-alignment=center]}}\\
\caption{C$_{60}$. Cartesian coordinates (Bohr) from a MP2/cc-pVTZ optimization in gas phase.}
\endfirsthead
\endhead
\hline
\hline
Atom  &  $X$ & $Y$ & $Z$  \\
\hline
C	&	0.000000000000	&	1.326606436	&	6.558847477	\\
C	&	0.000000000000	&	-1.326606436	&	6.558847477	\\
C	&	-2.206176587	&	-2.690098552	&	5.716163006	\\
C	&	-4.35267089	&	-1.363492116	&	4.896275139	\\
C	&	-4.35267089	&	1.363492116	&	4.896275139	\\
C	&	-2.206176587	&	2.690098552	&	5.716163006	\\
C	&	-5.716163006	&	-2.206176587	&	2.690098552	\\
C	&	-6.558847477	&	0.000000000000	&	1.326606436	\\
C	&	-5.716163006	&	2.206176587	&	2.690098552	\\
C	&	-4.896275139	&	-4.35267089	&	1.363492116	\\
C	&	-4.896275139	&	-4.35267089	&	-1.363492116	\\
C	&	-5.716163006	&	-2.206176587	&	-2.690098552	\\
C	&	-6.558847477	&	0.000000000000	&	-1.326606436	\\
C	&	-2.690098552	&	-5.716163006	&	2.206176587	\\
C	&	-1.363492116	&	-4.896275139	&	4.35267089	\\
C	&	-5.716163006	&	2.206176587	&	-2.690098552	\\
C	&	-4.896275139	&	4.35267089	&	-1.363492116	\\
C	&	-4.896275139	&	4.35267089	&	1.363492116	\\
C	&	-2.690098552	&	5.716163006	&	2.206176587	\\
C	&	-1.363492116	&	4.896275139	&	4.35267089	\\
C	&	2.206176587	&	2.690098552	&	5.716163006	\\
C	&	1.363492116	&	4.896275139	&	4.35267089	\\
C	&	-1.326606436	&	6.558847477	&	0.000000000000	\\
C	&	-2.690098552	&	5.716163006	&	-2.206176587	\\
C	&	-4.35267089	&	1.363492116	&	-4.896275139	\\
C	&	-4.35267089	&	-1.363492116	&	-4.896275139	\\
C	&	-2.690098552	&	-5.716163006	&	-2.206176587	\\
C	&	-1.326606436	&	-6.558847477	&	0.000000000000	\\
C	&	1.363492116	&	-4.896275139	&	4.35267089	\\
C	&	2.206176587	&	-2.690098552	&	5.716163006	\\
C	&	4.35267089	&	-1.363492116	&	4.896275139	\\
C	&	4.35267089	&	1.363492116	&	4.896275139	\\
C	&	2.690098552	&	5.716163006	&	2.206176587	\\
C	&	1.326606436	&	6.558847477	&	0.000000000000	\\
C	&	-1.363492116	&	4.896275139	&	-4.35267089	\\
C	&	-2.206176587	&	2.690098552	&	-5.716163006	\\
C	&	-2.206176587	&	-2.690098552	&	-5.716163006	\\
C	&	-1.363492116	&	-4.896275139	&	-4.35267089	\\
C	&	1.326606436	&	-6.558847477	&	0.000000000000	\\
C	&	2.690098552	&	-5.716163006	&	2.206176587	\\
C	&	4.896275139	&	-4.35267089	&	1.363492116	\\
C	&	5.716163006	&	-2.206176587	&	2.690098552	\\
C	&	5.716163006	&	2.206176587	&	2.690098552	\\
C	&	4.896275139	&	4.35267089	&	1.363492116	\\
C	&	2.690098552	&	5.716163006	&	-2.206176587	\\
C	&	1.363492116	&	4.896275139	&	-4.35267089	\\
C	&	0.000000000000	&	1.326606436	&	-6.558847477	\\
C	&	0.000000000000	&	-1.326606436	&	-6.558847477	\\
C	&	1.363492116	&	-4.896275139	&	-4.35267089	\\
C	&	2.690098552	&	-5.716163006	&	-2.206176587	\\
C	&	4.896275139	&	-4.35267089	&	-1.363492116	\\
C	&	6.558847477	&	0.000000000000	&	1.326606436	\\
C	&	4.896275139	&	4.35267089	&	-1.363492116	\\
C	&	2.206176587	&	2.690098552	&	-5.716163006	\\
C	&	2.206176587	&	-2.690098552	&	-5.716163006	\\
C	&	4.35267089	&	-1.363492116	&	-4.896275139	\\
C	&	5.716163006	&	-2.206176587	&	-2.690098552	\\
C	&	6.558847477	&	0.000000000000	&	-1.326606436	\\
C	&	5.716163006	&	2.206176587	&	-2.690098552	\\
C	&	4.35267089	&	1.363492116	&	-4.896275139	\\
\hline
\hline
\end{longtable}
\end{spacing}
\clearpage

\begin{spacing}{.7}
\begin{longtable}{c*{4}{S[round-mode=places, round-precision=6, table-format=1.8, table-number-alignment=center]}}\\
\caption{Acetaldehyde. Cartesian coordinates ({\AA}) from a MP2/cc-pVTZ optimization in gas phase. }\\
\hline
\hline
{Atom}  &  {$X$} & {$Y$} & {$Z$}  \\
\hline
C    &   -0.227983	&	 0.401184	&	-0.000001	\\
O    &   -1.234224	&	-0.276104	&	 0.000000	\\
C    &    1.165710	&	-0.149966	&	-0.000001	\\
H    &    1.701067	&	 0.216204	&	-0.876282	\\
H    &    1.701048	&	 0.216183	&	 0.876302	\\
H    &    1.144625	&	-1.235154	&	-0.000011	\\
H    &   -0.299307	&	 1.504289	&	 0.000002	\\
\hline
\hline
\end{longtable}
\end{spacing}

\begin{spacing}{.7}
\begin{longtable}{c*{4}{S[round-mode=places, round-precision=6, table-format=1.9, table-number-alignment=center]}}\\
\caption{Butane. Cartesian coordinates ({\AA}) from a MP2/cc-pVTZ optimization in gas phase.}\\
\hline
\hline
\hline
{Atom}  &  {$X$} & {$Y$} & {$Z$}  \\
\hline
C   &     0.702992	&	1.815058	&	0.0000000	\\
H   &     0.188707	&	2.201756	&	0.879596	\\
H   &     0.188707	&	2.201756	&	-0.879596	\\
H   &     1.714246	&	2.218440	&	0.0000000	\\
C   &     0.702992	&	0.292509	&	0.0000000	\\
H   &     1.243489	&	-0.077578	&	0.874153	\\
H   &     1.243489	&	-0.077578	&	-0.874153	\\
C   &    -0.702992	&	-0.292509	&	0.0000000	\\
H   &    -1.243489	&	0.077578	&	-0.874153	\\
H   &    -1.243489	&	0.077578	&	0.874153	\\
C   &    -0.702992	&	-1.815058	&	0.0000000	\\
H   &    -0.188707	&	-2.201756	&	0.879596	\\
H   &    -1.714246	&	-2.21844	&	0.0000000	\\
H   &    -0.188707	&	-2.201756	&	-0.879596	\\
\hline
\hline
\end{longtable}
\end{spacing}

\begin{spacing}{.7}
\begin{longtable}{c*{4}{S[round-mode=places, round-precision=6, table-format=1.9, table-number-alignment=center]}}\\
\caption{Ethane. Cartesian coordinates ({\AA}) from a MP2/cc-pVTZ optimization in gas phase. }\\
\hline
\hline
{Atom}  &  {$X$} & {$Y$} & {$Z$}  \\
\hline
C    &  0.000000	&	0.000000	&	0.761647	\\
H    &  0.000000	&	1.014477	&	1.155787	\\
H    &  0.878563	&	-0.507239	&	1.155787	\\
H    & -0.878563	&	-0.507239	&	1.155787	\\
C    &  0.000000	&	0.000000	&	-0.761647	\\
H    & -0.878563	&	0.507239	&	-1.155787	\\
H    &  0.878563	&	0.507239	&	-1.155787	\\
H    &  0.000000	&	-1.014477	&	-1.155787	\\
\hline
\hline
\end{longtable}
\end{spacing}

\begin{spacing}{.7}
\begin{longtable}{c*{4}{S[round-mode=places, round-precision=6, table-format=1.9, table-number-alignment=center]}}\\
\caption{Methane. Cartesian coordinates ({\AA}) from a MP2/cc-pVTZ optimization in gas phase. }\\
\hline
\hline
{Atom}  &  {$X$} & {$Y$} & {$Z$}  \\
\hline
C   &   0.000000	&	0.000000		&	0.000000		\\
H   &   0.626641	&	0.626641	&	0.626641	\\
H   &  -0.626641	&	-0.626641	&	0.626641	\\
H   &  -0.626641	&	0.626641	&	-0.626641	\\
H   &   0.626641	&	-0.626641	&	-0.626641	\\
\hline
\hline
\end{longtable}
\end{spacing}

\clearpage

\begin{spacing}{.7}
\begin{longtable}{c*{3}{S[round-mode=places, round-precision=6, table-format=1.9, table-number-alignment=center]}}\\
\caption{Formaldehyde. Cartesian coordinates ({\AA}) from a MP2/cc-pVTZ optimization in gas phase.}\\
\hline
\hline
{Atom}  &  {$X$} & {$Y$} & {$Z$}  \\
\hline
C  &     -0.532531	&	 0.000000	&	 0.000006	\\
O  &      0.677980	&	 0.000000	&	-0.000007	\\
H  &     -1.114325	&	 0.934145	&	-0.000007	\\
H  &     -1.114325	&	-0.934145	&	 0.000031	\\
\hline
\hline
\end{longtable}
\end{spacing}

\begin{spacing}{.7}
\begin{longtable}{c*{4}{S[round-mode=places, round-precision=6, table-format=1.9, table-number-alignment=center]}}\\
\caption{Pentane. Cartesian coordinates ({\AA}) from a MP2/cc-pVTZ optimization in gas phase. }\\
\hline
\hline
{Atom}  &  {$X$} & {$Y$} & {$Z$}  \\
\hline
C    &    0.000000	&	2.532869	&	0.327344	\\
H    &    0.879582	&	2.564774	&	0.970004	\\
H    &   -0.879582	&	2.564774	&	0.970004	\\
H    &    0.000000	&	3.433777	&	-0.284061	\\
C    &    0.000000	&	1.271809	&	-0.526283	\\
H    &    0.874204	&	1.269114	&	-1.181074	\\
H    &   -0.874204	&	1.269114	&	-1.181074	\\
C    &    0.000000	&	0.000000	&	0.310799	\\
H    &   -0.874875	&	0.000000	&	0.967441	\\
H    &    0.874875	&	0.000000	&	0.967441	\\
C    &    0.000000	&	-1.271809	&	-0.526283	\\
H    &    0.874204	&	-1.269114	&	-1.181074	\\
H    &   -0.874204	&	-1.269114	&	-1.181074	\\
C    &    0.000000	&	-2.532869	&	0.327344	\\
H    &   -0.879582	&	-2.564774	&	0.970004	\\
H    &    0.000000	&	-3.433777	&	-0.284061	\\
H    &    0.879582	&	-2.564774	&	0.970004	\\
\hline
\hline
\end{longtable}
\end{spacing}

\begin{spacing}{.7}
\begin{longtable}{c*{4}{S[round-mode=places, round-precision=6, table-format=1.9, table-number-alignment=center]}}\\
\caption{Propane. Cartesian coordinates ({\AA}) from a MP2/cc-pVTZ optimization in gas phase.}\\
\hline
\hline
{Atom}  &  {$X$} & {$Y$} & {$Z$}  \\
\hline
C  &    0.000000	&	1.262049	&	-0.260597	\\
H  &   -0.879450	&	1.293529	&	-0.90363	\\
H  &    0.879450	&	1.293529	&	-0.90363	\\
H  &    0.000000	&	2.162724	&	0.351008	\\
C  &    0.000000	&	0.000000	&	0.591572	\\
H  &   -0.873382	&	0.000000	&	1.245115	\\
H  &    0.873382	&	0.000000	&	1.245115	\\
C  &    0.000000	&	-1.262049	&	-0.260597	\\
H  &    0.879450	&	-1.293529	&	-0.90363	\\
H  &    0.000000	&	-2.162724	&	0.351008	\\
H  &   -0.879450	&	-1.293529	&	-0.90363	\\
\hline
\hline
\end{longtable}
\end{spacing}

\clearpage
\begin{spacing}{.7}
\begin{longtable}{c*{4}{S[round-mode=places, round-precision=6, table-format=1.9, table-number-alignment=center]}}\\
\caption{Acetone. Cartesian coordinates ({\AA}) from a MP2/cc-pVTZ optimization in gas phase.}\\
\hline
\hline
{Atom}  &  {$X$} & {$Y$} & {$Z$}  \\
\hline
C   &    0.000000	&	0.000000	&	0.183652	\\
O   &    0.000000	&	0.000000	&	1.401041	\\
C   &    0.000000	&	1.280895	&	-0.614646	\\
H   &    0.855237	&	1.30254	&	-1.289686	\\
H   &   -0.896713	&	1.33167	&	-1.23228	\\
H   &    0.034001	&	2.134113	&	0.054723	\\
C   &    0.000000	&	-1.280895	&	-0.614646	\\
H   &   -0.855237	&	-1.30254	&	-1.289686	\\
H   &    0.896713	&	-1.33167	&	-1.23228	\\
H   &   -0.034001	&	-2.134113	&	0.054723	\\
\hline
\hline
\end{longtable}
\end{spacing}

\begin{spacing}{.7}
\begin{longtable}{c*{3}{S[round-mode=places, round-precision=6, table-format=1.9, table-number-alignment=center]}}\\
\caption{Benzene. Cartesian coordinates ({\AA}) from a MP2/cc-pVTZ optimization in gas phase.}\\
\hline
\hline
{Atom}  &  {$X$} & {$Y$} & {$Z$}  \\
\hline
C   &    0.799830	&	1.141275	&	0.000001	\\
C   &    1.388288	&	-0.122035	&	-0.00001	\\
C   &    0.588458	&	-1.26331	&	0.000005	\\
C   &   -0.799829	&	-1.141275	&	-0.000002	\\
C   &   -1.388288	&	0.122036	&	-0.000004	\\
C   &   -0.588458	&	1.26331	&	0.000006	\\
H   &    1.420489	&	2.026893	&	-0.000002	\\
H   &    2.465586	&	-0.216736	&	0.000008	\\
H   &    1.045097	&	-2.243626	&	0.000013	\\
H   &   -1.420492	&	-2.026891	&	0.00001	\\
H   &   -2.465586	&	0.216732	&	-0.000013	\\
H   &   -1.045095	&	2.243627	&	0.000011	\\
\hline
\hline
\end{longtable}
\end{spacing}

\begin{spacing}{.7}
\begin{longtable}{c*{3}{S[round-mode=places, round-precision=6, table-format=1.9, table-number-alignment=center]}}\\
\caption{Ethene. Cartesian coordinates ({\AA}) from a MP2/cc-pVTZ optimization in gas phase.}\\
\hline
\hline
{Atom}  &  {$X$} & {$Y$} & {$Z$}  \\
\hline
C   &     0.000000	&	0.666027	&	0.000000	\\
H   &    -0.922813	&	1.227957	&	0.000000	\\
H   &     0.922807	&	1.227966	&	0.000000	\\
C   &     0.000000	&	-0.666027	&	0.000000	\\
H    &    0.922813	&	-1.227957	&	0.000000	\\
H    &   -0.922807	&	-1.227966	&	0.000000	\\
\hline
\hline
\end{longtable}
\end{spacing}

\begin{spacing}{.7}
\begin{longtable}{c*{3}{S[round-mode=places, round-precision=6, table-format=1.9, table-number-alignment=center]}}\\
\caption{Ethyne. Cartesian coordinates ({\AA}) from a MP2/cc-pVTZ optimization in gas phase.}\\
\hline
\hline
{Atom}  &  {$X$} & {$Y$} & {$Z$}  \\
\hline
C    &    0.000000	&	0.000000	&	0.605688	\\
H    &    0.000000	&	0.000000	&	1.667203	\\
C    &    0.000000	&	0.000000	&	-0.605688	\\
H    &    0.000000	&	0.000000	&	-1.667203	\\
\hline
\hline
\end{longtable}
\end{spacing}

\clearpage

\begin{spacing}{0.7}
\scriptsize
\begin{longtable}{c*{3}{S[round-mode=places, round-precision=6, table-format=1.9, table-number-alignment=center]}}\\
\caption{C$_{84}$. Cartesian coordinates ({\AA}) from a B3LYP/cc-pVDZ optimization in gas phase. }\\
\hline
\hline
{Atom}  &  {$X$} & {$Y$} & {$Z$}  \\
\hline
C	&	0.850824	&	1.191837	&	3.756218	\\
C	&	1.594109	&	2.203994	&	2.993324	\\
C	&	1.469193	&	3.814692	&	1.122555	\\
C	&	0.86032	&	3.248918	&	2.272221	\\
C	&	2.876388	&	1.834093	&	2.512569	\\
C	&	1.356289	&	3.8122	&	-1.240367	\\
C	&	0.684149	&	4.225482	&	-0.025745	\\
C	&	3.233044	&	2.214966	&	-1.587365	\\
C	&	2.625623	&	3.270783	&	-0.86288	\\
C	&	4.080531	&	0.123894	&	1.393874	\\
C	&	3.370843	&	0.471346	&	2.617396	\\
C	&	2.704546	&	3.31615	&	0.592302	\\
C	&	4.00146	&	1.193041	&	-0.855935	\\
C	&	4.089323	&	1.282723	&	0.561561	\\
C	&	3.402501	&	2.348801	&	1.282157	\\
C	&	1.385698	&	0.125692	&	-3.894543	\\
C	&	2.575691	&	1.842657	&	-2.796408	\\
C	&	2.647426	&	0.481673	&	-3.289782	\\
C	&	1.286425	&	2.342174	&	-3.142731	\\
C	&	0.607402	&	3.291394	&	-2.340422	\\
C	&	0.55464	&	1.285804	&	-3.836526	\\
C	&	-0.850824	&	1.191837	&	-3.756218	\\
C	&	-1.594109	&	2.203994	&	-2.993324	\\
C	&	-1.469193	&	3.814692	&	-1.122555	\\
C	&	-0.86032	&	3.248918	&	-2.272221	\\
C	&	-2.876388	&	1.834093	&	-2.512569	\\
C	&	-1.356289	&	3.8122	&	1.240367	\\
C	&	-0.684149	&	4.225482	&	0.025745	\\
C	&	-3.233044	&	2.214966	&	1.587365	\\
C	&	-2.625623	&	3.270783	&	0.86288	\\
C	&	-4.080531	&	0.123894	&	-1.393874	\\
C	&	-3.370843	&	0.471346	&	-2.617396	\\
C	&	-2.704546	&	3.31615	&	-0.592302	\\
C	&	-4.00146	&	1.193041	&	0.855935	\\
C	&	-4.089323	&	1.282723	&	-0.561561	\\
C	&	-3.402501	&	2.348801	&	-1.282157	\\
C	&	-1.385698	&	0.125692	&	3.894543	\\
C	&	-2.575691	&	1.842657	&	2.796408	\\
C	&	-2.647426	&	0.481673	&	3.289782	\\
C	&	-1.286425	&	2.342174	&	3.142731	\\
C	&	-0.607402	&	3.291394	&	2.340422	\\
C	&	-0.55464	&	1.285804	&	3.836526	\\
C	&	0.850824	&	-1.191837	&	-3.756218	\\
C	&	1.594109	&	-2.203994	&	-2.993324	\\
C	&	1.469193	&	-3.814692	&	-1.122555	\\
C	&	0.86032	&	-3.248918	&	-2.272221	\\
C	&	2.876388	&	-1.834093	&	-2.512569	\\
C	&	1.356289	&	-3.8122	&	1.240367	\\
C	&	0.684149	&	-4.225482	&	0.025745	\\
C	&	3.233044	&	-2.214966	&	1.587365	\\
C	&	2.625623	&	-3.270783	&	0.86288	\\
C	&	4.080531	&	-0.123894	&	-1.393874	\\
C	&	3.370843	&	-0.471346	&	-2.617396	\\
C	&	2.704546	&	-3.31615	&	-0.592302	\\
C	&	4.00146	&	-1.193041	&	0.855935	\\
C	&	4.089323	&	-1.282723	&	-0.561561	\\
C	&	3.402501	&	-2.348801	&	-1.282157	\\
C	&	1.385698	&	-0.125692	&	3.894543	\\
C	&	2.575691	&	-1.842657	&	2.796408	\\
C	&	2.647426	&	-0.481673	&	3.289782	\\
C	&	1.286425	&	-2.342174	&	3.142731	\\
C	&	0.607402	&	-3.291394	&	2.340422	\\
C	&	0.55464	&	-1.285804	&	3.836526	\\
C	&	-0.850824	&	-1.191837	&	3.756218	\\
C	&	-1.594109	&	-2.203994	&	2.993324	\\
C	&	-1.469193	&	-3.814692	&	1.122555	\\
C	&	-0.86032	&	-3.248918	&	2.272221	\\
C	&	-2.876388	&	-1.834093	&	2.512569	\\
C	&	-1.356289	&	-3.8122	&	-1.240367	\\
C	&	-0.684149	&	-4.225482	&	-0.025745	\\
C	&	-3.233044	&	-2.214966	&	-1.587365	\\
C	&	-2.625623	&	-3.270783	&	-0.86288	\\
C	&	-4.080531	&	-0.123894	&	1.393874	\\
C	&	-3.370843	&	-0.471346	&	2.617396	\\
C	&	-2.704546	&	-3.31615	&	0.592302	\\
C	&	-4.00146	&	-1.193041	&	-0.855935	\\
C	&	-4.089323	&	-1.282723	&	0.561561	\\
C	&	-3.402501	&	-2.348801	&	1.282157	\\
C	&	-1.385698	&	-0.125692	&	-3.894543	\\
C	&	-2.575691	&	-1.842657	&	-2.796408	\\
C	&	-2.647426	&	-0.481673	&	-3.289782	\\
C	&	-1.286425	&	-2.342174	&	-3.142731	\\
C	&	-0.607402	&	-3.291394	&	-2.340422	\\
C	&	-0.55464	&	-1.285804	&	-3.836526	\\
\hline
\hline
\end{longtable}
\end{spacing}

\clearpage

\begin{spacing}{0.7}
\scriptsize
\begin{longtable}{c*{3}{S[round-mode=places, round-precision=6, table-format=1.9, table-number-alignment=center]}}\\
\caption{C$_{82}$. Cartesian coordinates ({\AA}) from a B3LYP/cc-pVDZ optimization in gas phase.}\\
\hline
\hline
{Atom}  &  {$X$} & {$Y$} & {$Z$}  \\
\hline
C	&	0.77229	&	1.168739	&	3.713628	\\
C	&	0.627914	&	-1.331014	&	3.653903	\\
C	&	1.382567	&	-0.12321	&	3.770221	\\
C	&	1.33552	&	4.180822	&	-0.051153	\\
C	&	1.921455	&	3.641393	&	-1.265046	\\
C	&	1.117188	&	3.444169	&	-2.389943	\\
C	&	1.218836	&	2.210233	&	-3.107357	\\
C	&	2.880006	&	2.652654	&	-0.887293	\\
C	&	3.046453	&	1.457617	&	-1.616185	\\
C	&	2.161013	&	1.224025	&	-2.759872	\\
C	&	0.131965	&	-1.846836	&	-3.5539	\\
C	&	0.527935	&	-0.501337	&	-3.630185	\\
C	&	1.879648	&	-0.138266	&	-3.198339	\\
C	&	2.732448	&	-1.152297	&	-2.73401	\\
C	&	1.660029	&	2.06863	&	2.98232	\\
C	&	1.15421	&	3.14425	&	2.217226	\\
C	&	1.936896	&	3.510857	&	1.078569	\\
C	&	2.95837	&	2.64349	&	0.573182	\\
C	&	3.379414	&	1.502602	&	1.276988	\\
C	&	2.802023	&	1.322622	&	2.569494	\\
C	&	3.210702	&	-1.076977	&	2.47287	\\
C	&	2.665633	&	-0.003533	&	3.121502	\\
C	&	3.620864	&	-0.922489	&	-1.59129	\\
C	&	3.670347	&	0.321656	&	-0.933802	\\
C	&	3.850929	&	0.341494	&	0.520968	\\
C	&	3.905687	&	-0.900644	&	1.201968	\\
C	&	2.911895	&	-3.162847	&	-1.547696	\\
C	&	3.731036	&	-2.166323	&	-0.880588	\\
C	&	3.815494	&	-2.146149	&	0.509526	\\
C	&	2.160346	&	-4.068241	&	-0.797535	\\
C	&	0.272367	&	-3.860684	&	-2.369316	\\
C	&	1.030896	&	-2.872366	&	-3.116519	\\
C	&	2.323398	&	-2.527651	&	-2.710039	\\
C	&	0.817913	&	-4.420769	&	-1.212782	\\
C	&	0.000289	&	-4.52984	&	-0.011239	\\
C	&	2.986744	&	-3.05752	&	1.294643	\\
C	&	2.523762	&	-2.362083	&	2.458783	\\
C	&	1.189158	&	-2.481537	&	2.927198	\\
C	&	2.15897	&	-3.970453	&	0.656166	\\
C	&	0.817913	&	-4.170524	&	1.14291	\\
C	&	0.292896	&	-3.40993	&	2.222025	\\
C	&	-0.77229	&	-1.168739	&	3.713628	\\
C	&	-0.627914	&	1.331014	&	3.653903	\\
C	&	-1.382567	&	0.12321	&	3.770221	\\
C	&	-1.33552	&	-4.180822	&	-0.051153	\\
C	&	-1.921455	&	-3.641393	&	-1.265046	\\
C	&	-1.117188	&	-3.444169	&	-2.389943	\\
C	&	-1.218836	&	-2.210233	&	-3.107357	\\
C	&	-2.880006	&	-2.652654	&	-0.887293	\\
C	&	-3.046453	&	-1.457617	&	-1.616185	\\
C	&	-2.161013	&	-1.224025	&	-2.759872	\\
C	&	-0.131965	&	1.846836	&	-3.5539	\\
C	&	-0.527935	&	0.501337	&	-3.630185	\\
C	&	-1.879648	&	0.138266	&	-3.198339	\\
C	&	-2.732448	&	1.152297	&	-2.73401	\\
C	&	-1.660029	&	-2.06863	&	2.98232	\\
C	&	-1.15421	&	-3.14425	&	2.217226	\\
C	&	-1.936896	&	-3.510857	&	1.078569	\\
C	&	-2.95837	&	-2.64349	&	0.573182	\\
C	&	-3.379414	&	-1.502602	&	1.276988	\\
C	&	-2.802023	&	-1.322622	&	2.569494	\\
C	&	-3.210702	&	1.076977	&	2.47287	\\
C	&	-2.665633	&	0.003533	&	3.121502	\\
C	&	-3.620864	&	0.922489	&	-1.59129	\\
C	&	-3.670347	&	-0.321656	&	-0.933802	\\
C	&	-3.850929	&	-0.341494	&	0.520968	\\
C	&	-3.905687	&	0.900644	&	1.201968	\\
C	&	-2.911895	&	3.162847	&	-1.547696	\\
C	&	-3.731036	&	2.166323	&	-0.880588	\\
C	&	-3.815494	&	2.146149	&	0.509526	\\
C	&	-2.160346	&	4.068241	&	-0.797535	\\
C	&	-0.272367	&	3.860684	&	-2.369316	\\
C	&	-1.030896	&	2.872366	&	-3.116519	\\
C	&	-2.323398	&	2.527651	&	-2.710039	\\
C	&	-0.817913	&	4.420769	&	-1.212782	\\
C	&	-0.000289	&	4.52984	&	-0.011239	\\
C	&	-2.986744	&	3.05752	&	1.294643	\\
C	&	-2.523762	&	2.362083	&	2.458783	\\
C	&	-1.189158	&	2.481537	&	2.927198	\\
C	&	-2.15897	&	3.970453	&	0.656166	\\
C	&	-0.817913	&	4.170524	&	1.14291	\\
C	&	-0.292896	&	3.40993	&	2.222025	\\
\hline
\hline
\end{longtable}
\end{spacing}

\clearpage

\begin{spacing}{0.7}
\scriptsize
\begin{longtable}{c*{3}{S[round-mode=places, round-precision=6, table-format=1.9, table-number-alignment=center]}}\\
\caption{C$_{80}$. Cartesian coordinates ({\AA}) from a B3LYP/cc-pVDZ optimization in gas phase.}\\
\hline
\hline
{Atom}  &  {$X$} & {$Y$} & {$Z$}  \\
\hline
C	&	1.829633	&	1.720799	&	2.977323	\\
C	&	1.829704	&	0.370359	&	3.418718	\\
C	&	3.08073	&	-0.239365	&	2.99614	\\
C	&	3.080488	&	-1.55771	&	2.57002	\\
C	&	1.829857	&	-2.300156	&	2.556779	\\
C	&	1.829645	&	3.366193	&	0.704298	\\
C	&	3.080543	&	2.775546	&	1.153587	\\
C	&	3.080364	&	1.962942	&	2.275683	\\
C	&	3.872246	&	0.750923	&	2.296302	\\
C	&	3.872367	&	2.416625	&	-0.004525	\\
C	&	4.631034	&	1.236289	&	-0.002454	\\
C	&	4.630942	&	0.384257	&	1.174512	\\
C	&	1.829573	&	3.363631	&	-0.716533	\\
C	&	3.080454	&	2.771423	&	-1.163848	\\
C	&	1.829335	&	1.710142	&	-2.983667	\\
C	&	3.080133	&	1.954788	&	-2.283018	\\
C	&	4.630972	&	-1.001145	&	-0.724965	\\
C	&	4.630823	&	0.380048	&	-1.176364	\\
C	&	3.871994	&	0.742692	&	-2.29936	\\
C	&	3.872149	&	-1.957191	&	-1.417047	\\
C	&	3.080694	&	-2.923455	&	0.697867	\\
C	&	3.872303	&	-1.952111	&	1.423644	\\
C	&	4.63103	&	-0.998531	&	0.728061	\\
C	&	3.080628	&	-2.925929	&	-0.687723	\\
C	&	1.829827	&	-3.137396	&	1.408795	\\
C	&	1.829682	&	-3.142401	&	-1.397749	\\
C	&	1.829606	&	-2.309301	&	-2.548757	\\
C	&	3.080447	&	-0.250082	&	-2.995621	\\
C	&	1.82935	&	0.358124	&	-3.420203	\\
C	&	3.080238	&	-1.566907	&	-2.564785	\\
C	&	0.614607	&	2.372109	&	2.614633	\\
C	&	0.614693	&	-0.369576	&	3.510488	\\
C	&	0.614745	&	-1.75397	&	3.064259	\\
C	&	0.614525	&	3.224562	&	1.436086	\\
C	&	0.614383	&	3.219401	&	-1.447691	\\
C	&	0.614339	&	2.362761	&	-2.623188	\\
C	&	0.614613	&	-3.455161	&	-0.721044	\\
C	&	0.614684	&	-3.452573	&	0.733335	\\
C	&	0.61443	&	-1.764908	&	-3.058039	\\
C	&	0.614329	&	-0.382124	&	-3.5092	\\
C	&	-1.829633	&	-1.720799	&	-2.977323	\\
C	&	-1.829704	&	-0.370359	&	-3.418718	\\
C	&	-3.08073	&	0.239365	&	-2.99614	\\
C	&	-3.080488	&	1.55771	&	-2.57002	\\
C	&	-1.829857	&	2.300156	&	-2.556779	\\
C	&	-1.829645	&	-3.366193	&	-0.704298	\\
C	&	-3.080543	&	-2.775546	&	-1.153587	\\
C	&	-3.080364	&	-1.962942	&	-2.275683	\\
C	&	-3.872246	&	-0.750923	&	-2.296302	\\
C	&	-3.872367	&	-2.416625	&	0.004525	\\
C	&	-4.631034	&	-1.236289	&	0.002454	\\
C	&	-4.630942	&	-0.384257	&	-1.174512	\\
C	&	-1.829573	&	-3.363631	&	0.716533	\\
C	&	-3.080454	&	-2.771423	&	1.163848	\\
C	&	-1.829335	&	-1.710142	&	2.983667	\\
C	&	-3.080133	&	-1.954788	&	2.283018	\\
C	&	-4.630972	&	1.001145	&	0.724965	\\
C	&	-4.630823	&	-0.380048	&	1.176364	\\
C	&	-3.871994	&	-0.742692	&	2.29936	\\
C	&	-3.872149	&	1.957191	&	1.417047	\\
C	&	-3.080694	&	2.923455	&	-0.697867	\\
C	&	-3.872303	&	1.952111	&	-1.423644	\\
C	&	-4.63103	&	0.998531	&	-0.728061	\\
C	&	-3.080628	&	2.925929	&	0.687723	\\
C	&	-1.829827	&	3.137396	&	-1.408795	\\
C	&	-1.829682	&	3.142401	&	1.397749	\\
C	&	-1.829606	&	2.309301	&	2.548757	\\
C	&	-3.080447	&	0.250082	&	2.995621	\\
C	&	-1.82935	&	-0.358124	&	3.420203	\\
C	&	-3.080238	&	1.566907	&	2.564785	\\
C	&	-0.614607	&	-2.372109	&	-2.614633	\\
C	&	-0.614693	&	0.369576	&	-3.510488	\\
C	&	-0.614745	&	1.75397	&	-3.064259	\\
C	&	-0.614525	&	-3.224562	&	-1.436086	\\
C	&	-0.614383	&	-3.219401	&	1.447691	\\
C	&	-0.614339	&	-2.362761	&	2.623188	\\
C	&	-0.614613	&	3.455161	&	0.721044	\\
C	&	-0.614684	&	3.452573	&	-0.733335	\\
C	&	-0.61443	&	1.764908	&	3.058039	\\
C	&	-0.614329	&	0.382124	&	3.5092	\\
\hline
\hline
\end{longtable}
\end{spacing}
\clearpage

\begin{spacing}{0.7}
\scriptsize
\begin{longtable}{c*{3}{S[round-mode=places, round-precision=6, table-format=1.9, table-number-alignment=center]}}\\
\caption{C$_{78}$. Cartesian coordinates ({\AA}) from a B3LYP/cc-pVDZ optimization in gas phase.}\\
\hline
\hline
{Atom}  &  {$X$} & {$Y$} & {$Z$}  \\
\hline
C	&	0.701468	&	4.221475	&	1.180083	\\
C	&	0.724689	&	4.068204	&	-1.268176	\\
C	&	1.421456	&	4.114517	&	-0.066055	\\
C	&	2.354083	&	1.186371	&	-2.994727	\\
C	&	2.320231	&	2.383077	&	-2.201464	\\
C	&	1.172132	&	3.202407	&	-2.341767	\\
C	&	0.000000	&	2.723585	&	-3.034845	\\
C	&	0.000000	&	1.463792	&	-3.69989	\\
C	&	1.224721	&	0.734484	&	-3.719235	\\
C	&	3.024887	&	2.383817	&	-0.90868	\\
C	&	2.562862	&	3.228516	&	0.131481	\\
C	&	3.062488	&	0.000000	&	-2.545105	\\
C	&	3.696039	&	0.000000	&	-1.325787	\\
C	&	3.676284	&	1.184686	&	-0.488329	\\
C	&	3.620178	&	0.731681	&	0.856788	\\
C	&	2.96299	&	1.469665	&	1.87136	\\
C	&	2.542216	&	2.785175	&	1.501226	\\
C	&	1.416344	&	3.421872	&	2.154032	\\
C	&	0.727367	&	2.720694	&	3.130935	\\
C	&	2.255561	&	0.737148	&	2.929288	\\
C	&	1.177271	&	1.410041	&	3.555368	\\
C	&	0.000000	&	0.687574	&	3.961371	\\
C	&	-0.701468	&	4.221475	&	1.180083	\\
C	&	-0.724689	&	4.068204	&	-1.268176	\\
C	&	-1.421456	&	4.114517	&	-0.066055	\\
C	&	-2.354083	&	1.186371	&	-2.994727	\\
C	&	-2.320231	&	2.383077	&	-2.201464	\\
C	&	-1.172132	&	3.202407	&	-2.341767	\\
C	&	-1.224721	&	0.734484	&	-3.719235	\\
C	&	-3.024887	&	2.383817	&	-0.90868	\\
C	&	-2.562862	&	3.228516	&	0.131481	\\
C	&	-3.062488	&	0.000000	&	-2.545105	\\
C	&	-3.696039	&	0.000000	&	-1.325787	\\
C	&	-3.676284	&	1.184686	&	-0.488329	\\
C	&	-3.620178	&	0.731681	&	0.856788	\\
C	&	-2.96299	&	1.469665	&	1.87136	\\
C	&	-2.542216	&	2.785175	&	1.501226	\\
C	&	-1.416344	&	3.421872	&	2.154032	\\
C	&	-0.727367	&	2.720694	&	3.130935	\\
C	&	-2.255561	&	0.737148	&	2.929288	\\
C	&	-1.177271	&	1.410041	&	3.555368	\\
C	&	0.701468	&	-4.221475	&	1.180083	\\
C	&	0.724689	&	-4.068204	&	-1.268176	\\
C	&	1.421456	&	-4.114517	&	-0.066055	\\
C	&	2.354083	&	-1.186371	&	-2.994727	\\
C	&	2.320231	&	-2.383077	&	-2.201464	\\
C	&	1.172132	&	-3.202407	&	-2.341767	\\
C	&	0.000000	&	-2.723585	&	-3.034845	\\
C	&	0.000000	&	-1.463792	&	-3.69989	\\
C	&	1.224721	&	-0.734484	&	-3.719235	\\
C	&	3.024887	&	-2.383817	&	-0.90868	\\
C	&	2.562862	&	-3.228516	&	0.131481	\\
C	&	3.676284	&	-1.184686	&	-0.488329	\\
C	&	3.620178	&	-0.731681	&	0.856788	\\
C	&	2.96299	&	-1.469665	&	1.87136	\\
C	&	2.542216	&	-2.785175	&	1.501226	\\
C	&	1.416344	&	-3.421872	&	2.154032	\\
C	&	0.727367	&	-2.720694	&	3.130935	\\
C	&	2.255561	&	-0.737148	&	2.929288	\\
C	&	1.177271	&	-1.410041	&	3.555368	\\
C	&	0.000000	&	-0.687574	&	3.961371	\\
C	&	-0.701468	&	-4.221475	&	1.180083	\\
C	&	-0.724689	&	-4.068204	&	-1.268176	\\
C	&	-1.421456	&	-4.114517	&	-0.066055	\\
C	&	-2.354083	&	-1.186371	&	-2.994727	\\
C	&	-2.320231	&	-2.383077	&	-2.201464	\\
C	&	-1.172132	&	-3.202407	&	-2.341767	\\
C	&	-1.224721	&	-0.734484	&	-3.719235	\\
C	&	-3.024887	&	-2.383817	&	-0.90868	\\
C	&	-2.562862	&	-3.228516	&	0.131481	\\
C	&	-3.676284	&	-1.184686	&	-0.488329	\\
C	&	-3.620178	&	-0.731681	&	0.856788	\\
C	&	-2.96299	&	-1.469665	&	1.87136	\\
C	&	-2.542216	&	-2.785175	&	1.501226	\\
C	&	-1.416344	&	-3.421872	&	2.154032	\\
C	&	-0.727367	&	-2.720694	&	3.130935	\\
C	&	-2.255561	&	-0.737148	&	2.929288	\\
C	&	-1.177271	&	-1.410041	&	3.555368	\\
\hline
\hline
\end{longtable}
\end{spacing}

\clearpage
\begin{spacing}{0.7}
\scriptsize
\begin{longtable}{c*{3}{S[round-mode=places, round-precision=6, table-format=1.9, table-number-alignment=center]}}\\
\caption{C$_{70}$. Cartesian coordinates ({\AA}) from a B3LYP/cc-pVDZ optimization in gas phase.}\\
\hline
\hline
{Atom}  &  {$X$} & {$Y$} & {$Z$}  \\
\hline
C	&	0.000000	&	1.456704	&	-3.257344	\\
C	&	1.208195	&	0.718264	&	-3.398391	\\
C	&	0.000000	&	2.648096	&	-2.392587	\\
C	&	1.208057	&	3.010026	&	-1.733514	\\
C	&	2.446002	&	2.290621	&	-1.969677	\\
C	&	2.446382	&	1.165356	&	-2.787364	\\
C	&	3.227105	&	0.000000	&	-2.418839	\\
C	&	0.000000	&	3.547165	&	0.37902	\\
C	&	1.208048	&	3.453518	&	-0.367004	\\
C	&	2.445986	&	3.010865	&	0.246775	\\
C	&	3.226562	&	2.300124	&	-0.747273	\\
C	&	3.97558	&	1.176071	&	-0.382161	\\
C	&	3.975912	&	0.000000	&	-1.236999	\\
C	&	0.000000	&	3.093362	&	1.779581	\\
C	&	1.208349	&	2.578932	&	2.327487	\\
C	&	2.445576	&	2.580819	&	1.569697	\\
C	&	3.226237	&	1.421819	&	1.956461	\\
C	&	3.975464	&	0.726976	&	1.000574	\\
C	&	1.208232	&	1.416433	&	3.17202	\\
C	&	2.446207	&	0.695671	&	2.940336	\\
C	&	0.000000	&	0.73623	&	3.491232	\\
C	&	-1.208195	&	0.718264	&	-3.398391	\\
C	&	-1.208057	&	3.010026	&	-1.733514	\\
C	&	-2.446002	&	2.290621	&	-1.969677	\\
C	&	-2.446382	&	1.165356	&	-2.787364	\\
C	&	-3.227105	&	0.000000	&	-2.418839	\\
C	&	-1.208048	&	3.453518	&	-0.367004	\\
C	&	-2.445986	&	3.010865	&	0.246775	\\
C	&	-3.226562	&	2.300124	&	-0.747273	\\
C	&	-3.97558	&	1.176071	&	-0.382161	\\
C	&	-3.975912	&	0.000000	&	-1.236999	\\
C	&	-1.208349	&	2.578932	&	2.327487	\\
C	&	-2.445576	&	2.580819	&	1.569697	\\
C	&	-3.226237	&	1.421819	&	1.956461	\\
C	&	-3.975464	&	0.726976	&	1.000574	\\
C	&	-1.208232	&	1.416433	&	3.17202	\\
C	&	-2.446207	&	0.695671	&	2.940336	\\
C	&	0.000000	&	-1.456704	&	-3.257344	\\
C	&	1.208195	&	-0.718264	&	-3.398391	\\
C	&	0.000000	&	-2.648096	&	-2.392587	\\
C	&	1.208057	&	-3.010026	&	-1.733514	\\
C	&	2.446002	&	-2.290621	&	-1.969677	\\
C	&	2.446382	&	-1.165356	&	-2.787364	\\
C	&	0.000000	&	-3.547165	&	0.37902	\\
C	&	1.208048	&	-3.453518	&	-0.367004	\\
C	&	2.445986	&	-3.010865	&	0.246775	\\
C	&	3.226562	&	-2.300124	&	-0.747273	\\
C	&	3.97558	&	-1.176071	&	-0.382161	\\
C	&	0.000000	&	-3.093362	&	1.779581	\\
C	&	1.208349	&	-2.578932	&	2.327487	\\
C	&	2.445576	&	-2.580819	&	1.569697	\\
C	&	3.226237	&	-1.421819	&	1.956461	\\
C	&	3.975464	&	-0.726976	&	1.000574	\\
C	&	1.208232	&	-1.416433	&	3.17202	\\
C	&	2.446207	&	-0.695671	&	2.940336	\\
C	&	0.000000	&	-0.73623	&	3.491232	\\
C	&	-1.208195	&	-0.718264	&	-3.398391	\\
C	&	-1.208057	&	-3.010026	&	-1.733514	\\
C	&	-2.446002	&	-2.290621	&	-1.969677	\\
C	&	-2.446382	&	-1.165356	&	-2.787364	\\
C	&	-1.208048	&	-3.453518	&	-0.367004	\\
C	&	-2.445986	&	-3.010865	&	0.246775	\\
C	&	-3.226562	&	-2.300124	&	-0.747273	\\
C	&	-3.97558	&	-1.176071	&	-0.382161	\\
C	&	-1.208349	&	-2.578932	&	2.327487	\\
C	&	-2.445576	&	-2.580819	&	1.569697	\\
C	&	-3.226237	&	-1.421819	&	1.956461	\\
C	&	-3.975464	&	-0.726976	&	1.000574	\\
C	&	-1.208232	&	-1.416433	&	3.17202	\\
C	&	-2.446207	&	-0.695671	&	2.940336	\\
\hline
\hline
\end{longtable}
\end{spacing}

\begin{spacing}{0.65}
\scriptsize
\begin{longtable}{c*{3}{S[round-mode=places, round-precision=6, table-format=1.9, table-number-alignment=center]}}\\
\caption{C$_{60}$. Cartesian coordinates ({\AA}) from a B3LYP/cc-pVDZ optimization in gas phase.}\\
\hline
\hline
{Atom}  &  {$X$} & {$Y$} & {$Z$}  \\
\hline
C	&	1.177418	&	0.382566	&	3.332579	\\
C	&	0.000000	&	1.238011	&	3.332579	\\
C	&	2.308303	&	0.750013	&	2.597685	\\
C	&	0.000000	&	2.427094	&	2.597685	\\
C	&	1.177418	&	2.80966	&	1.832552	\\
C	&	2.308303	&	1.988024	&	1.832552	\\
C	&	3.035988	&	1.751585	&	0.594541	\\
C	&	3.485722	&	0.367447	&	0.594541	\\
C	&	0.727685	&	1.001572	&	-3.332579	\\
C	&	0.698925	&	2.965132	&	-1.832552	\\
C	&	1.42661	&	1.96356	&	-2.597685	\\
C	&	3.035988	&	0.251559	&	-1.832552	\\
C	&	2.604028	&	1.580994	&	-1.832552	\\
C	&	2.604028	&	2.346126	&	-0.594541	\\
C	&	1.42661	&	3.201571	&	-0.594541	\\
C	&	0.727685	&	3.428665	&	0.594541	\\
C	&	-1.177418	&	0.382566	&	3.332579	\\
C	&	-2.308303	&	0.750013	&	2.597685	\\
C	&	-1.177418	&	2.80966	&	1.832552	\\
C	&	-2.308303	&	1.988024	&	1.832552	\\
C	&	-3.035988	&	1.751585	&	0.594541	\\
C	&	-3.485722	&	0.367447	&	0.594541	\\
C	&	-0.727685	&	1.001572	&	-3.332579	\\
C	&	-0.698925	&	2.965132	&	-1.832552	\\
C	&	-1.42661	&	1.96356	&	-2.597685	\\
C	&	-3.035988	&	0.251559	&	-1.832552	\\
C	&	-2.604028	&	1.580994	&	-1.832552	\\
C	&	-2.604028	&	2.346126	&	-0.594541	\\
C	&	-1.42661	&	3.201571	&	-0.594541	\\
C	&	-0.727685	&	3.428665	&	0.594541	\\
C	&	1.177418	&	-0.382566	&	-3.332579	\\
C	&	0.000000	&	-1.238011	&	-3.332579	\\
C	&	2.308303	&	-0.750013	&	-2.597685	\\
C	&	0.000000	&	-2.427094	&	-2.597685	\\
C	&	1.177418	&	-2.80966	&	-1.832552	\\
C	&	2.308303	&	-1.988024	&	-1.832552	\\
C	&	3.035988	&	-1.751585	&	-0.594541	\\
C	&	3.485722	&	-0.367447	&	-0.594541	\\
C	&	0.727685	&	-1.001572	&	3.332579	\\
C	&	0.698925	&	-2.965132	&	1.832552	\\
C	&	1.42661	&	-1.96356	&	2.597685	\\
C	&	3.035988	&	-0.251559	&	1.832552	\\
C	&	2.604028	&	-1.580994	&	1.832552	\\
C	&	2.604028	&	-2.346126	&	0.594541	\\
C	&	1.42661	&	-3.201571	&	0.594541	\\
C	&	0.727685	&	-3.428665	&	-0.594541	\\
C	&	-1.177418	&	-0.382566	&	-3.332579	\\
C	&	-2.308303	&	-0.750013	&	-2.597685	\\
C	&	-1.177418	&	-2.80966	&	-1.832552	\\
C	&	-2.308303	&	-1.988024	&	-1.832552	\\
C	&	-3.035988	&	-1.751585	&	-0.594541	\\
C	&	-3.485722	&	-0.367447	&	-0.594541	\\
C	&	-0.727685	&	-1.001572	&	3.332579	\\
C	&	-0.698925	&	-2.965132	&	1.832552	\\
C	&	-1.42661	&	-1.96356	&	2.597685	\\
C	&	-3.035988	&	-0.251559	&	1.832552	\\
C	&	-2.604028	&	-1.580994	&	1.832552	\\
C	&	-2.604028	&	-2.346126	&	0.594541	\\
C	&	-1.42661	&	-3.201571	&	0.594541	\\
C	&	-0.727685	&	-3.428665	&	-0.594541	\\
\hline
\hline
\end{longtable}
\end{spacing}